\documentclass[onecolumn]{emulateapj}

\newcommand{\be}{\begin{equation}}
\newcommand{\ee}{\end{equation}}
\newcommand{\kms}{\mbox{km s$^{-1}$}}
\newcommand{\hh}{h$^{-1}$}

\begin{document}


\title{The density field of the 10k zCOSMOS galaxies\footnotemark[1]}


\author{
K.~Kova\v{c}\altaffilmark{2},
S. J.~Lilly\altaffilmark{2},
O.~ Cucciati\altaffilmark{3},
C.~ Porciani\altaffilmark{2,18},
A.~ Iovino\altaffilmark{3},
G.~ Zamorani\altaffilmark{4},
P.~ Oesch\altaffilmark{2},
M.~Bolzonella\altaffilmark{4},
C. ~Knobel\altaffilmark{2},
A. ~Finoguenov\altaffilmark{5},
Y.~ Peng\altaffilmark{2},
C. M.~Carollo\altaffilmark{2},
L.~Pozzetti\altaffilmark{4},
K.~Caputi\altaffilmark{2},
J.~D.~Silverman\altaffilmark{2},
L.~ Tasca\altaffilmark{6},
M.~ Scodeggio\altaffilmark{7},
D.~ Vergani\altaffilmark{7},
E.~ Zucca\altaffilmark{4},
T.~Contini\altaffilmark{8},
J.-P.~ Kneib\altaffilmark{6},
O.~ Le F\`{e}vre\altaffilmark{6},
V.~Mainieri\altaffilmark{9},
A.~ Renzini\altaffilmark{10},
N.~Z.~ Scoville\altaffilmark{11},
P.~ Capak\altaffilmark{12,11},
S.~Bardelli\altaffilmark{4},
A.~Bongiorno\altaffilmark{5},
G.~ Coppa\altaffilmark{4},
S.~ de la Torre\altaffilmark{6},
L.~ de Ravel\altaffilmark{6},
P.~ Franzetti\altaffilmark{7},
B.~ Garilli\altaffilmark{7},
L.~ Guzzo\altaffilmark{3},
P.~ Kampczyk\altaffilmark{2},
F.~Lamareille\altaffilmark{8},
J.-F.~ Le Borgne\altaffilmark{8},
V.~ Le Brun\altaffilmark{6},
C.~Maier\altaffilmark{2},
M.~ Mignoli\altaffilmark{4},
R.~ Pello\altaffilmark{6},
E.~ Perez Montero\altaffilmark{6},
E.~ Ricciardelli\altaffilmark{10},
M.~ Tanaka\altaffilmark{9},
L.~ Tresse\altaffilmark{6},
U.~ Abbas\altaffilmark{6,19},
D.~ Bottini\altaffilmark{7},
A.~ Cappi\altaffilmark{4},
P.~ Cassata\altaffilmark{6},
A.~ Cimatti\altaffilmark{13},
M.~ Fumana\altaffilmark{7},
D.~ Maccagni\altaffilmark{7},
C.~ Marinoni\altaffilmark{14},
H.~ J. McCracken\altaffilmark{15},
P.~ Memeo\altaffilmark{7},
B.~ Meneux\altaffilmark{5,16},
R.~ Scaramella\altaffilmark{17}
A. M. ~Koekemoer\altaffilmark{20}
}

\footnotetext[1]{Based on observations
   obtained at the European Southern Observatory (ESO) Very Large
   Telescope (VLT), Paranal, Chile, as part of the Large Program
   175.A-0839 (the zCOSMOS Spectroscopic Redshift Survey)}
\altaffiltext{2}{Institute of Astronomy, ETH Zurich, 8093 Zurich, Switzerland}
\altaffiltext{3}{INAF Osservatorio Astronomico di Brera, Milan, Italy}
\altaffiltext{4}{INAF Osservatorio Astronomico di Bologna, via  
Ranzani 1, I-40127, Bologna, Italy}
\altaffiltext{5}{Max-Planck-Institut f\"ur extraterrestrische Physik,  
D-84571 Garching, Germany}
\altaffiltext{6}{Laboratoire d'Astrophysique de Marseille, Marseille,  
France}
\altaffiltext{7}{INAF - IASF Milano, Milan, Italy}
\altaffiltext{8}{Laboratoire d'Astrophysique de l'Observatoire Midi- 
Pyr\'en\'ees, Toulouse, France}
\altaffiltext{9}{European Southern Observatory, Karl-Schwarzschild- 
Strasse 2, Garching, D-85748, Germany}
\altaffiltext{10}{Dipartimento di Astronomia, Universita di Padova,  
Padova, Italy}
\altaffiltext{11}{California Institute of Technology, MS 105-24,
Pasadena, CA 91125, USA}
\altaffiltext{12}{Spitzer Science Center, 314-6 Caltech, Pasadena, CA 91125, USA}
\altaffiltext{13}{Dipartimento di Astronomia, Universit\'a di Bologna,  
via Ranzani 1, I-40127, Bologna, Italy}
\altaffiltext{14}{Centre de Physique Theorique, Marseille, Marseille,  
France}
\altaffiltext{15}{Institut d'Astrophysique de Paris, UMR 7095 CNRS,  
Universit\'e Pierre et Marie Curie, 98 bis Boulevard Arago, F-75014  
Paris, France.}
\altaffiltext{16}{Universitats-Sternwarte, Scheinerstrasse 1, D-81679  
Muenchen, Germany}
\altaffiltext{17}{INAF, Osservatorio di Roma, Monteporzio Catone  
(RM), Italy}
\altaffiltext{18}{Argelander Institut f\"ur Astronomie, Auf dem H\"ugel 71, D-53121 Bonn,Germany}
\altaffiltext{19}{ELSA Marie Curie Postdoctoral Fellow, INAF - Osservatorio Astronomico di Torino, 10025 Pino Torinese, Italy}
\altaffiltext{20}{Space Telescope Science Institute, 3700 San Martin Drive, Baltimore, MD 21218, USA}




\begin{abstract} 

We use the  current sample of $\sim$10,000 zCOSMOS  spectra of sources
selected with $I_{\rm  AB} < 22.5$ to define the  density field out to
z$\sim$1, with  much greater resolution  in the radial  dimension than
has been  possible with either photometric redshifts  or weak lensing.
We apply new  algorithms that we have developed  (ZADE) to incorporate
objects  not   yet  observed  spectroscopically   by  modifying  their
photometric redshift probability distributions using the spectroscopic
redshifts  of nearby  galaxies.  This  strategy allows  us to  probe a
broader range of  galaxy environments and reduce the  Poisson noise in
the  density field.  The reconstructed  overdensity field  of  the 10k
zCOSMOS  galaxies  consists  of  cluster-like patterns  surrounded  by
void-like  regions,  extending  up  to  $z  \sim  1$.  Some  of  these
structures are very  large, spanning the $\sim$ 50  \hh Mpc transverse
direction of the COSMOS field and extending up to $\Delta z \sim 0.05$
in  redshift. We present  the three  dimensional overdensity  maps and
compare  the  reconstructed  overdensity  field to  the  independently
identified virialised groups of  galaxies and clusters detected in the
visible and in X-rays. The distribution of the overdense structures is
in general well traced by these virialised structures. A comparison of
the  large  scale structures  in  the zCOSMOS  data  and  in the  mock
catalogues reveals an excellent agreement between the fractions of the
volume enclosed in  structures of all sizes above  a given overdensity
between the data and the mocks in $0.2<z<1$.

\end{abstract}






\section{Introduction}

Although on  scales larger than $\sim$ 100  \hh Mpc the  universe is homogeneous
and  isotropic,  on smaller  scales  structures  in  the universe  are
organised hierarchically.   The full view of the  cosmic structure was
revealed for the first time in the first CfA-II slice \citep{deLapparent.etal.1986},  with  galaxies  distributed in  bubble  like
structures   surrounding  empty   regions.  Later  surveys,
particularly the Sloan Digital Sky Survey \citep[SDSS,][]{York.etal.2000} and
the  two-degree  Field  Galaxy  Redshift Survey  \citep[2dFGRS,][]{Colless.etal.2001}, have confirmed the highly  complex hierarchical picture  of the
universe at $z \sim 0$. Our  universe exhibits a  range of cellular  structures, the
cosmic-web \citep{Bond.etal.1996},  where galaxies define  the structures
 which form patterns  of dense  compact clusters, elongated  filaments and
sheet-like  walls outlining large  and almost  empty voids.  While the
network  of filaments  and  clumps sitting  in  the intersection  of
filaments dominate  the visual impression  of the cosmic web,  most of
the volume is in the underdense regions (voids).

The picture emerging from the  surveys of the higher redshift universe
is limited by  small fields and cosmic variance,  but the richness and
complexity of the cosmic web appears (at least visually) to match that
of the local universe out to z $\sim$ 1.5 \citep[e.g.][]{Scoville.etal.2007a, Marinoni.etal.2008}.   As  redshift  increases,  the peak  in  the
probability  distribution  function of  the  galaxy overdensity  field
shifts  towards lower overdensity values,  and the fraction of the volume in the underdense regions decreases \citep{Marinoni.etal.2008}.

The galaxy  distribution is  believed to be  a (biased) tracer  of the
underlying  smooth and continuous  matter density  field. It  has been
shown by  numerical simulations and by analytical work  that the observed
large scale structure (LSS) pattern  is a natural manifestation of the
gravitational structure formation process through the amplification of
density fluctuations and their subsequent collapse.  The presence of the cosmic web
can be  explained with  the tendency of  the matter  concentrations to
contract and  collapse gravitationally in an  anisotropic manner \citep{Bond.etal.1996, vandeWeygaert&Bertschinger.1996}.

One of the main characteristics  of the cosmic-web is its hierarchical
nature. The  LSS structures cover  a broad range in (over)density
values,  physical  scales and  geometrical  shapes.  The
reconstruction of  the density field from an  observed galaxy sample
should ideally  preserve all  these  features. However, the  samples of  the 
galaxies  are  incomplete, particularly  at  high  redshifts, and  the
sampling is often irregular.  Different galaxy populations evolve with
redshift  in different  ways, and  they  may  trace the underlying mass distribution  in different ways
at different redshifts.  On top of this, one has  to add an uncertainty
in the measured redshifts, magnitudes and other galaxy properties. All of
these points make the reconstruction of the density field in the Universe non trivial.

The reconstruction of a continuous,  presumably smooth field
from a set of  measured discrete data points usually involves the
interpolation and smoothing  of the  data  with some 
 filtering function  into a
continuous map. Unlike the velocity  or temperature fields for which we
sample the field values in  the observed points, for the density field
reconstruction there are no field samples available. The measured data
points (galaxies) are used to define the density field itself. A common feature of all the available approaches to reconstruct the density field is that the results of the reconstructed density field depend on the method used to carry out the reconstruction. Usually a particular scientific application guides the choice of the method to reconstruct the density field, and we briefly review these here.

\subsection{Applications of the density field}

\subsubsection{Extraction of the components of the cosmic web}

An immediate application of the reconstructed density field is the extraction of the four components of the cosmic web (clumps, filaments, sheets and voids). While the human eye can easily distinguish between these shapes, there does not exist an ideal algorithm yet to extract these features in an algorithmically well-defined way. A lot of progress has been made recently, by applying a geometrical classification directly to the density field, such as the multi-scale analysis of the Hessian matrix of the density field \citep{AragonCalvo.etal.2007} or the skeleton analysis of the density field \citep{Novikov.etal.2006, Sousbie.etal.2008}. Alternatively, a classification of the cosmic web can be done using either a linear \citep{Lee&Lee.2008} or non-linear \citep{Hahn.etal.2007a, Hahn.etal.2007b, Forero-Romero.etal.2008} gravitational potential. Some of these methods have been applied so far only on numerical simulations and result in the classification of the dark matter structures. The reconstruction of the density field from observed galaxy data is based on a very much smaller number of objects, and therefore it lacks the spatial resolution of the simulations. However, probably the most difficult task to tackle is that the observations provide information on the position of galaxies in redshift space, while for the proper reconstruction of the cosmic web galaxy positions need to be measured in real space \citep{Lee&Lee.2008, Lee&Li.2008}. This is not of course a problem in simulations.

\subsubsection{Correlation between galaxies and environment in which they reside}

Several correlations between galaxy properties and the environments in
which they reside are observed in the local  Universe. In this context, the environment of a galaxy is given by the number density (or overdensity) of the neighbouring galaxies. It has been known for many years that elliptical and
lenticular  galaxies reside in  more  dense environments  than
spiral galaxies  \citep[e.g.][]{Dressler.1980}.  Red, older, less  star forming
galaxies live  in  more  dense regions than blue,  younger galaxies
with high star formation rates \citep[e.g.][]{Kauffmann.etal.2004, Blanton.etal.2006}.  These trends are  observed in environments  ranging from
clusters  to  voids.

Are  the  early  conditions of galaxy
formation  (the so-called nature  scenario) or  the environment  in which
they subsequently reside  (the so-called nurture  scenario), or both  of these,    responsible  for  the   observed  dependences   of  galaxy
properties on environment?  An obvious  way to tackle this question is
to  go to  higher redshifts  and to  establish at  which  redshift the
 relations in question are in  place. Not  long  ago, high  redshift
observations were  limited to the most dense  structures (i.e. clusters) or
targeted  to detect a  specific galaxy  population (e.g.   Lyman break
galaxies or Ly$\alpha$  emitters).  Large and deep  redshift surveys are
the best probes of the  (over-)density field delineated by galaxies at
all redshifts.   The recent  spectroscopic high redshift  surveys, the
DEEP2  Galaxy Redshift Survey  \citep{Davis.etal.2003}  and the  VVDS \citep{LeFevre.etal.2005}  and now zCOSMOS \citep{Lilly.etal.2007, Lilly.etal.inprep}, sample  the broad
range of galaxy population  and allow the continuous reconstruction of
galaxy environments. Most of the galaxy property-environment relations observed at $z \sim 0$ seems to be at place already at $z \sim 1$ \citep[e.g.][]{Cucciati.etal.2006, Cooper.etal.2007}.

Even in the local Universe, 
there is an ongoing debate on whether the observed galaxy-environment relations
depend on the particular scale at which the environment  has been measured. This would potentially give  a clue to their origin. 
\citet{Kauffmann.etal.2004} and \citet{Blanton.etal.2006} both find  that only the
environment measured  on the  small scales (of  1 \hh Mpc)  appears to
affect the star formation histories of galaxies. In the context of the
CDM scenario, this scale corresponds to the scale of the individual dark
matter halo  in which galaxy  resides, whereas the larger  scale would
probe more the location in the cosmic web.

In the current state
of the  art of the  high redshift surveys,  it is very  challenging to
reconstruct environments  on such a small  scale.  Galaxies targeted for the observations at higher redshifts are generally more luminous, and their mean inter-galaxy separations are thus larger. Furthermore, only a
fraction  of  galaxies is  targeted  for  the  observations, and  this
fraction may be  substantially smaller  than locally.

\subsubsection{Galaxy density field as biased tracer of the matter density field}

In the current cosmological picture it is almost a paradigm that galaxies are a biased tracer of the underlying matter distribution.  The biasing factor can be a non-trivial function of the scale, of the redshift and of the type of galaxies used to reconstruct the density field. Using a statistical approach, one of the ways to infer the biasing function is from a comparison of the probability distribution functions (PDF) of the galaxy and matter overdensities \citep{Sigad.etal.2000, Marinoni.etal.2005}. While the first is based on observations, the PDF of the matter comes from theory and is dependent on the assumed cosmological parameters. In the case of the three dimensional overdensities reconstructed with a top-hat filter, the PDF of the matter is well described by the log-normal distribution \citep{Coles&Jones.1991}.

\subsection{Goals of this study}

The main goal of this paper is to reconstruct the galaxy overdensity field in the zCOSMOS region. 
We reconstruct  the overdensity field using the first $\sim$ 10,000 spectra of galaxies from  the zCOSMOS survey,  up to
redshift  $z \sim$  1 \citep{Lilly.etal.inprep}, the so-called 10k sample.
High sampling rate and measurements
of redshifts  with a  precision of about  100 kms$^{-1}$ enable  us to
delineate the environments  of galaxies from  the 100 kpc scale  of galaxy
groups --  the scale of  environment expected to dominate  the various
galaxy  evolutionary processes  -- up  to the  100 Mpc  scales  of the
cosmic web.  It  is one  of  the major
scientific  drivers  of  the  zCOSMOS  survey to  study  the  role  of
environment on galaxy evolution up to high redshifts. We therefore pay particular attention to reconstruct the density field on scales that are as small as possible. With this aim, a new method \citep[ZADE,][]{Kovac.etal.ZADE} has been developed to reconstruct the density field  using  both galaxies with spectroscopic  redshifts, and those with only photometric redshifts. The photometric probability distribution functions are modified depending on the proximity of galaxies with high quality spectroscopic redshifts. We utilise this method to reconstruct the overdensity field at the positions of the 10k zCOSMOS galaxies and in any random point in the zCOSMOS volume. The reconstruction has been carried out in a number of ways so as to facilitate a wide range of scientific explorations as discussed above.  We present the three dimensional  overdensity maps, and compare the LSS structures in the overdensity field to the independently estimated virialised structures in the volume of the zCOSMOS survey. Extensive use is made of COSMOS mock catalogues \citep{Kitzbichler&White.2007} to estimate the errors on the reconstructed overdensity field and to compare the zCOSMOS overdensity field to those obtained from the mock catalogues. Throughout this paper, we assume a flat cosmology described with $\Omega_{m,0}=0.25$ and $H_0=70$ \kms. However, we express the results related to the overdensity field using $h$, $H_0= 100 h$, while stellar masses and absolute magnitudes of galaxies  are quoted with an explicitly incorporated $H_0=70$ \kms.

In  follow-up
papers,  we  study  the   dependence  of  star  formation  properties
\citep{Cucciati.etal.inprep}  and morphology  \citep{Tasca.etal.inprep}  on the environments
of   the  10k  zCOSMOS   galaxies  presented   in  this   paper.   The
environmental  dependence  of the  luminosity  and  mass functions  is
discussed in \citet{Zucca.etal.inprep}  and \citet{Bolzonella.etal.inprep}, respectively. The biasing function between the zCOSMOS and matter overdensity fields is presented in \citet{Kovac.etal.bias}. Other studies cover environmental dependence of specific galaxy populations, such as IR galaxies \citep{Caputi.etal.2008}, AGN \citep{Silverman.etal.submitted} or post-starburst galaxies \citep{Vergani.etal.inprep}.

\section{(Galaxy) density field reconstruction}

A  number  of methods  exist  to  evaluate  the density  field,  often
developed  for numerical  simulations, e.g.  grid  based  methods,  smooth
particle  hydrodynamics  (SPH)  like  methods,  Voronoi  and  Delaunay
tessellation field estimators (VTFE and DTFE). Here, we concentrate on
the  problem   of  reconstruction  of  the  density   field  from  the
observational data sample of galaxies.

\label{sec_densrec}

In  practice, reconstruction of  the galaxy  density field  (or galaxy
environments) reduces  to the (weighted) count of  objects within some
aperture  around  a set of positions  where   the  density  field  is  to  be
evaluated.  In  the  general  case,  the  density at  an observationally
 defined  position
${\bf r}={\bf r}(\alpha, \delta, z)$ can be estimated as

\be \rho({\bf r}) = \Sigma_i \frac{m_i W(|{\bf r}-{\bf r_i}|;R)}{\phi({\bf r_i})}. \label{eq_rhodef} \ee

\noindent The above equation gives the value of a mean density at a position {\bf r} in the redshift space (averaged over the aperture in which the density field is measured), with $W$ a spatial window function, $m_i$ a weight based on the astrophysical properties for each galaxy and $\phi$ a function correcting for  various observational issues. The difference ${| \bf r}-{\bf r_i}| $ refers to any arbitrarily defined distance between observationally defined points.


The following points should be noted.

\begin{itemize}

\item
The form of equation~\ref{eq_rhodef} which we use to define the density at a point is similar to the density estimate in the SPH simulations \citep[e.g.][]{Hernquist&Katz.1989}, with the exception of the selection function $\phi$. 
The density can be evaluated at a position of a galaxy \citep[the so-called scatter approach, e.g.][]{Hernquist&Katz.1989}, or at any  chosen point in space \citep[the so-called gather approach, e.g.][]{Hernquist&Katz.1989}. The points can be of course chosen such that they form a regularly spaced grid.


\noindent
\item The summation in  equation~\ref{eq_rhodef} goes over those galaxies in
the sample that have been chosen to define the  density field, which we  refer to as
{\it tracer} galaxies. These  tracer galaxies might be all galaxies
detected  in the survey  (which is  commonly flux  limited) or  only a
subsample of those satisfying some selection criteria, e.g. a criterion
that is as much as possible independent of redshift, so as to form a ``volume
limited sample''.

\item The function
$W(|{\bf r}-{\bf r_i}|;R)$ is  the kernel  used to weight  the tracer
galaxies, which is a  spatial smoothing function (e.g. a  top-hat or a Gaussian function) and
$R$ is the smoothing length. The $W$ function is typically chosen such
that  it   weights  tracer   galaxies  depending  on   their  distance
$|{\bf r}-{\bf r_i}|$ from  the position  where the density  is being
reconstructed. The distance between two points can be defined in various ways. In most of the applications the smoothing function $W$ shows some symmetry with respect to the reconstruction point. Commonly, it is of fixed shape  and normalised to unity, i.e. $\int W dV = 1$.


\item The smoothing length $R$ defines the  aperture  within which  the
environment is  measured. To obtain a well defined value at every point, the field needs to be filtered over a large aperture, such that the shot-noise effects are suppressed. However, all properties of the density field on scales smaller than the smoothing filter $R$ will be smoothed away. The smoothing length $R$ in a given reconstruction procedure can  be defined to be of fixed or adaptive size. The fixed scale is then the same at every point and should be chosen such that there are enough tracer galaxies within the smoothing kernel to be able to reliably reconstruct the density at a given point. Unfortunately, in order to keep the number of galaxies in low density regions reasonably large to obtain statistically meaningful results, the smoothing length $R$ is then required to be rather large. The statistical fluctuations of the measured density will depend on the number of tracer galaxies within the smoothing aperture, i.e. on the density and will be statistically more accurate in the regions with higher density. On the other hand, the adaptive scale varies over the volume in which the density field is being reconstructed. It is given by the density of the neighbouring tracer galaxies, typically defined by the distance from the point of the density reconstruction (which can be a galaxy or any point in space) to the fixed Nth nearest neighbour. The use of an adaptive scale leads to a larger dynamical range and higher spatial resolution (in the dense regions) with respect to the fixed scale and the purely Poisson noise can also be made constant for a given population of tracer galaxies.

\item $m_i$  is the astrophysical  weight (``mass'')  of the
tracer  galaxy. In the reconstruction of galaxy density fields it usually has a  value  of  unity for each tracer galaxies, producing  a ``number-density''. However, $m_i$ can be  any
measured property of the tracer  galaxies (e.g. stellar mass). In the ideal
 case, it would
probably be the total mass  of a tracer galaxy. The resulting ``mass'' weighted  density is supposed to be related to the underlying matter density via the bias $b$ factor. However, the bias can be a nonlinear and stochastic function, depending on the scale, redshift and type of the tracer galaxies.

\item The reconstructed density field is supposed to be the density field based on the total population of chosen  tracer galaxies in the observed volume.  
The function $\phi$ may be introduced to correct for the fact that in reality only a fraction of the full population of tracer galaxies is at our disposal when reconstructing the density field.  The function $\phi$ should account for the observational restrictions such as the non-uniform sampling over the observed area, a radial selection function or the redshift success rate of the observed sources in the spectroscopic surveys. All of these constraints may depend on the intrinsic properties of galaxies (e.g. the luminosity or the morphological type) and as a result it may be very challenging  to model all possible dependences into the $\phi$ function.

\end{itemize}

Unavoidably, the  properties of  the reconstructed density  field will
depend   on  the  adopted   method.   The   choice  of   a  particular
reconstruction method, given by the exact functional forms of $W(|{\bf
r}-{\bf r_i}|; R)$, $\phi({\bf r_i})$ and $m_i$, and the choice of the
tracer galaxies, is  guided primarily by the scientific  goals, but it
will  always contain  some level  of user-specified  arbitrariness. In
particular,  one  has to  optimise  between  the  scale at  which  the
environment is measured and the error of the environment reconstructed
on that scale,  arising because we are using  discrete tracers.  While
it may be desirable to measure environments on a very small scale, the
statistical  errors  for the  smallest  scales  are  also the  largest
\citep[e.g.][see also  our results on  the mock catalogues  in Section
4]{Blanton.etal.2003}.  For  some  interpretations of  the  scientific
results  it may  be  desirable to  use  a fixed  scale  to define  the
environment,  because the adaptive  scales are  smaller in  the denser
regions  and larger  in empty  regions. However,  the  adaptive scales
prove to be  superior over the fixed scales  in preserving the complex
morphology of the density field \citep[e.g.][]{Park.etal.2007}.

One limitation of the outlined scheme is its lack of sensitivity to the geometry of the mass distribution: the used kernels are commonly of  fixed shape and isotropic. Even smoothing with the adaptive scale will smear out the smallest structures, particularly the ones of anisotropic shapes \citep[e.g.][]{Romano-Diaz&vandeWeygaert.2007}. \citet{Shapiro.etal.1996} and \citet{Owen.etal.1998} introduced in simulations an elliptical kernel with axis ratio dependent on the geometrical distribution of particles.

The VTFE and DTFE are other methods to reconstruct the continuous
density fields \citep{Bernardeau&vandeWeygaert.1996,  Schaap&vandeWeygaert.2000, Schaap.2007}. Both VTFE and DTFE  are fully adaptive
and  volume  covering  methods   based  on  the  Voronoi  and  Delaunay
tessellation of  the point  sample, respectively, which  divide space
into   a   space-filling   network   of  polyhedral   cells  \citep[Voronoi
tessellation,][]{Dirichlet.1850,  Voronoi.1908} or  mutually  disjunct
tetrahedral  cells in three  dimensions \citep[Delaunay  tessellation,][]{Delone.1934}  according to  the local  density and  geometry of  the sampling
points. Therefore  the VTFE and  DTFE methods have the  advantage over
the  methods which use  kernels of  fixed size  to better  recover the
anisotropic structures in density fields, such as filaments and walls.

However, direct applications of the VTFE or DTFE to reconstruct the density field in the current $z \sim 1$ and higher redshift surveys also have some drawbacks. First, the spectroscopic surveys at these redshifts have usually a small angular size on the sky (1-2 deg$^2$), and with the current number density of the tracer galaxies the volumes of the Voronoi and Delaunay cells will be greatly affected at the edges. \citet{Cooper.etal.2005} used the Voronoi tessellation to estimate the local density of galaxies in the DEEP2-like mock catalogues. They conclude that more than 45$\%$ of the sample galaxies are affected by the edges when using the Voronoi tessellation to reconstruct densities around galaxies, compared with $\sim 15\%$ when using a cylindrical kernel with $R=1$ \hh Mpc. Second, in $z \sim 1$ spectroscopic surveys the fraction of galaxies with a high quality spectrum, which leads to a reliable measure of a redshift used to delineate the density field, often is not greater than $\sim 50\%$. It has not been investigated which effects this sampling fraction will have on an asymmetric density estimator such as DTFE. Moreover, it is not clear how to deal with probabilistic objects in the VTFE or DTFE \citep[see][who use a Monte Carlo approach to sample the photometric redshift probability function of galaxies in the Voronoi tessellation method]{vanBreukelen.etal.2006}. The DTFE patterns also contain some artefacts, the most prominent one being the triangular imprint of the smoothing kernel \citep{Schaap.2007, Romano-Diaz&vandeWeygaert.2007}.

For some  scientific applications, such as biasing,  the density field
needs to be reconstructed on a fixed scale. In a flux limited survey,
the mean  separation between galaxies will increase  with the redshift
for a  given smoothing scale and  increase the shot  noise. Even a given
population  of  galaxies will be characterised  by  some average inter-galaxy
separation, and  therefore with any given galaxy  population the density
field on  scales smaller  than this separation will be dominated by
the  shot  noise.  For  fixed  apertures,  the  Wiener  filtering
technique can be  used to deconvolve the noise  from the reconstructed
density  field \citep[e.g.][]{Lahav.etal.1994, Hoffman.1994, Zaroubi.etal.1995}.  This technique is based  on the  minimum variance
reconstruction of  the density field,  which requires the  noise model
and power spectrum of the density field to be known a priori.

It is  common to express  the resulting measurement of density
 as a dimensionless  density  contrast $\delta({\bf r})$
(which we interchangeably refer to also as ``overdensity'') defined as

\be \delta({\bf r}) = \frac{\rho({\bf r}) - \rho_m(z)}{\rho_m(z)} ,  \label{eq_deltadef} \ee

\noindent
where $\rho_m(z)$ is the mean density at a given redshift. In most of the applications $\rho_m(z)$ is evaluated as a volume average, but it can be estimated also as an average over galaxies (e.g. Cooper et al. 2006). Although we generally refer to $\delta$ also as ``overdensity'', we will characterise regions  with $\delta > 0$ as being overdense, and regions with $\delta < 0$ as being underdense.

\section{ZADE methodology}
\label{subsub_ZADE}

The photometric  redshift technique enables  us to obtain  redshifts of
large number of galaxies in relatively modest amount of observing time
with respect  to  spectroscopic surveys. The  obtained photometric
redshifts can be used for a variety of applications, such as luminosity and
mass functions. Clearly, spectroscopic redshifts offer a major improvement over even a high quality photometric redshifts. For instance, \citet{Cooper.etal.2005} find that for the density field reconstruction, if only
the photometric  redshifts are used, even with  uncertainties as small
as 0.02(1+z), the reconstructed environment in the line-of-sight direction
is  smeared  out on   small  scales.   Only  for
uncertainties  of photometric  redshifts smaller  than 0.005  and when
measuring  densities in projection, photometric  redshift surveys can
become  comparable  to   spectroscopic  surveys in  terms  of  
environment reconstruction  \citep{Cooper.etal.2005}. Including a properly devised background correction the projected density estimator based on photometric redshifts becomes substantially unbiased, even though with a large scatter. This background corrected estimator is robust for projected densities larger than 10 galaxies per (\hh Mpc)$^2$ \citep{Guzzo.etal.2007}. However, 
the spectroscopic surveys with redshift uncertainties smaller than the
velocity dispersion of the small group (i.e. below 200-250 \kms) are the preferred 
source of  data with redshift of sufficient precision to reconstruct
the {\it small-scale environments} of galaxies.

To reconstruct the density field on the given scale and with acceptable error one needs to balance between two requirements: precise spectroscopic redshifts to accurately trace the overdense and underdense regions, but as large number of objects as possible to lower the noise and enable reconstruction of the density on the smaller scales.

We have  developed a new algorithm  \citep[ZADE,][]{Kovac.etal.ZADE}
which brings together the accuracy of spectroscopic redshifts with the
number statistics  of the photometric  redshifts, in order  to broaden
the scale of the  reliable reconstructed environments. Motivated by the literature results of the importance of the scales at which the density is measured, our main effort
has   been   made   to    reliably   reconstruct   the   small   scale
environments. The  algorithm  modifies the
individual redshift probability distributions  P(z) that are output by
a photometric redshift  code (e.g. ZEBRA)  based on the spectroscopic  redshifts of
galaxies located nearby on the sky.

It is well
 known that galaxies are highly correlated on scales up to 10 \hh Mpc. Therefore,
 a  galaxy  that has  only  a  photometric  redshift probability  function
 $P(z)$, is much more likely to  lie at some redshifts than at others,
 depending  on the  accurately determined  spectroscopic  redshifts of
 other galaxies that lie near to the same line of sight.

As described in more detail in \citet{Kovac.etal.ZADE}, 
 the ZADE algorithm modifies the initial $P(z)$ that is output of the  photometric redshift programme,  by counting the number of objects
 with  spectroscopic redshift within some radius (R$_{ZADE}$) as a function of
$z$ along the line of sight, to yield a 
modified probability $P_{ZADE}(z)$:

\be
   P_{ZADE}(z) = \frac{N( \le R_{ZADE}, z)\times P(z)}{\int{N( \le R_{ZADE}, z)\times P(z)dz}} .
\ee
\noindent

\begin{figure}
\centering
\includegraphics[width=0.45\textwidth]{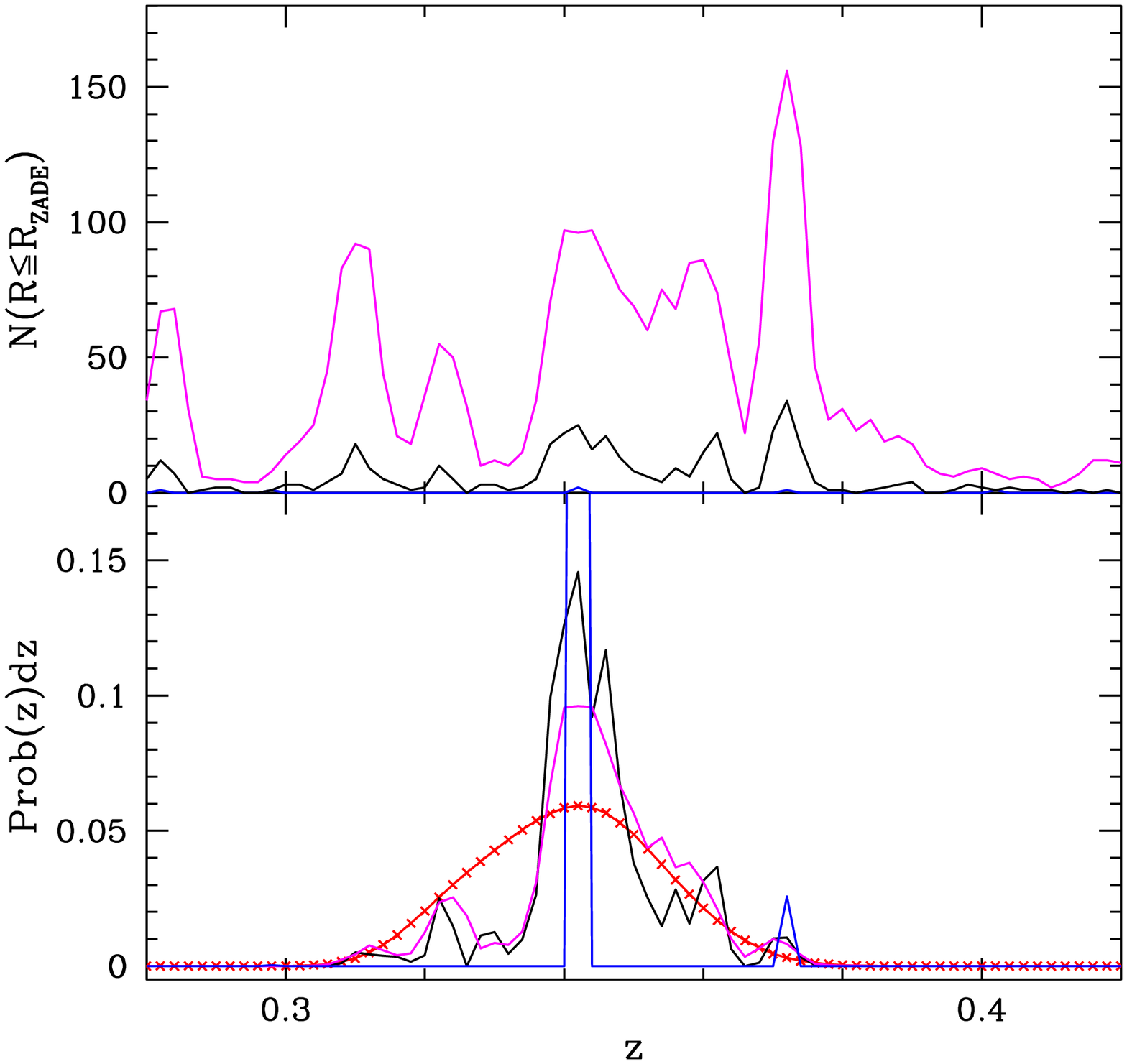}   
\includegraphics[width=0.45\textwidth]{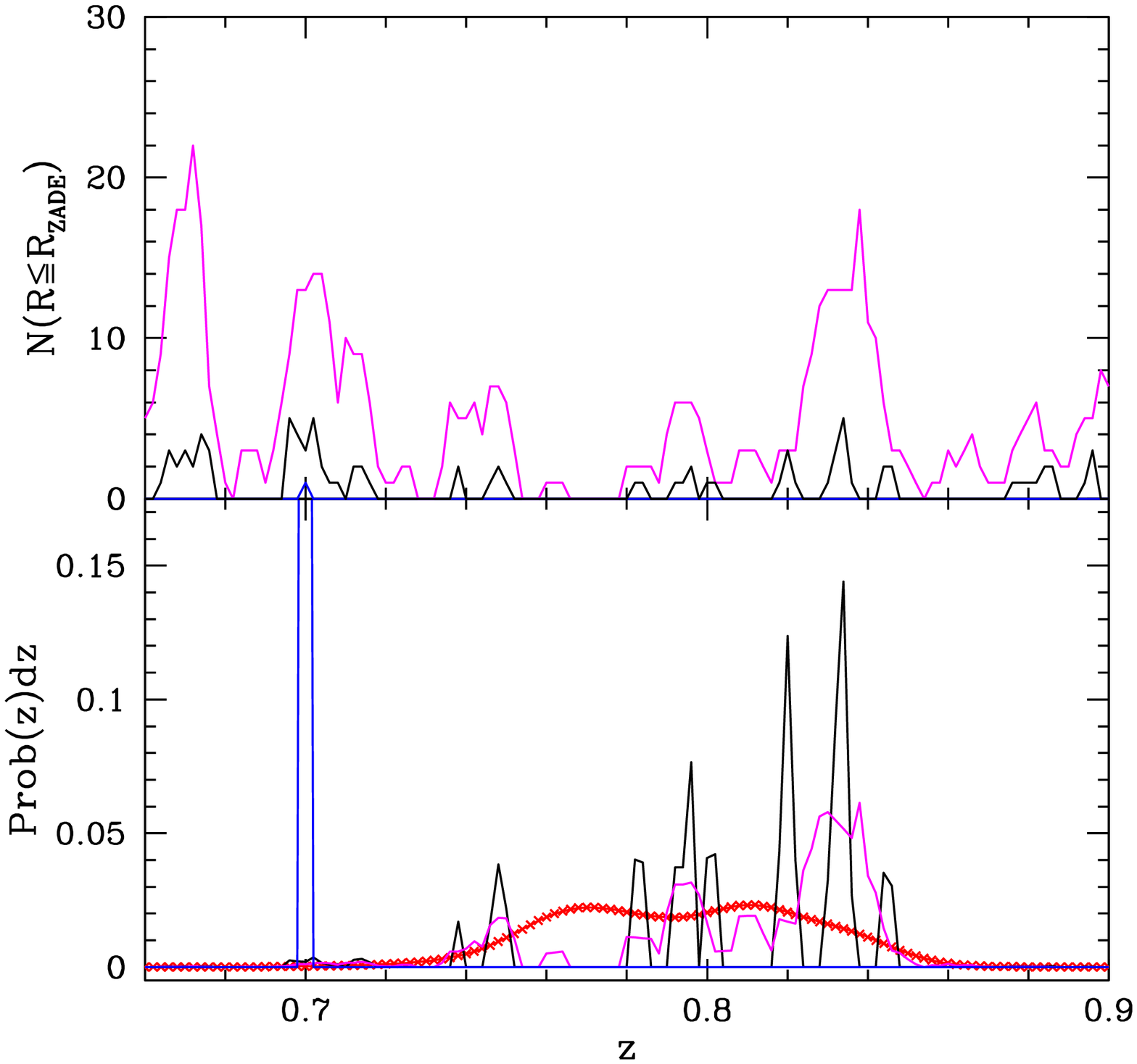}
\caption{ZADE approach to modify the initial $P(z)$, discretised in $\Delta z = 0.002$. The photometric redshift probability  functions are presented in the lower panels. The $P(z)$ output from ZEBRA is presented in red and the ZADE-modified $P_{ZADE}(z)$ are presented in blue (R$_{ZADE} \le$ 1 \hh Mpc), 
black (R$_{ZADE} \le$ 5 \hh Mpc) and magenta(R$_{ZADE} \le$ 10 \hh Mpc). The number 
counts of objects with spectroscopic redshifts within $R_{ZADE}$ radii at a given $z$ are presented in the upper panels, using the same colour coding as in the lower panels.
The left panel is for an object with  I$_{\rm AB}$=21.94 mag, the right panel is for an object 
with I$_{\rm AB}$=22.38 mag. Note that the number of objects with spectroscopic redshift within $R_{ZADE} \le$ 1 \hh Mpc is very small, and therefore $N(R \le R_{ZADE}$=1 \hh Mpc) is zero at almost all $z$ (upper panels). See text for more details.
\label{fig_zadeexmpl}}
\end{figure}

In  Figure~\ref{fig_zadeexmpl}  we  show  two  examples  of  the  ZADE
approach  to  modify the  initial  $P(z)$,  obtained  using the  ZEBRA
code \citep{Feldmann.etal.2006}. For  the computing purposes,  we discretise
the ZADE-modified $P(z)$ of every galaxy in $\Delta z = 0.002$ (this does not match the ZEBRA precision). At the current state  of  the uncertainty  in the  photometric
redshifts,  there  are  typically  a few  structures  within  the
 $P(z)$  along a given line of  sight (effectively distribution
of galaxies within $R_{ZADE}$), so the ZADE approach is only statistical.

ZADE has been extensively tested  on mock catalogues.  We tested our method using five different $R_{ZADE}$ values: 1, 2.5, 5, 7.5 and 10 \hh Mpc, even though in Figure~\ref{fig_zadeexmpl} we show results only for three of them (1, 5 and 10 \hh Mpc). The value
$R_{ZADE}=5$ \hh Mpc is chosen as an optimal ZADE radius to modify the
photometric probability functions in  order to reconstruct the density
field without any systematics.

In \citet{Kovac.etal.ZADE} we conclude that the use of the combination of objects
with  the  spectroscopic   and  ZADE-modified  photometric  redshifts
reconstructs  the overdensity  field  at a  given  scale with  smaller
errors  than the traditional approach of using  only  the  spectroscopic  sample of  objects
weighted  with some  $\phi({\bf r_i})$  function in equation~\ref{eq_rhodef} (which  in our  tests
accounts  for a non uniform $RA-DEC$ sampling).  Therefore, this 
enables us to  use smaller smoothing scales  in the reconstruction
process, and thereby broadens  the dynamical range of the reconstructed
environments.

For  every density  reconstruction within the zCOSMOS survey we  use this
combination   of  galaxies  with   spectroscopic  and   ZADE  modified
photometric  redshifts. The  summation in  equation~\ref{eq_rhodef} is
therefore over all galaxies of  a given population in the survey volume
(and not only over a  fraction of galaxies with reliable spectroscopic
redshifts) equivalent to setting $\phi({\bf r_i})
= 1$  for every tracer galaxy.   This is one  of the major  advantages of
using  galaxies  with both  spectroscopic  and photometric  redshifts,
because  in  practice it is  impossible  to  model the  $\phi({\bf r_i})$
function in  such a  way to take  into account all the complex selection
effects of most spectroscopic surveys. In equation~\ref{eq_rhodef}, galaxies with only photometric redshift will be counted with their ZADE-modified $P_{ZADE}(z)$ according to a smoothing filter $W$.

\section{Reconstructing the density field within the zCOSMOS volume}

\subsection{zCOSMOS survey}

zCOSMOS \citep{Lilly.etal.2007, Lilly.etal.inprep} is a redshift survey   undertaken in
the  1.7   deg$^2$  COSMOS  field   \citep{Scoville.etal.2007b}.   The
observations are  carried out with  the VLT using VIMOS,  a multi-slit
spectrograph. In zCOSMOS-bright a magnitude selection I$_{\rm AB} <$
22.5 has been  applied to select galaxies in the  redshift range up to
$\sim$ 1.4. The  final sampling rate of the  zCOSMOS-bright survey should
be a uniform 60-70$\%$. This paper is based on  the first 10,000 spectra,
which have a rather non-uniform  sampling pattern, 30$\%$ on average [see the Figure with
the 10k  zCOSMOS sampling in \citet{Lilly.etal.inprep}].   This sample of all spectra yields to the so-called
``10k sample'' of galaxies with only secure redshifts.

Thanks to  the high-quality multi-wavelength photometry  of the COSMOS
survey \citep[see][]{Taniguchi.etal.2007,  Capak.etal.2007},  practically all
galaxies  in  the  zCOSMOS  field can also  have  a  photometric
redshift. We employ  the ZEBRA  code in  the  Maximum-Likelihood mode
\citep{Feldmann.etal.2006}  to calculate the  full  probability  distribution  functions $P(z)$ and photometric redshifts [the  maximum likelihood of $P(z)$] of galaxies
in the  COSMOS field,  using typically 10  COSMOS broad bands.   The
uncertainty in the photometric redshifts  that we use in this paper is
$\sigma_z$=0.023(1+z) at  I$_{\rm AB} <$ 22.5.   The accuracy in
photometric redshifts will further  improve, when new photometric data
become available \citep[e.g.][]{Ilbert.etal.2008, Salvato.etal.2008}.

In our  analysis, we  complement the ``10k  sample'' of  galaxies with
$\sim$ 25,000 I$_{\rm AB} <$ 22.5 galaxies which do not have, yet,
spectroscopic  redshifts,  but  for  which photometric  redshifts  are
available.   We  refer  to  this  composite sample  as  the  ``10k+30k
sample''. We  additionally refer to the ``40k  sample'' as the set  of all galaxies
with I$_{\rm AB} <$ 22.5 in the area of  the zCOSMOS survey. In
the analysis  below, we  will frequently make  use of  mock catalogues
\citep[based  on][]{Kitzbichler&White.2007} for various tests, in which  all galaxies  in the
``40k sample'' have accurately known redshifts.

We use rest-frame  absolute magnitudes  of  galaxies obtained using  the
ZEBRA  code  as  the  best  fit template  normalised  to  each  galaxy
photometry  and  best available redshift \citep[spectroscopic or photometric,][]{Oesch.etal.inprep}. For the galaxies with only photometric redshift,  a prior which
requires $M_B$ to be in the  range between -13 and -24 mag is imposed.

Stellar  masses $M_*$  are  obtained from  fitting stellar  population
synthesis models to the multicolour spectral energy distribution (SED)
of  the observed  magnitudes \citep[see][]{Bolzonella.etal.inprep, Pozzetti.etal.inprep}. In this work, we use stellar masses calculated using the
\citet{Bruzual&Charlot.2003} libraries,  with the Chabrier  initial mass
function \citep{Chabrier.etal.2003}. The star  formation history  (SFH) is
assumed  to  decrease  exponentially  with a time  scale  $\tau$,  where
$0.1<\tau<30$ Gyr. The \citet{Calzetti.etal.2000} extinction law was used with $0 < A_V <
3$ and  solar metallicities.  The final stellar  mass is  obtained by
integrating the  SFH over  the galaxy age  and subtracting from  it the so
called  ``return fraction'',  which is  the mass  of gas  processed by
stars   and  returned   to  the   interstellar  medium   during  their
evolution.  The   stellar  masses   are  calculated  using  the
best available redshifts for each galaxy.

\subsection{The choices to measure the density field with zCOSMOS}
\label{sub_densitychoice}

Given the diversity of the scientific applications of the density field reviewed above, we have  produced a large variety of environment
measures that are appropriate for different applications. In the following text we  discuss  the  particular choices  used  for  the
measurement      of     the      zCOSMOS     density      field.  For  practical purposes, we  use the
symbol $\rho$ for all types of densities.

\subsubsection{Tracer galaxies: flux limited or volume limited}
\label{sub_sample}

For the  10k zCOSMOS overdensity reconstructions,  all galaxies within
the  zCOSMOS region  with I$_{\rm AB}  <$ 22.5  are  used  as possible
tracer galaxies.  Galaxies with  high confidence redshift class (flag $\ge$ 1.5) and
the broad-line emitters with  high confidence redshift class (flag $\ge$ 10+1.5) are
counted  as  one  object  at  the  spectroscopic  redshift  \citep{Lilly.etal.inprep}. Up to $z \sim 1$, we do not detect a significant dependence of the assigned redshift class on the type of a galaxy. For  the rest of galaxies within  the zCOSMOS survey limits
we use  the ZADE-modified ZEBRA photometric redshift  output. For the
few galaxies without a photometric  redshift estimate, we use the
ZADE-modified  $N(z)$ distribution of  all galaxies  in the  fields as
initial $P(z)$, normalised to unity  within the redshift interval $0 <
z < 1.4$ (only a  negligible fraction of galaxies with high confidence redshift
class  has a  redshift  $z  > 1.4$).   For  the few  QSOs  with a  low
confidence redshift we use  the same ZADE-modified $N(z)$ distribution
to be  their probability function, but normalised  to $f_{QSO}$, which
is the  fraction of the broad-line objects  with high-confidence class
below  $z=1.4$   in  the  sample   of  the  broad-line   objects  with
high-confidence class  detected at all redshifts. For  the 10k zCOSMOS
sample, $f_{QSO}  = 0.43$.

In summary, we use spectroscopic redshift information of galaxies with
 a  spectroscopic redshift  reliability $>  99.2\%$ (which  make about
 85\%  of   the  spectroscopic  sample),  combining   these  with  the
 ZADE-modified photometric  redshifts for  all of the  galaxies either
 without secure redshift or those not yet observed spectroscopically.

We construct three different samples of tracer galaxies to measure the
zCOSMOS over-density field. For  the ``flux limited sample'', all galaxies
within the zCOSMOS region with I$_{\rm AB}  <$ 22.5 are used as tracer
galaxies.  In total,  there are 33,211 such galaxies,  of which 8341
have  a reliable spectroscopic  redshift. In  addition, we  define two
volume  limited  samples of  tracer  galaxies, which should ideally include the same galaxies at every redshift. Understanding of galaxy evolution does not run deep enough to provide a recipe of galaxy evolution which includes all physical processes for every single galaxy (e.g. star formation episodes of various activity and merging of galaxies). Therefore, to constrain the volume limited samples, we just  take  into account  a
passive  evolution   of  galaxies  of one  magnitude   per  unit  redshift interval, which should on average account for most of the evolution for the majority of galaxies. The two volume samples satisfy the  criteria $M_B \le -19.3-z$ and $z \le
0.7$, and $M_B  \le -20.5-z$ and $0.4<z\le 1.0$.   To include galaxies
with only photometric  redshifts in the volume limited  samples we use
their maximum likelihood
redshift and  the  $M_B$  magnitude   at   the same
 redshift. However, in the density field reconstruction, we use the $P_{ZADE}(z)$ distribution of these galaxies. Galaxies without photometric  redshift or $M_B$ are not used
to  construct  the volume  limited  samples  of  tracer galaxies.  The
defined  volume limited  samples are  complete samples,  and  they are
selected to be the largest possible samples at these redshifts.

\subsubsection{Filter $W$ and smoothing length $R$}

Starting from the  work of \citet{Dressler.1980}, when studying
the galaxy properties as a  function of their environment, it is common
to project the tracer galaxies within some velocity interval, and then
estimate  the  surface  density.   The  main advantage  of  using  the
projected distances instead of the full three dimensional distances is
to minimise  the effects of  peculiar velocities. The  aperture, perpendicular to z, within
which the environment is measured can  be of fixed or adaptive
size. \citet{Dressler.1980}  for example used the projected  distance to the
10th nearest neighbours to define a rectangular aperture.

Following  the  adopted  formalism   to  describe  the  density  field
(equation~\ref{eq_rhodef}), the filter $W$ in this approach is:

\be
\label{eq_wproj}
W(\left|{\bf r}-{\bf r_i} \right|;D_p) = \left\{
                                  \begin{array}{lr}
                                  \frac{1}{\pi D_p^2} P_{i,ZADE} \ (z \pm \delta z) & \it{if} \ \left|r_p-r_{i,p} \right|\le D_p , \left|z_i - z \right| \le \delta z \ (\delta z = {\rm 1000~km~s^{-1}})  \\
                                  0 & \it{otherwise}.
                                  \end{array}
                                 \right.
\ee

\noindent 
This smoothing kernel is of cylindrical shape, with a radius of $D_p$ in the RA-DEC plane and a length of $\pm \delta z$ along the line of sight. The distance $|r_p-r_{p,i}|$ is measured between the point where we are reconstructing the density field
and the individual tracer galaxy projected to the redshift of that 
point.  The P$_{i,ZADE}(z\pm \delta z)$ is  the integrated probability  of the
tracer   galaxy   to   be   within the  $z \pm \delta z$   interval. For galaxies with spectroscopic redshift within this interval P$_{i,ZADE}(z\pm \delta z) = 1$. We use the interval  $\delta z$ given  by distance of $\pm$  1000 \kms\ to
the redshift $z$ of the point where the density field is measured. Varying  $\delta z$ by $\pm$ 500 \kms\ around the chosen value  of 1000 \kms\ does
not make a significant difference.

Using  the spatial  filter  $W$ given  in equation~\ref{eq_wproj},  we
measure  the density  field  at the  positions  of galaxies  and on  a
regularly spaced grid.  For the former, we include  the central galaxy
into the counts of objects, if this galaxy passes the criterion to be a
tracer galaxy.

We  use both fixed and adaptive  apertures to estimate
the density at a galaxy  position.  We define the adaptive aperture as
the projected  distance to the Nth nearest  neighbour $D_N$, excluding
the central galaxy from the nearest neighbours search, but including it
again when estimating  the density. For the measurement  on a grid we
use the  adaptive projected  scale $D_p  = D_N$  defined  by the
distance to the Nth nearest neighbour.  Given that we are working also
with the probability functions  (and therefore fractional objects), in
practice we measure the adaptive aperture defined by the distance to  the Nth nearest  neighbour as the minimum  distance for  which $\Sigma_i
P_{i,ZADE}(z \pm \delta z) \ge N$. This is valid for both estimates of
the density centred on a galaxy and on a grid point.


Some of the previous studies of the galaxy evolution as a function of their environment are based on  the density field reconstructed using the three dimensional Gaussian filter \citep[e.g.][]{Cucciati.etal.2006}. To enable a proper comparison with such  results, we also use a Gaussian filter to reconstruct the density field on scale $R_G$ at the positions of zCOSMOS galaxies:

\be
W(\left| {\bf r}-{\bf r_i} \right|;R_{G}) = \left \{
                                  \begin{array}{lr}\frac{1}{(2\pi R_G^2)^{(3/2)}} P_{i,ZADE}(z_i) \exp \left[-\frac{1}{2}(\frac{{\bf r}-{\bf r_i}}{R_G})^2 \right]  & \it{if} \  |{\bf r}-{\bf r_i}| \le R_G  \\

                                  0 & \it{otherwise},
                                  \end{array}
\                                 \right.
\label{eq_gauss}
\ee

The $|{\bf r}-{\bf r_i}|$ is a three dimensional separation between positions of a tracer
galaxy and a galaxy where the density is being evaluated. If the central galaxy  belongs to the sample of tracer galaxies used, it is also included in the counts.

For the biasing analysis \citep{Kovac.etal.bias}, we use a different scheme to reconstruct the over-density field. We want to compare the overdensity field traced by galaxies to the underlying matter overdensity, and therefore we need the overdensity field estimated on the same scale as the mass overdensity as available in the literature. Thus we use the spherical top-hat filter to reconstruct the density field of galaxies:

\be
\label{eq_wrth}
W(\left| {\bf r}-{\bf r_i} \right|;R_{TH}) = \left \{
                                  \begin{array}{lr}
                                  \frac{1}{\frac{4}{3} \pi R_{TH}^3} P_{i,ZADE}(z_i) &  \it{if} \  \left|{\bf r}-{\bf r_i} \right| \le R_{TH}   \\

                                  0 & \it{otherwise},
                                  \end{array}
\                                 \right.
\ee

\noindent
where $R_{TH}$  is the  smoothing scale of  the top-hat  filter. Here,
$|{\bf r}-{\bf r_i}|$ is a three dimensional distance between a tracer
galaxy and a point where the density is being estimated. An obvious shortcoming of this reconstruction is that it is affected by peculiar motions. We include this into consideration for the biasing analysis.

When the smoothing length $R$ of a  filter $W$ is defined using the projected distances (as in Equation~\ref{eq_wproj}) we will denote the reconstructed overdensity with $\delta_p$. In the case when $R$ is defined using the three dimensional distance (as in Equations~\ref{eq_gauss} and \ref{eq_wrth}), we will denote the reconstructed overdensity with $\delta$.

\subsubsection{Weighting $m_i$ of galaxies}

In addition to the number overdensity reconstruction, we also calculate the B-band luminosity and stellar mass $M_*$ weighted densities. Following equation~\ref{eq_rhodef}, these three types of reconstructed environment correspond to the cases of $m_i=1$, $m_i=L_B$ and $m_i=M_*$, respectively. We assume that the physical properties of galaxies are constant, i.e. they do not change along redshift with $P_{ZADE}(z_i)$. $L_B$ and $M_*$ are estimated at the best available redshift for each galaxy.

Weighting each galaxy with its physical property will lead to a galaxy density field still mainly determined by the spatial distribution of galaxies, with some additional discrimination based on the properties of galaxies. For example, for the same number of galaxies in two identical apertures, the measured number density will obviously be the same, while weighting galaxies with their $L_B$ or $M_*$ will add a finer differentiation between the measured densities.

A shortcoming of the density estimate with $m_i \neq 1$  is that an obtained relation between galaxy properties and environment  can contain some degree of correlation  between the studied galaxy property and the property of the galaxy used to weight the galaxy counts for the density reconstruction. This is particularly pertinent when including the central galaxy itself in the measurement of the environment. For example, the colour-density relation will become (artificially) more significant when using $M_*$-weighted overdensities compared to the number overdensities \citep[see][]{Cucciati.etal.inprep}, and the differences between the stellar mass functions in the overdense and underdense regions will appear more pronounced \citep[see][]{Bolzonella.etal.inprep}. Nevertheless, $m_i \neq 1$ weighting can be useful in defining a physically better motivated density estimator.

\subsection{Estimating mean density}

To convert  the density to an  overdensity an estimate  of the average
density  is required. Typically, galaxies in  the survey itself  are used to  define the
 mean density,  a value  which is  supposed to be  valid in  the whole
 Universe (at a given redshift). Therefore one needs to assume that these surveys are a fair sample of the Universe.

There are  various ways  to determine  the mean
density of a survey. One of  the commonly used methods is smoothing 
the  observed  number  distribution   of  the  tracer  galaxies  along
redshift, or fitting  this distribution with some assumed functional
form.  Sometimes only  the  pure  number density  of  objects in  some
redshift  intervals is  used, but  in  this case  a part  of the  true
variations in the overdensity with  redshift will be washed away.  The
other commonly used method is deriving mean densities of galaxies from
the galaxy  luminosity function, in which case  the redshift evolution
of the luminosity function needs to  be modelled.  We use here   a  non-parametric  treatment  of  the
luminosity function approach.

\begin{figure}
\centering
\includegraphics[width=0.60\textwidth]{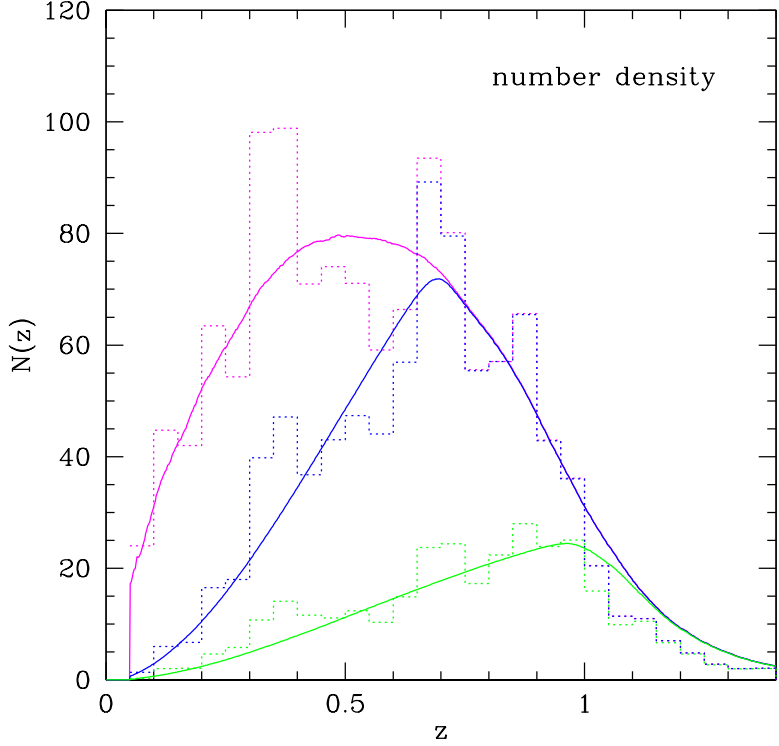}
\caption{\label{fig_nz} Redshift distribution of galaxies in the zCOSMOS area.  The continuous lines correspond to the smoothed N(z) distributions obtained by weighting galaxies according to their $\Delta  V(z)/V_{max}$ contribution in $\Delta z=0.002$ intervals as described in the text. The dotted lines correspond to the histogram distributions of tracer galaxies in the redshift bins of $\Delta z = 0.05$ and they are divided by 25 to match the redshift bin of the smoothed N(z). Magenta represents the flux limited sample of tracer galaxies, blue represents the sample of galaxies with $M_B < -19.3-z$ and green represents the sample of tracer galaxies with $M_B < -20.5-z$. The last two samples are volume limited up to $z < 0.7$ (blue curve) and $z < 1$ (green curve), respectively.}
\end{figure}

  We  use the  full population  of  tracer galaxies  from the  zCOSMOS
survey (selected from the 10k+30kZADE  sample) to define  the mean
volume density  as a function of  redshift. For each  galaxy, we first
calculate the maximum  volume $V_{max}$ in which this  galaxy could be
detected in the  zCOSMOS survey, including for completeness an assumed passive evolution of $\Delta m = \Delta z$ mag. In the next  step the contribution of
this galaxy  to the density in  each redshift bin, up  to $z_{max}$ is
calculated as $\Delta V(z)/V_{max}$, where $\Delta V(z)$ is the volume
of the individual redshift bins.  $z_{max}$ is the maximum redshift up
to  which  a  galaxy  of  a  given luminosity  could  be  detected  in a
 $I_{\rm AB}<22.5$ survey; $V_{max}$ is the volume of the zCOSMOS survey up to redshift $z_{max}$. This is carried out for every tracer galaxy and then the
contribution of all galaxies is  summed to obtain N(z). We present the
N(z)  distributions  for  the  three  samples of  the  zCOSMOS  tracer
galaxies in Figure~\ref{fig_nz}. Similarly, one can calculate also the
luminosity  and stellar  mass  weighted number of objects by  simply
weighting  $\Delta  V(z)/V_{max}$ contributions  by  $L_B$ and  $M_*$,
respectively.

As desired, the procedure outlined above introduces  a smoothing in the very peaky
observed   distribution    of   objects,    as it  can   be    seen   in
Figure~\ref{fig_nz}. With respect to  a more standard smoothing on the
observed N(z) distribution, it has  the advantage of not requiring any
assumption  on   the  smoothing   length. The kernel in this smoothing procedure is of quadrilateral shape, where one side is defined by the distance  between $z_{min}$ and $z_{max}$, and the vertical sides are given by $\Delta  V(z)/V_{max}$ at $z_{min}$ and $z_{max}$. The remaining side of this kernel is curved, defined by $\Delta  V(z)/V_{max}$ in redshifts between $z_{min}$ and $z_{max}$. The exact shape of the kernel depends on $I_{\rm AB}$ of the individual galaxy. Due to the curved side of the kernel, a large number of galaxies at $z \sim 0.85$ produces a somewhat overestimated N(z) above redshift of 1. The   produced  N(z)
distributions  using  the  spectroscopic  redshifts  or  only  maximum
probability  redshifts  for  the  same  group of  objects  are  almost
identical.

Finally, we  obtain the mean density $\rho_m(z)$ by normalising N(z) or  its weighted
counterparts by  the ``volume''  in which N(z)  is estimated.  When the  density  is  calculated using the three dimensional distances, this is obtained dividing N(z) by
the volume of the zCOSMOS survey in a redshift bin centred at redshift
z of that  galaxy (or a grid point). To  obtain the surface mean density
we first integrate N(z) in  the interval of $\pm$ 1000 \kms\ centred on
the  redshift  of a  galaxy  or  a grid  point  of  interest to  obtain
the corresponding number of  objects, which we afterwards divide  by the 
zCOSMOS area.

\subsection{Other issues - edge effects}

For all the  different types of the density  estimators, the effect of
the edges  of the survey on  the estimated density  is to artificially
lower the  density, since  galaxies are not  detected outside  of these
edges. One of  the ways to avoid this is to  simply exclude all (grid)
points at which  the aperture size is larger than  the distance to the
nearest  edge of  the survey.  However, in  the case of a  density
estimated  using the adaptive  aperture, this  approach will  bias the
density estimator  by excluding regions  of the lower  density (larger
adaptive  aperture)  in  larger  proportion than  the  denser  regions
(smaller adaptive aperture).

A common way to correct for  the edge effects is to scale the measured
density  with the  fraction  of  the aperture  which  lies within  the
geometrical limits of the survey, assuming that the tracer galaxies in
the  part of  the cell  within and  outside of  the survey  limits are
distributed in the same manner.

For the adaptive density estimator the criteria to define the aperture
close to  the edges of the  survey are nevertheless  still affected by
the edge. For example, in the case that the aperture is defined by the
distance to the Nth nearest neighbour, only galaxies within the survey
limits will be used to define this distance. Effectively, for the same density, the smoothing length will be larger  for the points close to the edges
than for  the points  within the survey.  This would artificially lower the
 densities  near the edge due to the greater smoothing.  We tested an algorithm  in which  the  number  of required neighbours  is lowered  close  to  the  edges,
according   to  the  fraction   of  the   volume  within   the  survey
limits. However,  based on the tests  on the mock  catalogues, we find that  correcting the densities at the points
close  to the  survey limits  by simply  scaling the   density  by the
fraction of  the volume (area)  within the survey  limits works better,  and is
certainly much easier to implement.

For all types of the density reconstruction we therefore simply define the
edge-corrected density-estimate using the following equation:

\be  \rho_c = \rho / f \ee

\noindent
where $\rho$ is estimated using equation~\ref{eq_rhodef} and $f$ is the fraction of the adopted aperture that lies within the survey region. Depending on  the scientific  goals,  one can also then exclude  points with low values of $f$, if desired.

\section{Testing the reconstruction method on mock catalogues}
\label{sec_mocks}

We use mock catalogues to  assess the performance of the density field
estimator based on the  combination of galaxies with the spectroscopic
and  ZADE-modified  photometric  redshifts  when applied  on  the  10k
zCOSMOS  survey.   The mock  catalogues  for  the  zCOSMOS survey  are
extracted   from   the    lightcone-mock   catalogues   described   in
Kitzbichler\&White  (2007). These  lightcones  are based  on the  dark
matter N-body  Millenium Simulation run  (Springel et al.   2005) with
$\Omega_m=  0.25$, $\Omega_b=0.045$, $h  = 0.73$,  $\Omega_{\Lambda} =
0.75$, $n=1$  and $\sigma_8=0.9$.   The semi-analytic galaxy modelling that is applied  to the
mock     catalogues     is     described     in     Kitzbichler\&White
(2007). Essentially, it follows the  model of Croton et al.  (2006) as
updated by de Lucia \& Blaizot (2007).  The only change is in the dust
model,  developed  to  better  match  the  observations  of  the  mass
functions at higher redshifts  (Kitzbichler\&White 2007). There are 24
independent mock  catalogues of  an area $1.4^{\circ} \times 1.4^{\circ}$  and $z
\lesssim 7$, with a flux  limit $r \le  26$ mag.  We use  12 of
these  mocks,  randomly chosen,  to  perform  various  tests.

The mocks are used in the following way. All mocks are cut to match the  exact zCOSMOS area and redshift limits. We define the ``parent'' catalogue (equivalent to the 40k
catalogue, I$_{\rm AB}<22.5$), in  which we assume all tracer galaxies
in the zCOSMOS volume have  a high quality spectroscopic redshift, and
the  ``observed''  catalogue,  in  which  about  10,000  galaxies  have
spectroscopic redshifts and the rest  of the tracer galaxies have only
photometric redshifts  (analogous to the 10k+30kZADE  catalogue in the
real  data).  The  distribution  of  objects  with  the  spectroscopic
redshifts in the observed catalogues  is modelled \citep[see][]{Knobel.etal.inprep} to  resemble  that  of  the  true  10k  zCOSMOS  catalogue  \citep{Lilly.etal.inprep}.

 In the mock catalogues, we model the photometric redshift probability function
P(z) as a Gaussian of dispersion $\sigma_z$ depending on the selection
I$_{\rm AB}$ magnitude.  For galaxies  with $I_{\rm AB} < 22.5$ mag we use
$\sigma$  = 0.023(1+z) for the  purpose  of the
simulations. Galaxies are randomly displaced from their true redshifts
by an amount selected from  this same distribution.  We then apply the
ZADE  algorithm  on  those  galaxies  without  reliable  spectroscopic
redshifts in the mock catalogues using $R_{ZADE}=5$ \hh Mpc.  We have done
a simple  check of the robustness  of our results with  respect to the
functional form of the  probability function by  randomly assigning actual probability functions, obtained  from ZEBRA  for the  real zCOSMOS
galaxies, to   the  galaxies  in   the  mock
catalogue. Our results did not  change significantly. The shape of the
ZADE-modified $P_{ZADE}(z)$ is determined mostly by the neighbouring galaxies
and not by its initial form.

For the test on the mock catalogues, the (over)density field  is calculated on a grid  equally spaced in the
angular units $\Delta RA$ = $\Delta  DEC$ = 2 arcmin and in the
redshift   $\Delta$z  =  0.002. 
The  density $\rho$  in the  individual  grid points  is estimated using equation~\ref{eq_wproj}, with the aperture defined by the distance to the Nth nearest neighbours D$_{p,N}$. As we mention earlier, because  we  are also work with
``fractional'' objects, in this  reconstruction we count the number of
objects  until $\Sigma_i  P_i(\Delta z)  \ge N$.

To obtain  overdensities, the $\rho$ estimate in  the individual grid
points   is   divided  by   $\rho_{mean}(z)$. In this way, 
the reconstructed  measure of  environment is
given in  the units  of   $1+\delta_p=\rho/\rho_{mean} $
(equation~\ref{eq_deltadef}).  For the individual mock catalogues,
$\rho_{mean}(z)$ is obtained from  the smooth number distribution N(z)
of the  full 40k sample of  tracer galaxies. We obtain the smooth $N(z)$ distributions separately in each mock to closely resemble the uncertainties in $\rho_{mean}(z)$ arising from the limited volume of the real survey.

We  present  here  only   the  number  overdensities, obtained using
  $m_i=1$   for  all   the   tracer
galaxies.  The tracer  galaxies  are selected  from  the flux  limited
sample of galaxies. In the following plots, the survey area is limited
to  the   central  0.8$\times$0.8  degrees  in   right  ascension  and
declination to minimise the edge effects.

\begin{figure}[h]
  \centering
  \includegraphics[width=0.95\textwidth]{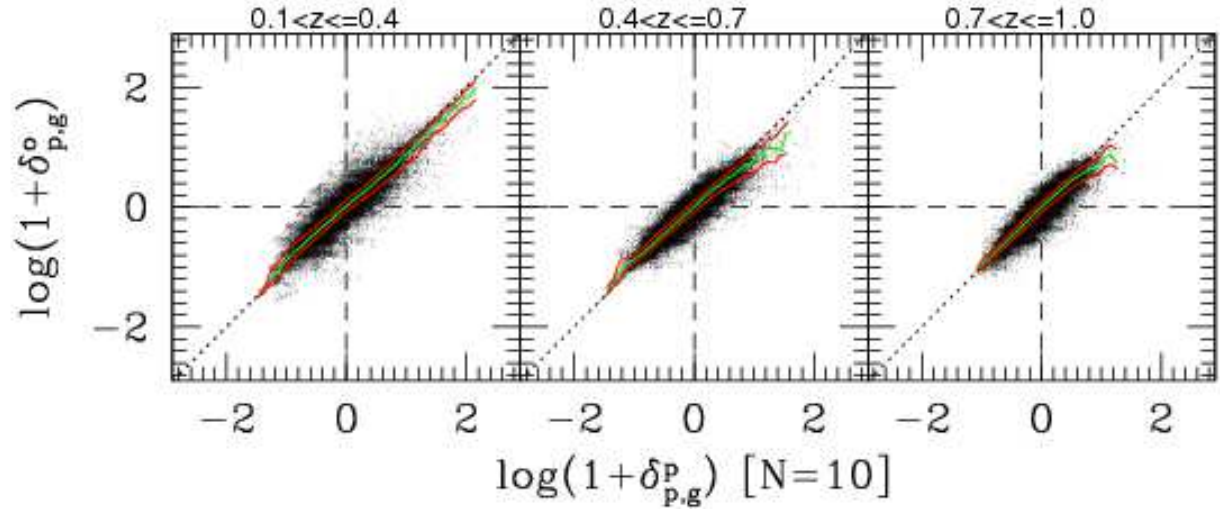}
 \caption{Comparison between the 10k+30kZADE observed and 40k parent overdensities.  The green line represents the median
for the  10kspec+30kZADE observed, the red  lines correspond to
the    25th   and   75th    percentiles   for    the   10kspec+30kZADE
observed. Binning  is  done along the parent  overdensity axis. Overdensities are obtained by using  the  distance to  the  10th   nearest
neighbour. Note that 50\% of all points are contained between the red lines. The lower indices $p$ and $g$ indicate that the overdensity has been reconstructed using the projected distances to define a smoothing length $R$ and has been reconstructed on the grid, respectively. The upper index refers to the type of the mock catalogue; ``p'' stands for the parent and ``o'' for the observed catalogue. See more details in the text.}
\label{compodens10}
\end{figure}

\begin{figure}[h]
  \centering
  \includegraphics[width=0.95\textwidth]{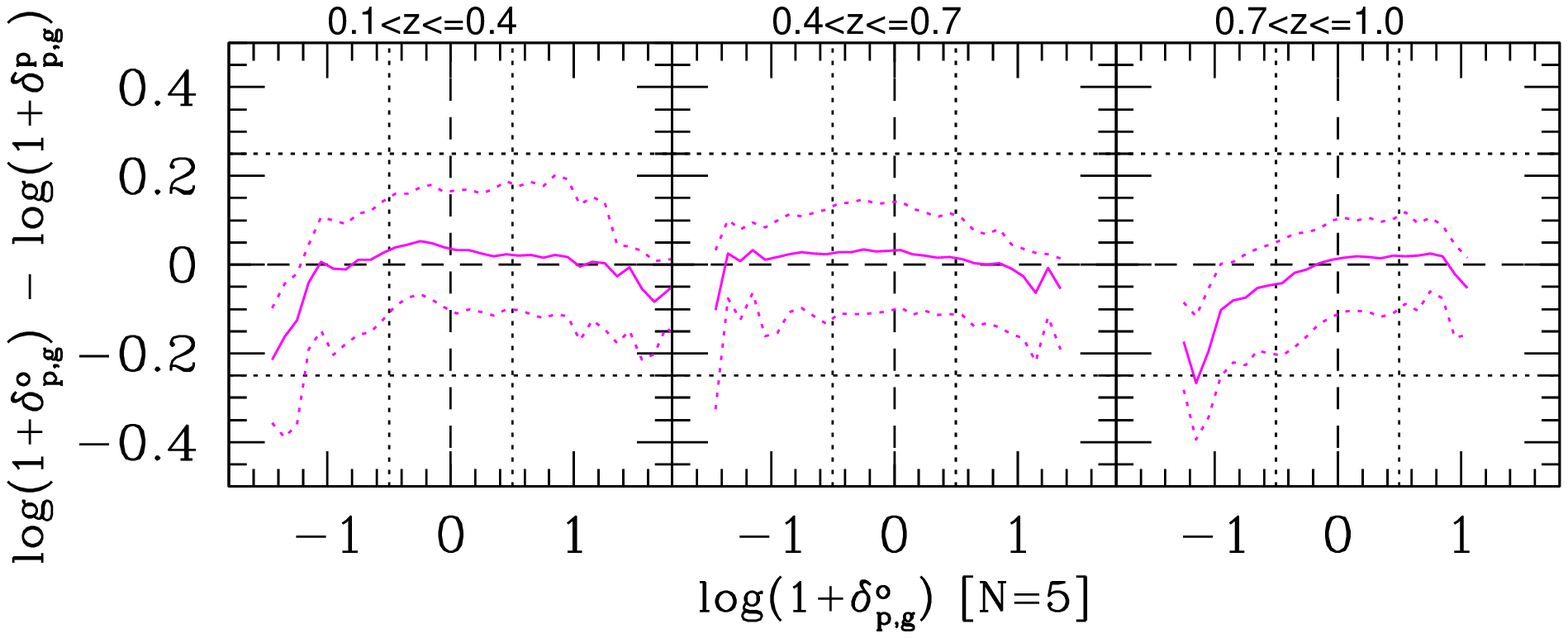}
  \includegraphics[width=0.95\textwidth]{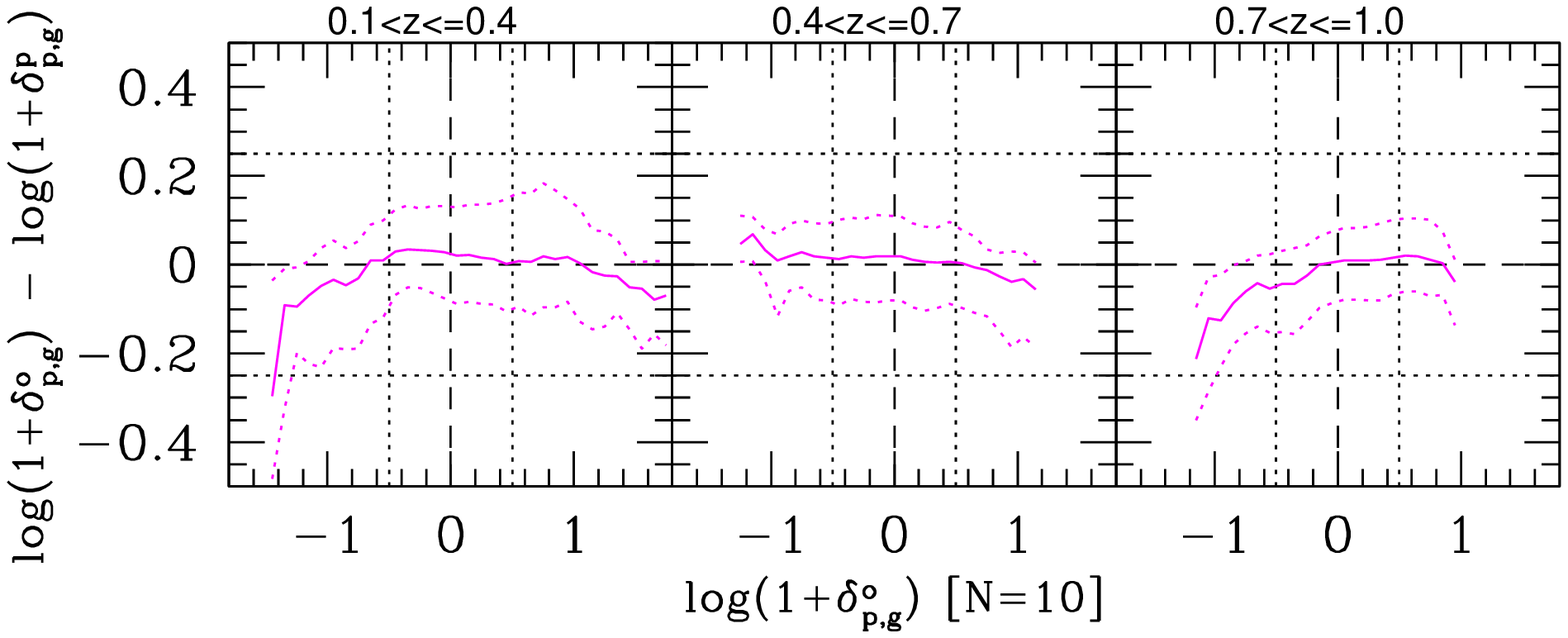}
  \includegraphics[width=0.95\textwidth]{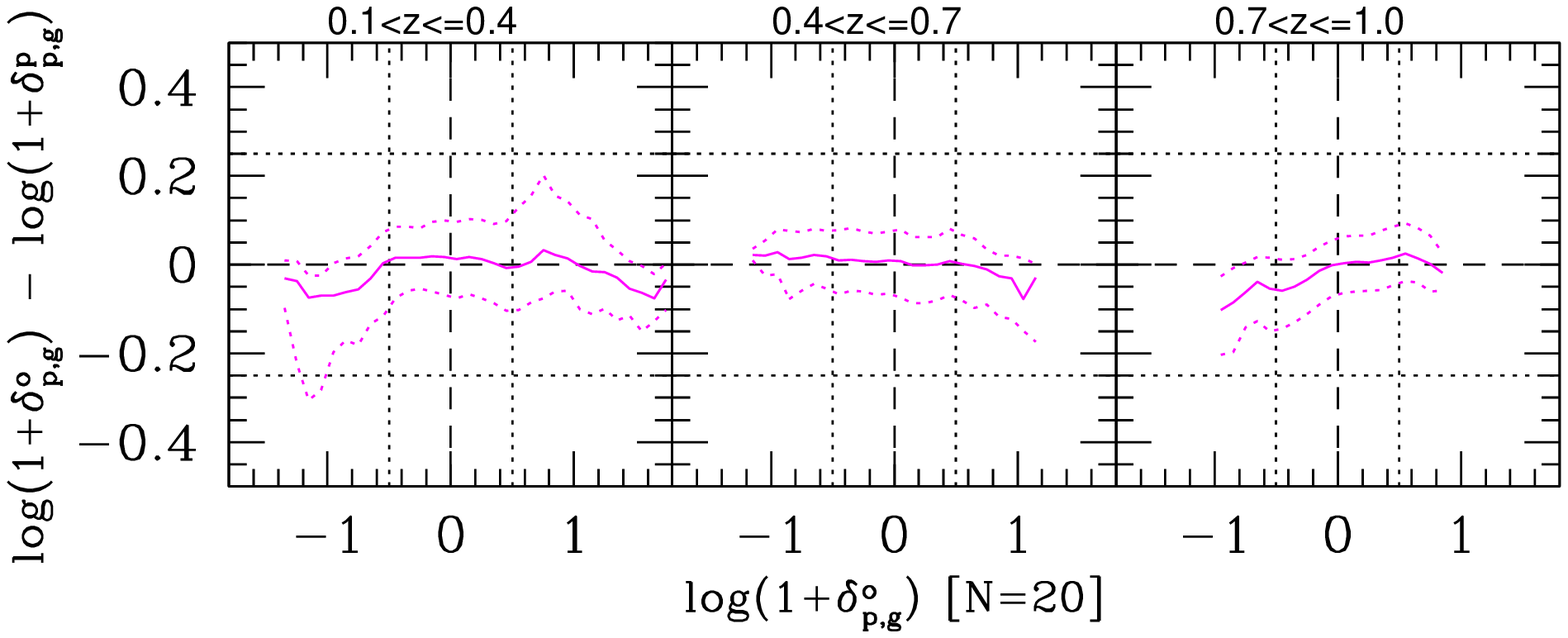}
\caption{The  difference between  overdensities reconstructed  in the
observed catalogue,  using the  ZADE formalism, and  in the parent catalogue,
plotted  as  a  function  of  the observed  overdensity. The density  is
obtained by measuring distance to  the 5th, 10th and 20th nearest neighbour
in the  top, middle and  bottom panel, respectively.  The continuous
line   represents  median  for   the  measured difference between the two reconstructed overdensities, the dotted lines correspond  to the 25th and 75th  percentiles for this difference. The indices are defined in Figure~\ref{compodens10}.}  
\label{complogodens_ZADE}
\end{figure}

Figure~\ref{compodens10}  presents  a  comparison of  the  10k+30kZADE
``observed'' and the 40k  ``parent'' overdensities for the individual
grid  points, using  the distance  to  the 10th  nearest neighbour  to
obtain  densities.  The  green  line  represents the  median  for  the
10kspec+30kZADE  observed, the red  lines correspond  to the  25th and
75th     percentiles      for     the     10kspec+30kZADE     observed
overdensities. Binning is done as a function of parent overdensity. Obviously, the ZADE method is not perfect - it slightly overestimates the
density  in the  most underdense  regions, and  it  underestimates the
density  in   most  overdense  regions, no doubt because of residual smearing of the initial $P(z)$.   However,  the  reconstructed
overdensity  reproduces the parent  overdensity without  a significant
systematic effect  in the majority  of regions, within the  errors. This
 is shown in Figure~\ref{complogodens_ZADE}. In
all  the three  panels of Figure~\ref{complogodens_ZADE} we  present a  difference between  the observed
10k+30kZADE and  the parent 40k overdensities for  the individual grid
points, plotted as a function of the observed overdensity. The density
field  is  reconstructed  using  three  different  adaptive  apertures
defined  by   the  distance  to   the  5th,  10th  and   20th  nearest
neighbour. Note that due to the ZADE approach we can compare overdensities obtained in the apertures defined by the same $N$ (number of nearest neighbours) in the ``parent'' and ``observed'' catalogues since they both contain same total number of objects.

Based on  the difference  between the 25th  and 75th  percentiles, the
error (statistical  scatter) in the  reconstructed overdensities using
the  ZADE method  is roughly  $0.1-0.15$ dex  in $1+\delta_p$.   It is
clear  that the error  on the  reconstructed overdensity  gets smaller
when using a  larger number of neighbours, i.e.   the density measured
at larger  apertures.  Depending  on the scientific  goal, one  has to
balance between the  smoothing scale and the statistical  error on the
reconstructed  overdensity. The  presented statistical  errors include
only those  uncertainties arising  from the fact  that we  are dealing
with a population  of galaxies with very different  qualities of their
measured  redshifts (i.e.  10k  galaxies with  spectroscopic redshifts
and 30k  galaxies with the  ZADE modified photometric  redshifts).  We
did  not include  any errors  arising  from the  uncertainties in  the
spectroscopic  redshift or  in the  measured physical  property. Also,
these  errors do  not  account  for any  differences  between the  40k
overdensities  and  the true  matter  overdensities.  If  we redo  the
presented  comparison using  the full  survey area,  and not  only the
inner   part,  the   median   and  lower   and   upper  quartiles   in
Figures~\ref{compodens10} and  ~\ref{complogodens_ZADE} remains almost
unchanged.  For  example,  in  Figure~\ref{compodens10}  there  is  an
indication  for  the  larger  underestimation of  the  most  overdense
regions.

\begin{figure} 
  \centering
  \includegraphics[width=0.95\textwidth]{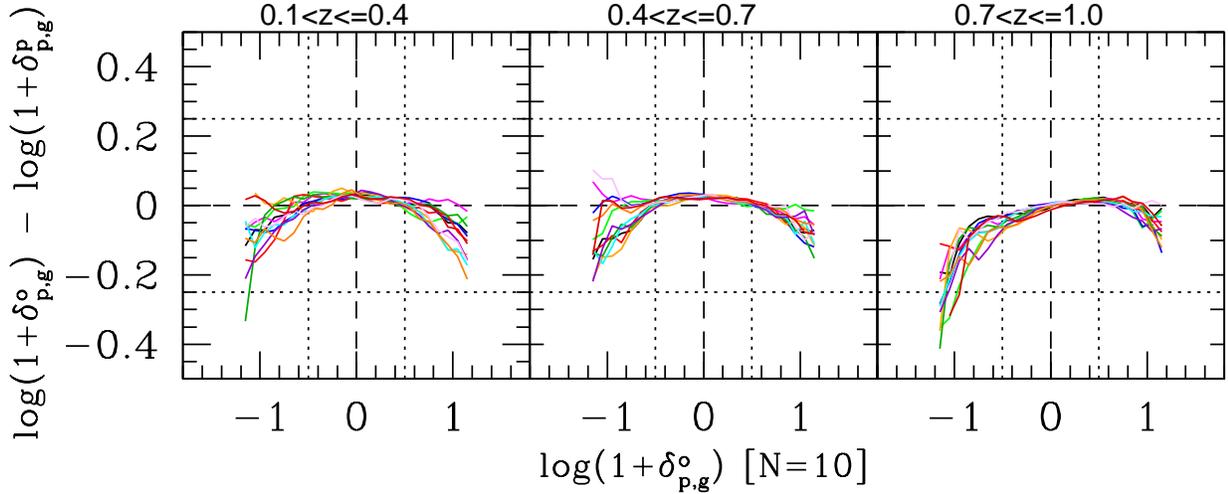}
\caption{The  difference between  overdensities reconstructed  in the
observed catalogue,  using the  ZADE formalism, and  parent catalogue,
plotted  as  a  function  of  the observed  overdensity. The density  is
obtained by measuring distance to  the 10th nearest neighbour. Each continuous
line corresponds to the median of  the difference between the two reconstructed overdensities (observed and parent), in the bins of the observed overdensity, for a single mock catalogue. Individual panels include results for 12 mock catalogues. The observed overdensities are presented in the range $-1.2 < \log(1+\delta_{p,g}^p) < 1.2$. The indices are defined in Figure~\ref{compodens10}.}
\label{complogodens_12mocks}
\end{figure}

Figure~\ref{compodens10}    and   Figure~\ref{complogodens_ZADE}   are
obtained      by      using      one     mock      catalogue.       In
Figure~\ref{complogodens_12mocks}  we present  the  difference between
the observed  and parent overdensities  as a function of  the observed
overdensities for all 12 mock  catalogues. The errors (i.e. the 25th and 75th percentiles of the difference
between  the observed  and  parent overdensity)  at  a given  observed
overdensity are  similar for all the individual  mock catalogues (and
we do not show them in Figure~\ref{complogodens_12mocks}).  The errors
in Figure~\ref{complogodens_ZADE} can therefore be  taken as a reference for the
errors  in  the observed  overdensities,  when  using this  particular
method of the reconstruction of the density field.

We  have chosen  to present  the errors  on the  reconstructed density
field for this particular method (projected nearest neighbour and flux
limited  tracer galaxies)  as the  majority of  our studies  of galaxy
properties  are   based  on  the  density  field   obtained  by  using
equation~\ref{eq_wproj}   and    the   apertures   probed    by   this
reconstruction method  are smallest.   We  have  carried   out  similar  tests  to  obtain  the
uncertainties  on  the reconstructed  overdensity  when  centred on  a
galaxy,  using  both  fixed  and  adaptive  apertures,  following
equation~\ref{eq_wproj}, and flux  and volume limited tracer galaxies,
weighting them with $m_i=1$, $m_i=L_B$ and $m_i=M_*$ . The statistical
errors  are  comparable  to  the  presented errors  for  the  adaptive
apertures defined  by the same  number of neighbours as  discussed here
and the fixed apertures starting from 3 \hh Mpc.

\section{The 10k zCOSMOS over-density field}

\subsection{Reconstructed environments of the zCOSMOS 10k galaxies}

We present here the reconstructed environments  of the 10k 
 zCOSMOS galaxies obtained following the choices presented in Subsection~\ref{sub_densitychoice}. We  discuss the obtained dynamical  range, the dependence  on  the chosen
aperture  and  on  the   weighting function  $m_i$  of  overdensity values quantified following equation~\ref{eq_wproj}, where the overdensity is estimated centred on a galaxy. The apertures are defined by projecting galaxies within $\pm 1000$ \kms, and we omit direct reference to this projection in the following text.

We obtain the broadest dynamical range of reliably reconstructed local
environments  of the  10k  zCOSMOS  sample of  galaxies  by using  the
apertures defined by the distance  to the 5th nearest neighbour in the
flux limited sample  of tracer galaxies. Based on the tests on the mock catalogue, 5 is the smallest number of neighbours which can be used to reliably reconstruct density at all redshifts probed. When using the larger number
of  objects to  define the  aperture (e.g.  10 or  20), both  the most
overdense   and  underdense  regions are  smoothed out,
particularly  the  most dense  ones which have a smaller physical size.

When  using  the volume  limited
samples  of  the tracer  galaxies,  the  distance  to the  Nth  nearest
neighbour is equal to or larger than the distance when using  the flux limited sample
of  tracer galaxies.  Therefore, the  obtained overdensities  with the
volume limited samples of tracer galaxies are also smoothed with respect to
the overdensities reconstructed with the flux limited sample of tracer
galaxies.

The distances to the 5th nearest neighbour used to define
the aperture for the reconstruction  of the density centred at a zCOSMOS galaxy for a
set     of     $\log(1+\delta_p)$     values    are     presented     in
Figure~\ref{fig_nndist},  for all  three samples  of tracer  galaxies. 
Rescaling  this plot one  can easily
obtain  the distances  for the other values of $N$,  since it  will  vary as
$(N+1)^{(1/2)}$.

If we use fixed apertures, the smallest
scale at which we are able to reliably reconstruct environments of the
10k zCOSMOS  galaxies is about  3 \hh Mpc  (based on the tests  on the
mock    catalogues),   using   the    density   estimate    given   by
Equation~\ref{eq_wproj}.   On this  and larger  scales,  the dynamical
range of  the overdensities  is smaller than  when using  the adaptive
approach. For  example, in the redshift range  $0.4<z\le0.7$ where we
are  complete  for both  $M_B<-19.3-z$  and  $M_B<-20.5-z$ samples  of
tracer galaxies, the local environments of  the 10k zCOSMOS  galaxies are
estimated  using adaptive apertures  smaller than or  equal to 3 \hh  Mpc for
94\%,  88\% and  62\% galaxies  for  the flux,  $M_B<-19.3-z$  and
$M_B<-20.5-z$    volume   limited    samples   of    tracer   galaxies
respectively. The  main point  to be  taken is that  when using  the fixed
aperture to  measure environments  we are
not  able   to  differentiate  between  the   most  overdense  regions
 that can be reconstructed at  the adaptive  scales.

\begin{figure}
\includegraphics[width=0.45\textwidth]{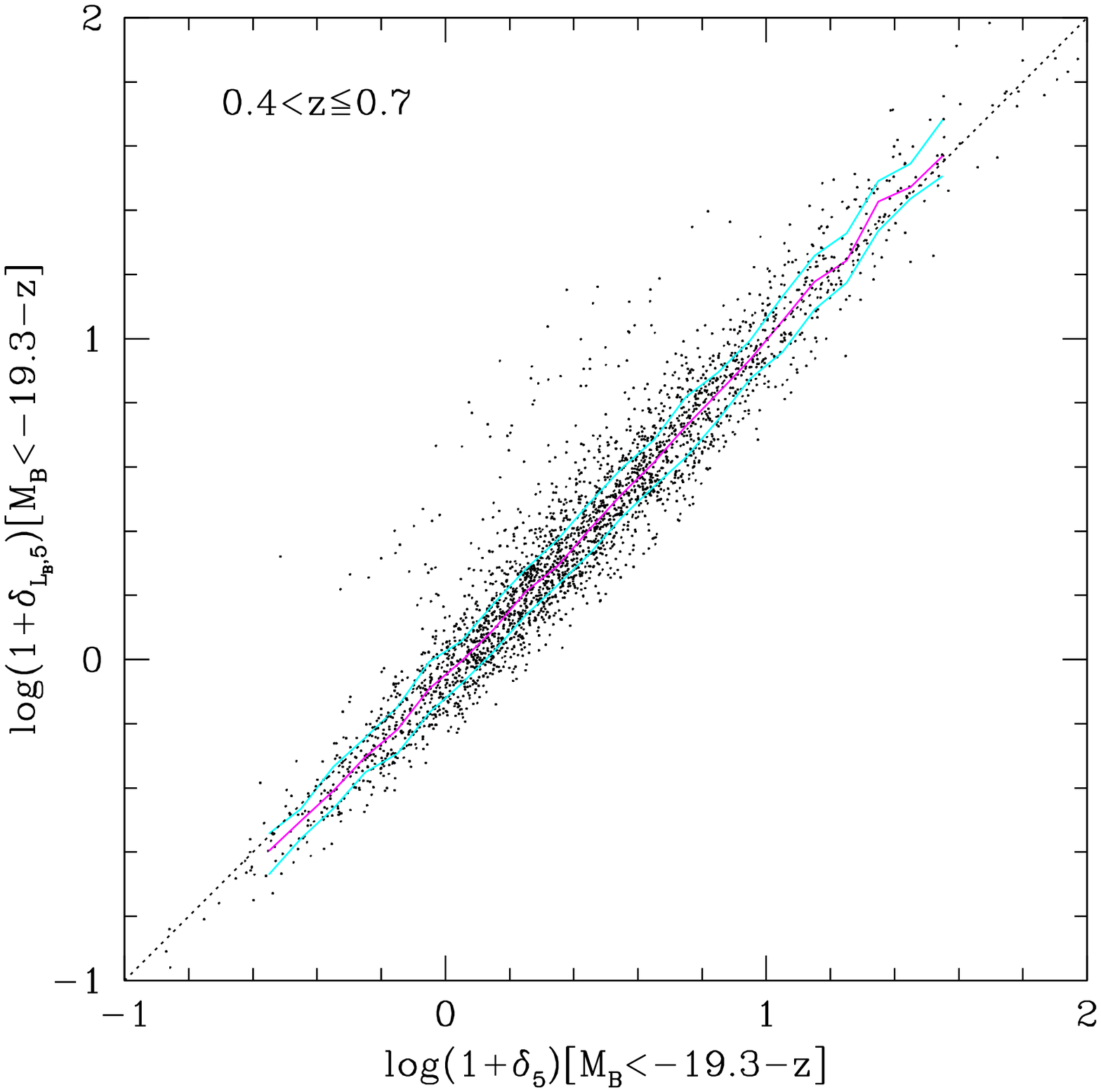}
\includegraphics[width=0.45\textwidth]{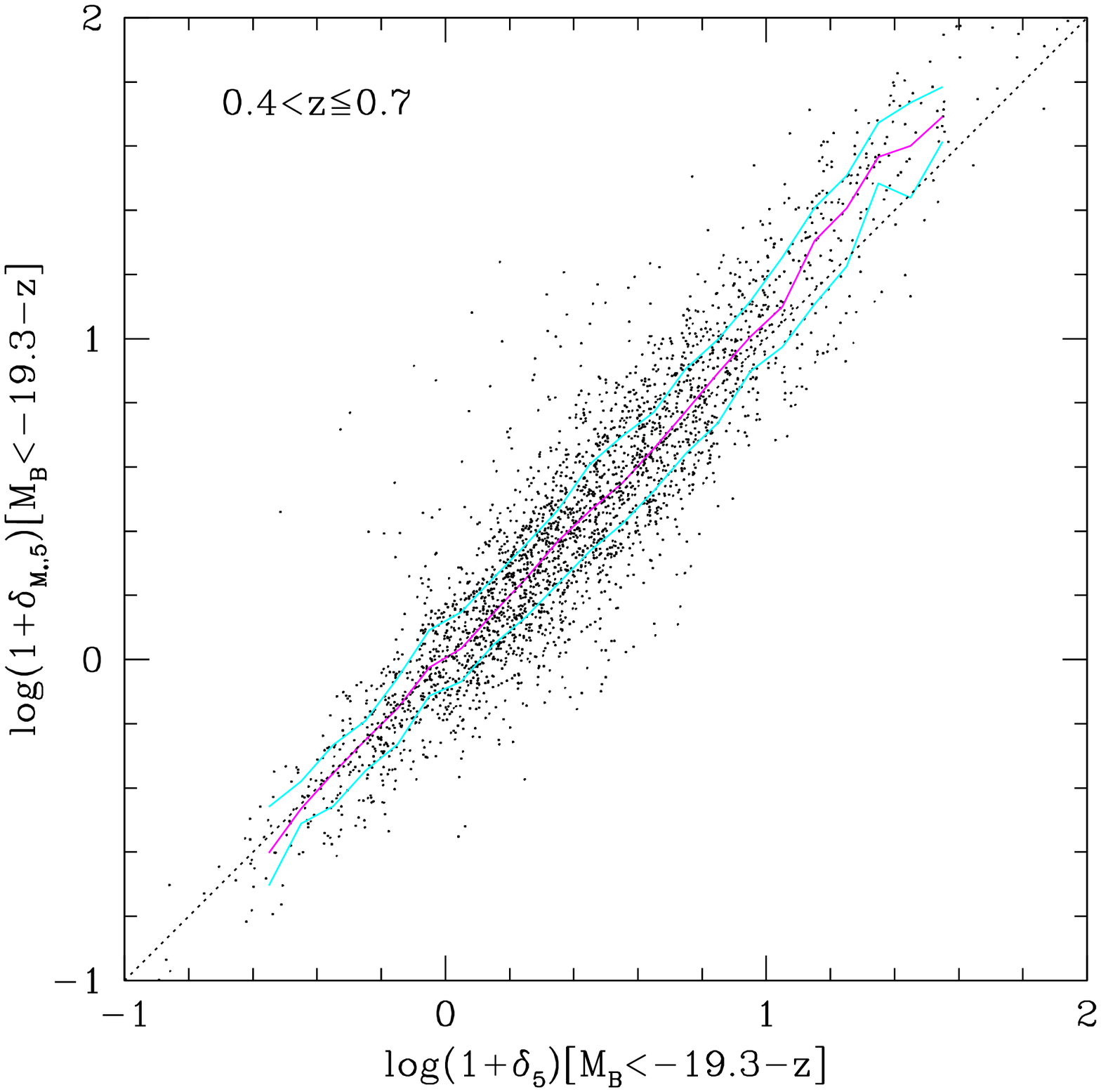}

\caption{\label{fig_compodens_nn5mB_numvsw}Comparison of the overdensities measured around the 10k zCOSMOS galaxies using equation~\ref{eq_wproj} with the $M_B<-19.3-z$ volume limited sample of tracer galaxies and within the apertures defined by the distance to the 5th nearest neighbour. We compare the B-band luminosity and stellar mass weighted overdensities to the unity weighted overdensities in the left and right panels, respectively. The magenta line is the median and the cyan lines are the lower and upper quartiles of the presented distributions, where the binning is carried out along the abscissa axis.  There is a good correlation between the number    overdensities   and   the    $L_B$   and    $M_*$   weighted overdensities.}
\end{figure}

The ranges of overdensities  discussed above are obtained by weighting
tracer galaxies with $m_i=1$.  There is a good correlation between the
number    overdensities   and   the    $L_B$   and    $M_*$   weighted
overdensities, because all are primarily set by the number of objects. However, the  most overdense regions become  even more overdense
when  using   both  $m_i=L_B$  and  $m_i=M_*$,   allowing  even  finer
differentiation  of the  most overdense  regions. When using the $M_*$ weighted counts of galaxies
the dynamical  range of overdensities  is broader than when  using the
number or $L_B$ weighted galaxy counts.  As  an  example, in
Figure~\ref{fig_compodens_nn5mB_numvsw}   we  compare   a   subset  of
differently  $m_i$-weighted   reconstructed  overdensities  using  the
$M_B<-19.3-z$ tracer galaxies in $0.4<z\le0.7$, and aperture defined by the 5th nearest neighbour. Two effects are responsible for the observed scatter: the noise of the mapping between the number of galaxies and their $L_B$ luminosities or stellar masses $M_*$, and different luminosity \citep{Zucca.etal.inprep} or stellar mass \citep{Bolzonella.etal.inprep} functions in different environments.

\begin{figure}
\includegraphics[width=0.3\textwidth]{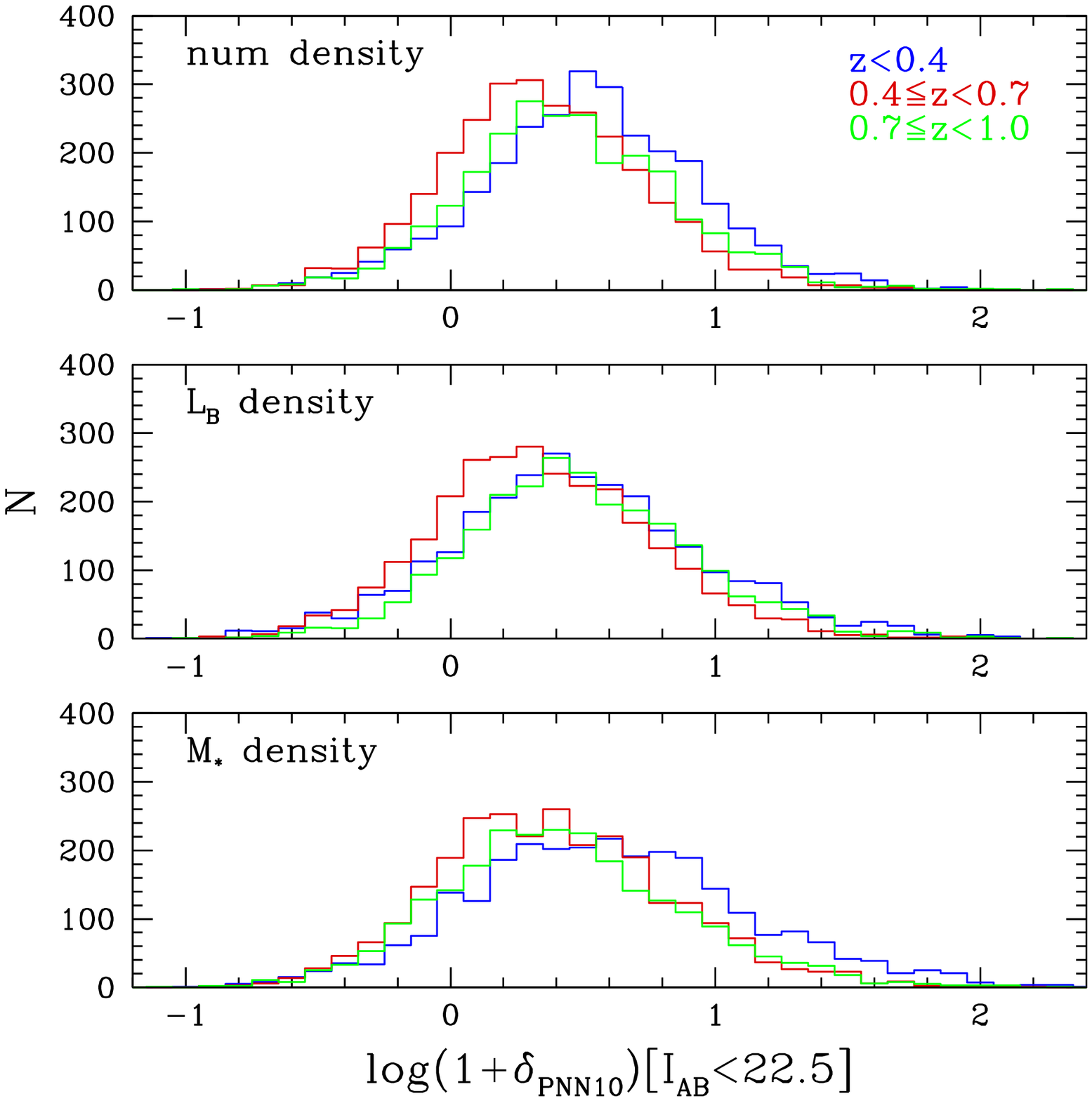}
\includegraphics[width=0.3\textwidth]{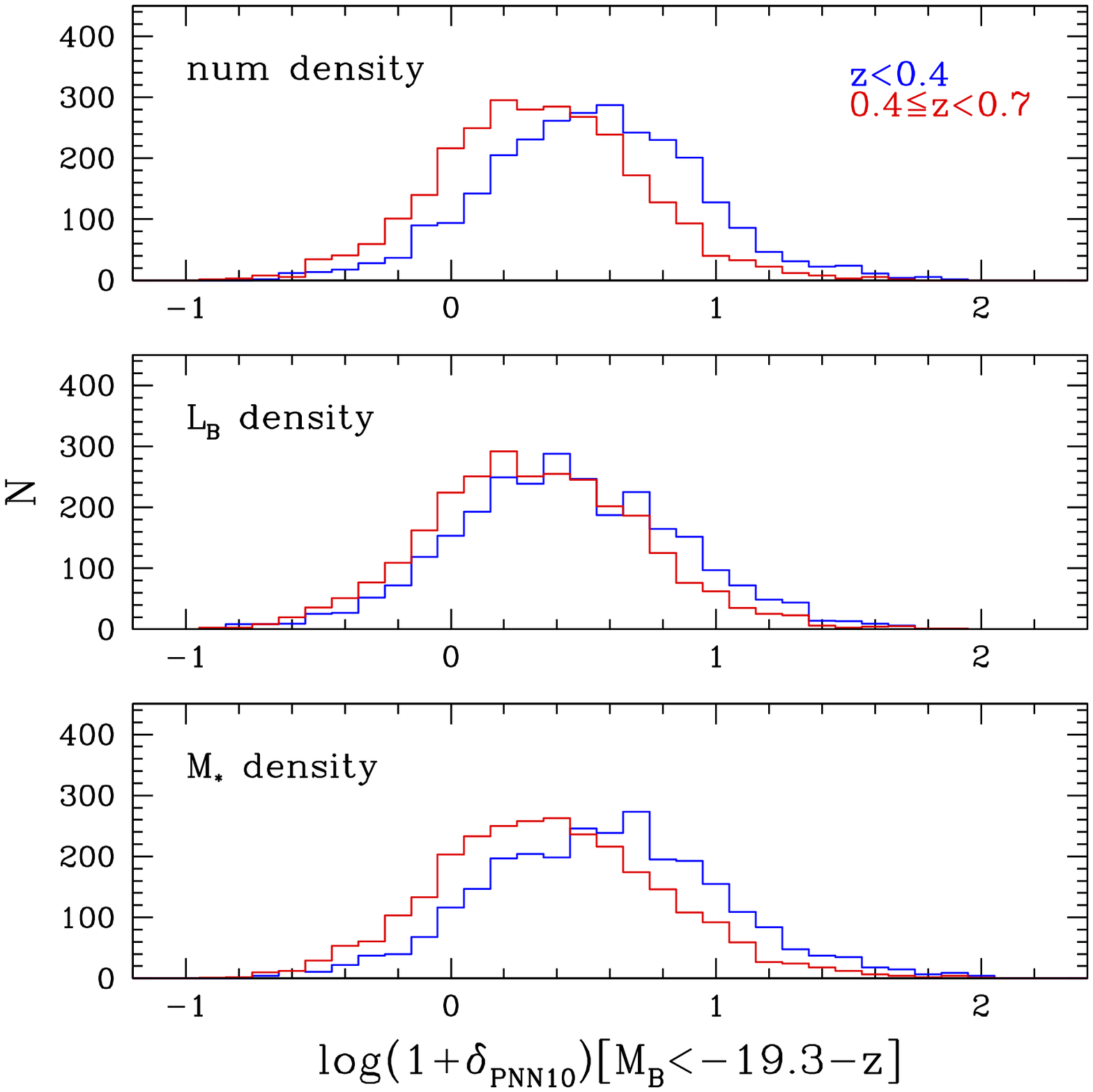}
\includegraphics[width=0.3\textwidth]{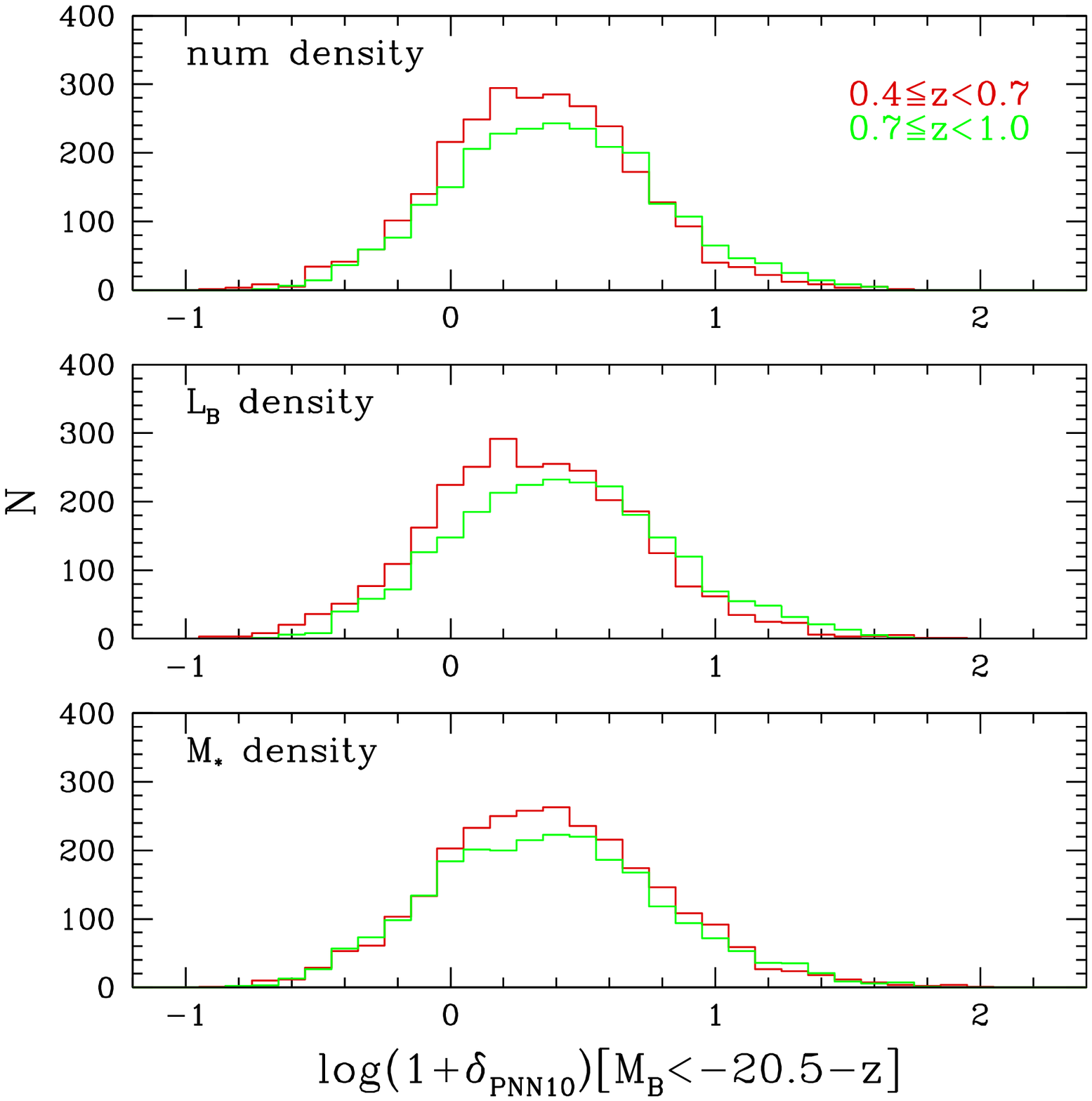}
\caption{\label{fig_cnn10}Histogram distributions of number of galaxies in the bins of overdensities $\log(1+\delta_p)$. The overdensities are reconstructed at positions of the 10k zCOSMOS galaxies, using the cylindrical adaptive filter $W$ (equation~\ref{eq_wproj}) with smoothing length defined by the 10th nearest neighbour. The distributions are  shown separately for three different redshift bins: $z<0.4$ (blue), $0.4 \le z < 0.7$ (red) and $0.7 \le z < 1$ (green). We use mass-weighting $m_i=1$, $m_i=L_B$ and $m_i=M_*$ for the reconstructions presented in the upper, middle and lower panels, respectively. A set of these three distributions of overdensities reconstructed for the  $I_{\rm AB}<22.5$, $M_B<-19.3-z$ and $M_B<-20.5-z$ samples of tracer galaxies is presented in the left, middle and right, respectively.}
\end{figure}

\begin{figure}
\includegraphics[width=0.3\textwidth]{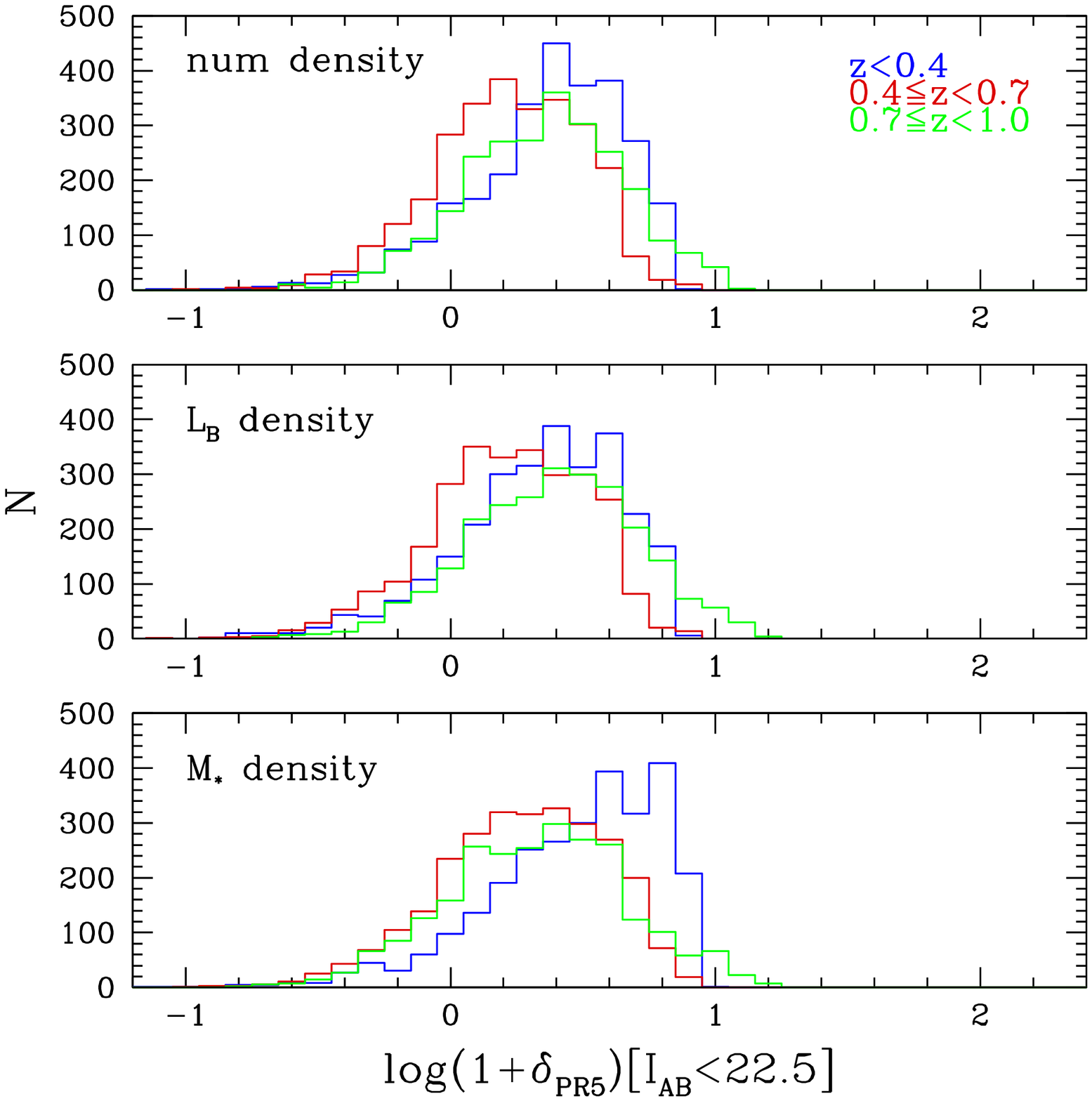}
\includegraphics[width=0.3\textwidth]{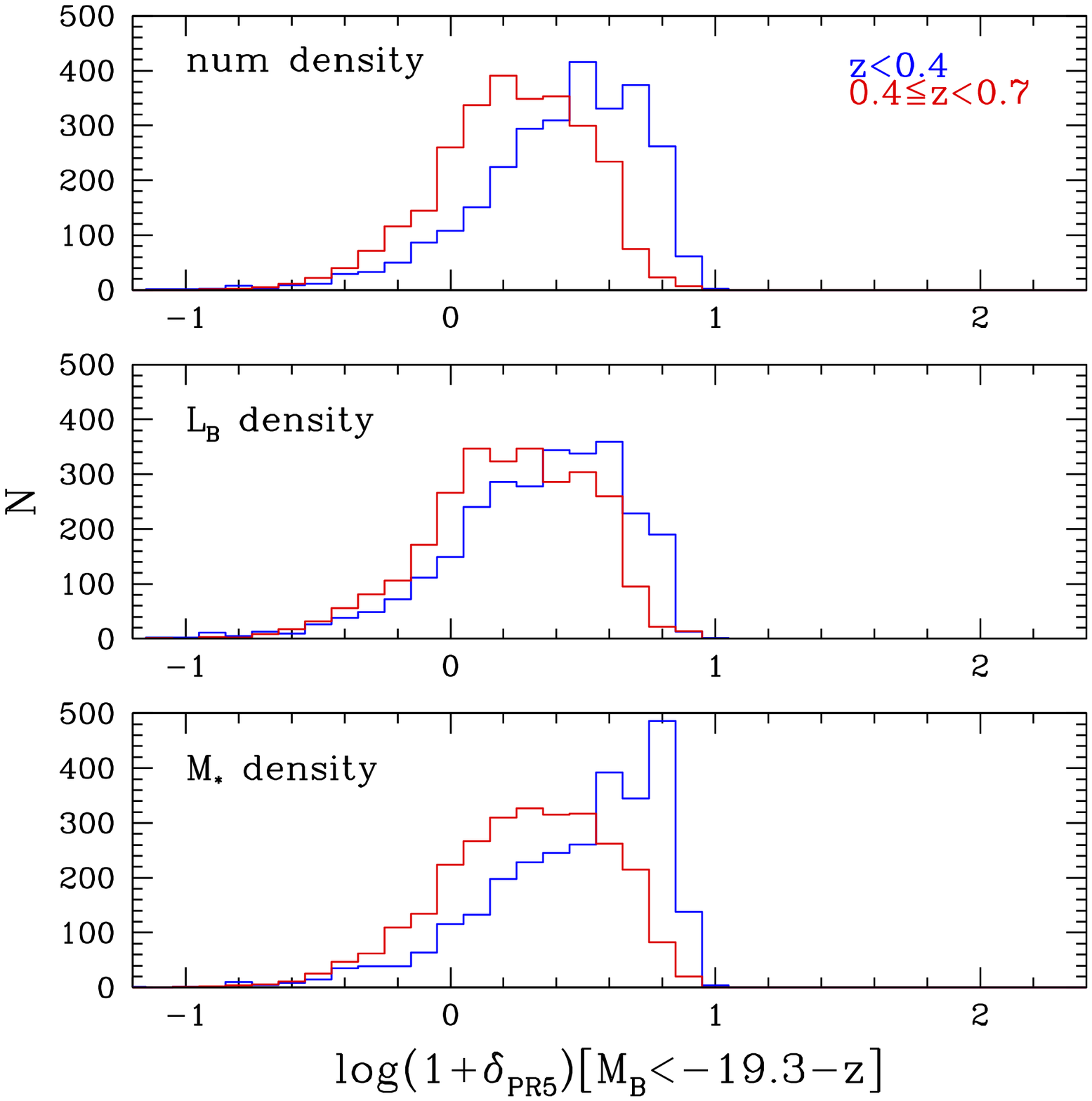}
\includegraphics[width=0.3\textwidth]{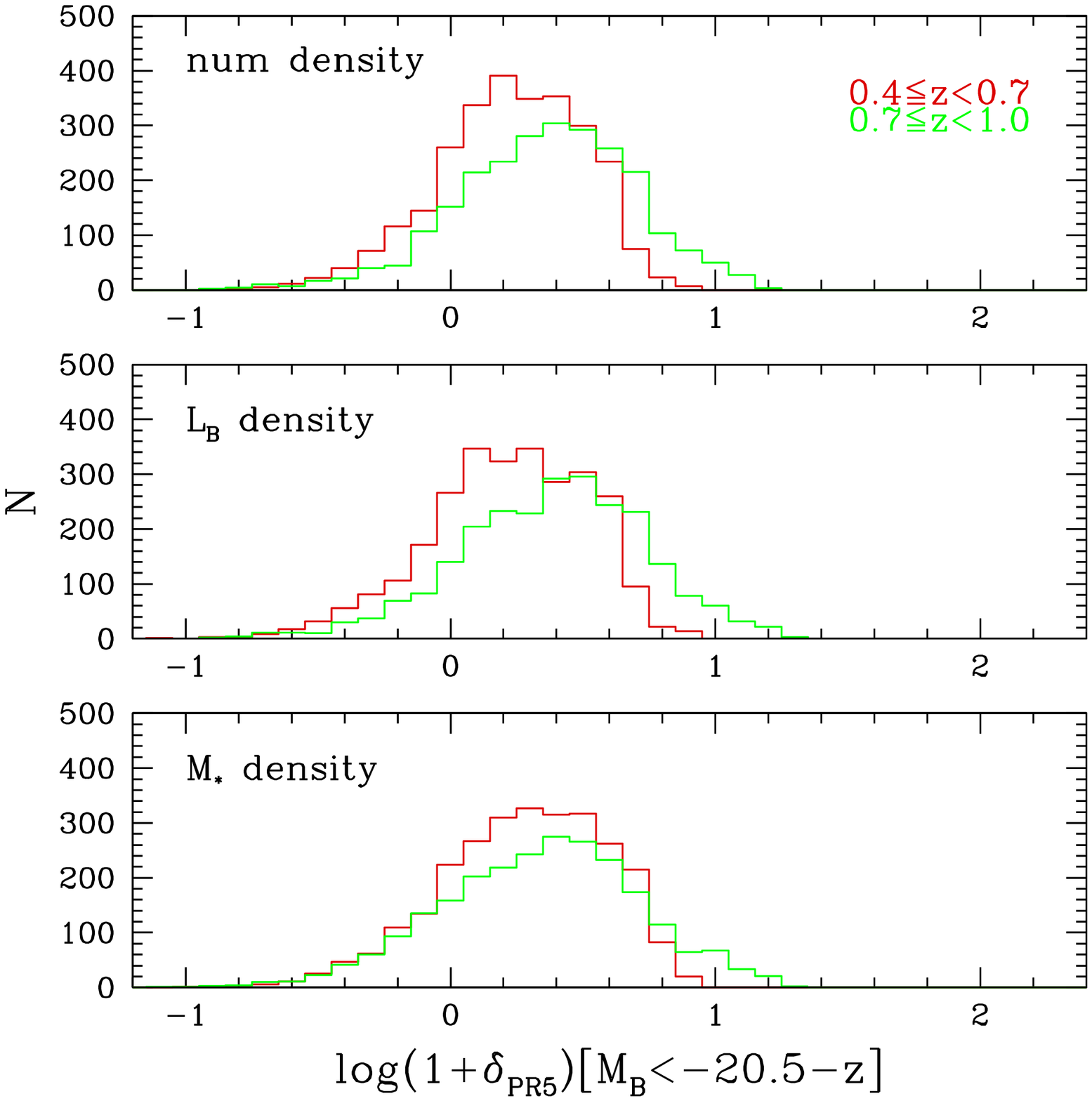}
\caption{\label{fig_cr5}Histogram distributions of number of galaxies in the bins of overdensities $\log(1+\delta_p)$. The overdensities are reconstructed at positions of the 10k zCOSMOS galaxies, using the cylindrical fixed filter $W$ (equation~\ref{eq_wproj}) with smoothing length defined to be 5 \hh Mpc. The distributions are  shown separately for three different redshift bins: $z<0.4$ (blue), $0.4 \le z < 0.7$ (red) and $0.7 \le z < 1$ (green). We use mass-weighting $m_i=1$, $m_i=L_B$ and $m_i=M_*$ for the reconstructions presented in the upper, middle and lower panels, respectively. A set of these three distributions of overdensities reconstructed for the  $I_{\rm AB}<22.5$, $M_B<-19.3-z$ and $M_B<-20.5-z$ samples of tracer galaxies is presented in the left, middle and right, respectively.}
\end{figure}

Many of these effects can be seen in Figures~\ref{fig_cnn10}  and  ~\ref{fig_cr5}, where we  present the histogram  distributions of
overdensities  reconstructed  around   the  10k  zCOSMOS  galaxies.   The overdensities  are
reconstructed in  the apertures defined by the  10th nearest neighbour
(Figure~\ref{fig_cnn10})  and  by  the   fixed  scale  of  5  \hh  Mpc
(Figure~\ref{fig_cr5}), with three types  of tracer galaxies: $I_{\rm AB}<22.5$
(left panels), $M_B<-19.3-z$  (middle panels) and $M_B<-20.5-z$ (right
panels)  and  three types  of  weighting:  $m_i=1$  (top row  panels),
$m_i=L_B$ (middle row panels) and $m_i=M_*$ (bottom row panels).

The advantage  of  the use  of the  volume
limited sample is that, at every  redshift,  we are using the same
type of objects  to define the environment, even  though the properties of
these  tracers can  slowly  change  with redshift  due  to effects  of
evolution,  which we  cannot  (yet)  take out  completely. The
number density  of these tracer  galaxies is roughly constant, thus
the Poisson  noise and smoothing scale (if relevant) in the estimated  density is on average the same at all redshifts.

On  the other hand, when selecting  galaxies in the total flux
limited sample, the number of  available tracers is larger, and one can
reliably reconstruct environments on  smaller scales than when using
the  volume limited  sample. This  can  be critical  for studying 
galaxy  properties as  a function  of environment.  However, different
populations  of galaxies  will be  used to  define the  environment at
different  redshifts and the  typical
smoothing scale  will change with  redshift in the adaptive approach. 
If  using the
fixed scale, systematically a different number of objects will be used
to measure the environment  at different redshifts and therefore the noise
component   in   the  estimate   will   be   different  at   different
redshifts.  Normalisation to the  mean overdensity  (at a  redshift of
consideration) can overcome some of these effects.

\subsection{Cosmographical tour of the 10k zCOSMOS survey}

The global  picture of  the LSS structures  traced by the  10k zCOSMOS
galaxies is obtained by the reconstruction of the density field on the
grid filling the zCOSMOS volume.  For the zCOSMOS data the overdensity
is reconstructed  on a regular grid with  spacing of 1 \hh  Mpc in all
three directions (along RA, DEC and z axis).  All tracer galaxies  within $\pm$ 1000
\kms\ are first  set to the redshift of the grid  point, and then used
to obtain  the projected  distance to the  5th, 10th and  20th nearest
neighbour (as  for the  mocks). The presented  overdensity field  is the projected surface  overdensity field $\delta_p$, and the values of the full
three  dimensional overdensity  field $\delta$ at the same smoothing
scale $R$ would be larger, $1+\delta = (1+\delta_p)^{(3/2)}$. The overdensity field reconstructed in this way is
not suited for the cosmological  analysis, it is more a compilation of
local environments as they would be experienced by an arbitrary object
residing in the zCOSMOS volume.  We use this overdensity field for the
cosmographical tour of the zCOSMOS survey.

\begin{figure}
\centering
\includegraphics[width=0.43\textwidth]{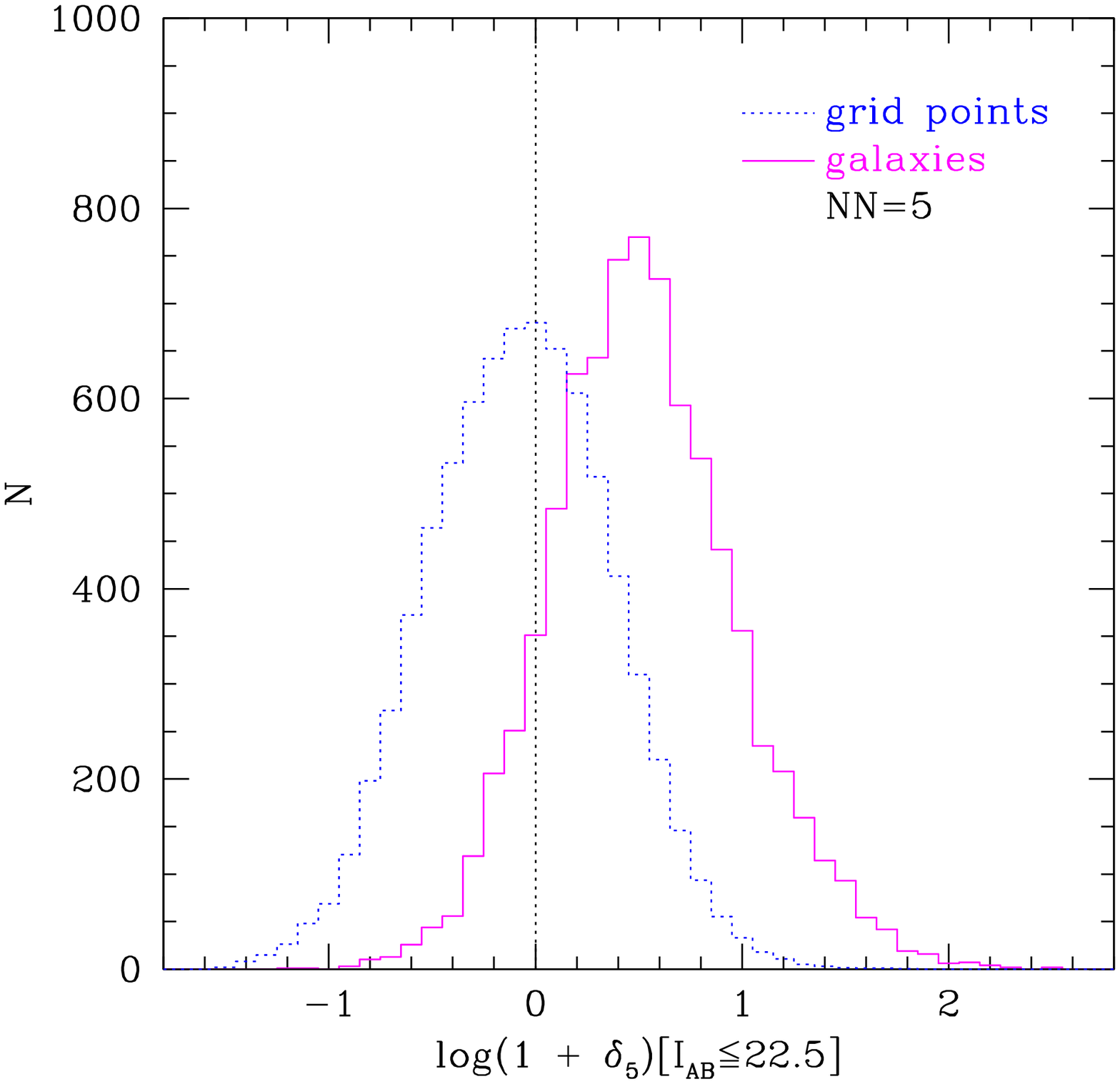}
\includegraphics[width=0.43\textwidth]{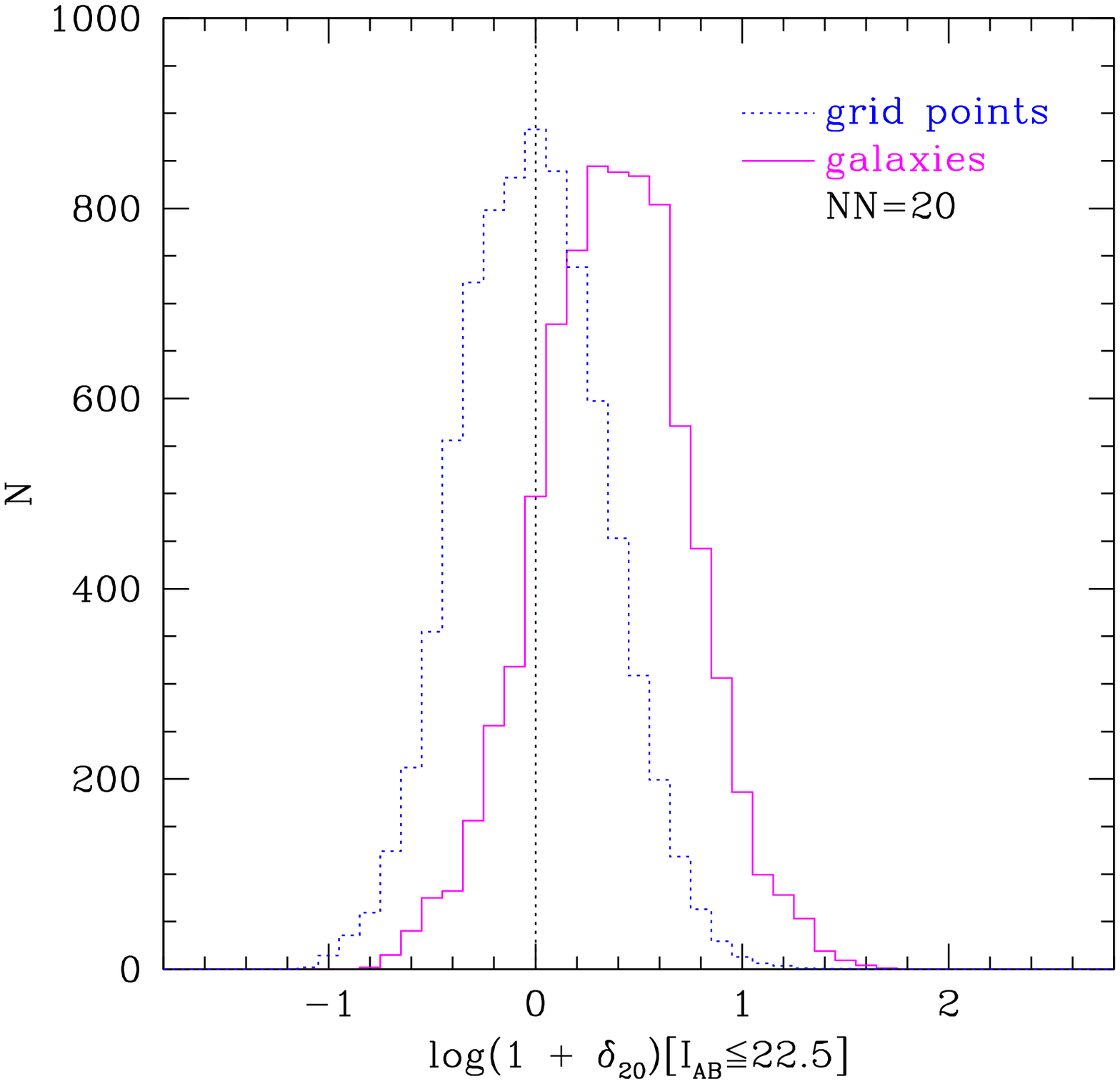}
\caption{\label{fig_comp_odensgalgr_nnf}Histogram of the overdensities in the zCOSMOS volume calculated on the grid points (blue) and at the positions of zCOSMOS galaxies (magenta). The histogram of the grid points is scaled to the number of galaxies with high quality redshifts. Galaxies are not distributed homogeneously in the zCOSMOS volume, they form clumps of the overdense regions leaving some of the volume empty.}
\end{figure}

We first  compare the  distributions of  overdensities reconstructed with the flux  limited tracer
galaxies  on the
grid, and at  the positions of  zCOSMOS galaxies,  in  Figure~\ref{fig_comp_odensgalgr_nnf}.  The
number  of  grid points  is  scaled to  match  the  number of  zCOSMOS
galaxies with  a high quality  redshift in $0.1<z<1$. The smoothing of
the overdensity  structures is clearly  visible in  both the overdense
and underdense  tails  when comparing
the overdensity  values reconstructed  using the apertures  defined by
the  20th  with respect  to  the  5th  nearest neighbour, as might be expected.   The  most underdense regions seen on the grid
 are more underdense than the lowest $1+\delta_p$ values obtained centred on 
galaxies. The  peak in
distribution  of overdensities  at  positions of  galaxies is also shifted
towards positive overdensity values [by $\log(1+\delta_5) \sim 0.5$
and  $\log(1+\delta_{20}) \sim  0.4$] with  respect to  the overdensity
field reconstructed  on the grid  points.  Again, the  distribution of
overdensity  values on  the grid  does not  represent  the universal
volume distribution of overdense  and underdense regions. In \citet{Kovac.etal.bias}, we will present the distribution of overdensity
 values
obtained from  the full three dimensional density reconstruction  on the fixed
scale.

\begin{figure}
\includegraphics[width=0.45\textwidth]{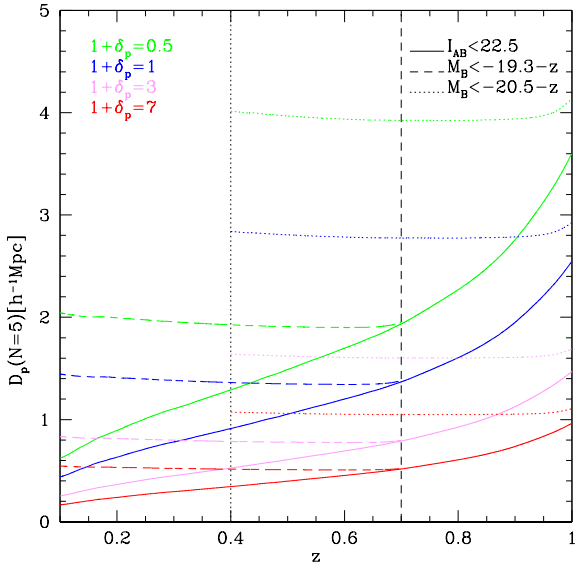}
\caption{\label{fig_nndist}Projected distances, defined by the distance to the 5th nearest neighbour, used to define the aperture for the reconstruction of densities centred on the 10k zCOSMOS galaxies. The continuous, long-dashed and short-dashed lines are for the $I_{\rm AB}<22.5$, $M_B<-19.3-z$ and $M_B<-20.5-z$ samples of tracer galaxies. The lines for the overdensities of 0.5, 1, 3 and 7 in $1+\delta_p$ units are presented in green, blue, pink and red, respectively.}
\end{figure}

\begin{figure}
\centering
\includegraphics[width=0.23\textwidth]{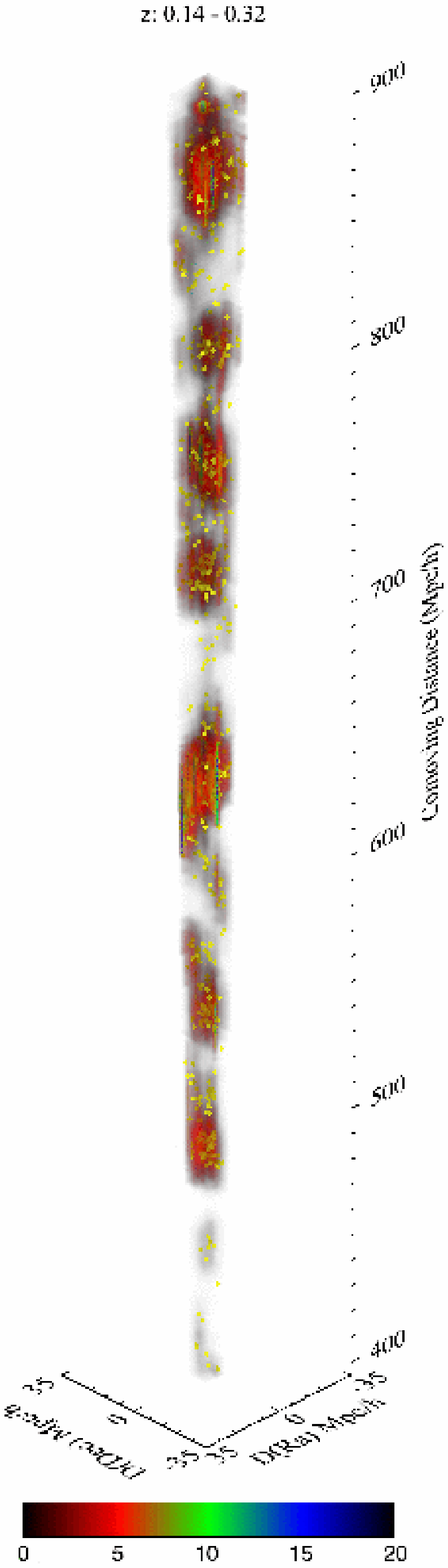}
\includegraphics[width=0.23\textwidth]{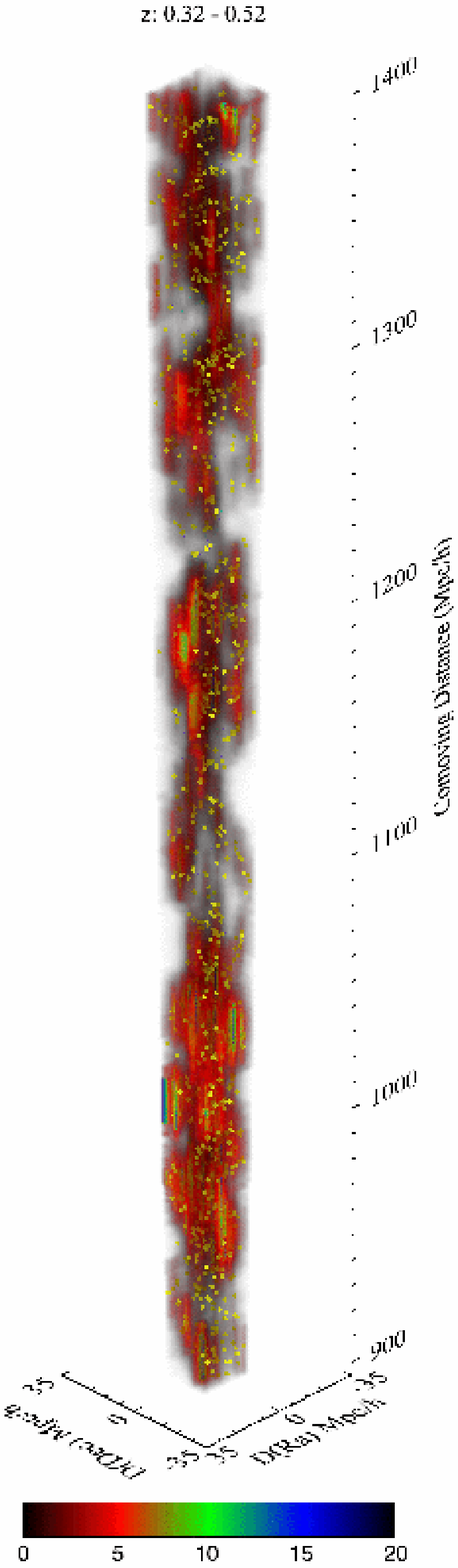}
\includegraphics[width=0.23\textwidth]{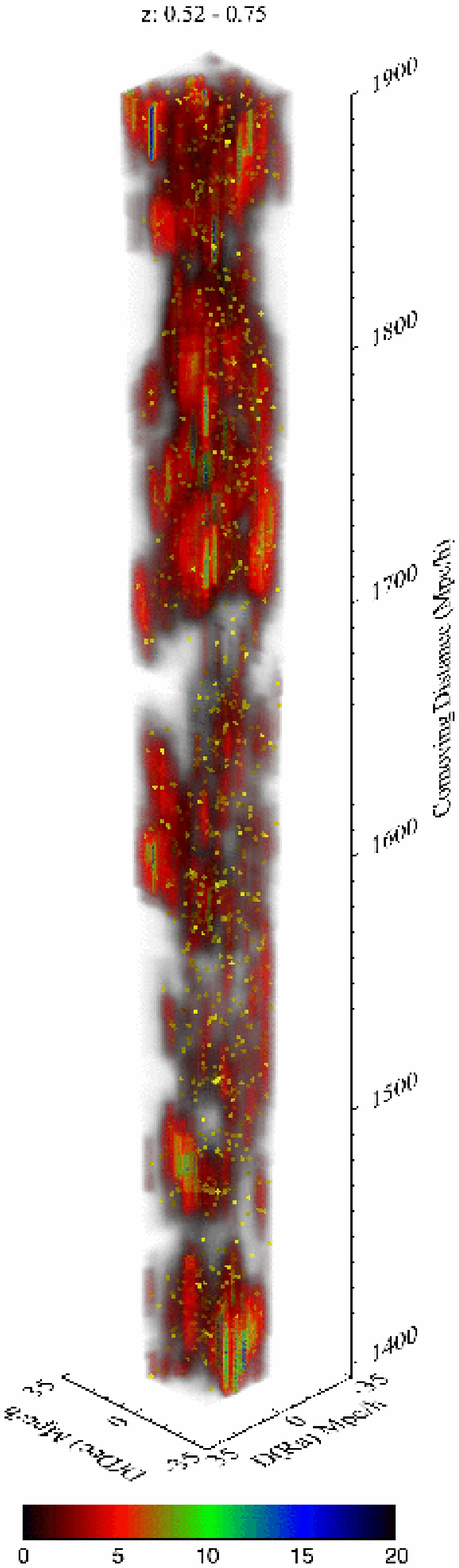}
\includegraphics[width=0.23\textwidth]{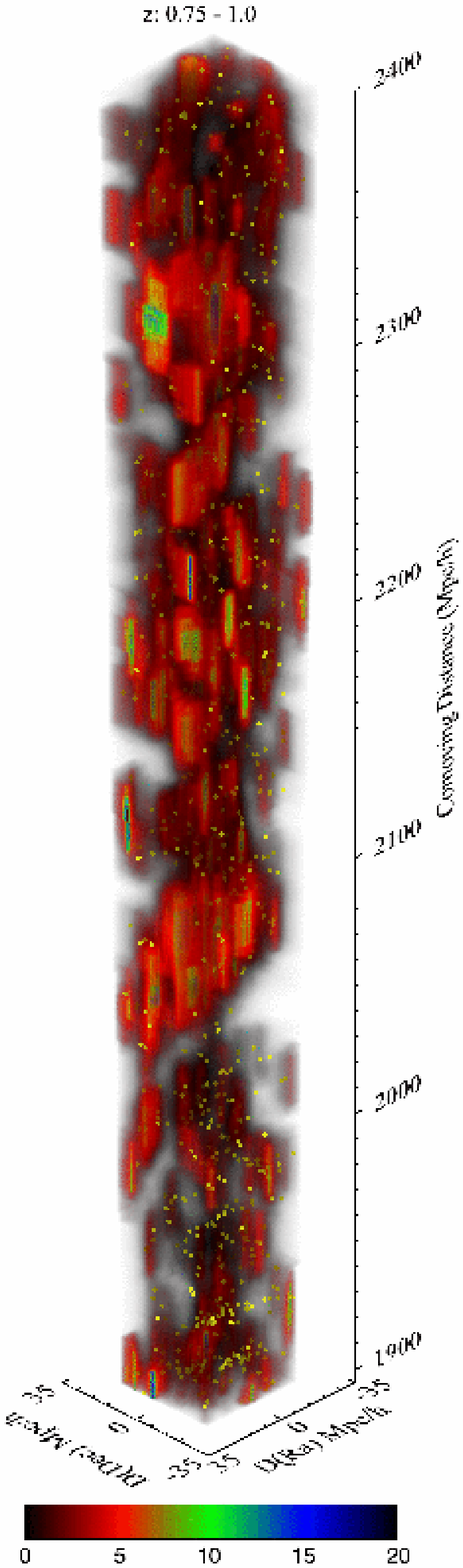}
\caption{\label{fig_10kgrayscale} Cosmographical tour of the zCOSMOS survey. The 10k zCOSMOS overdensity field is reconstructed on the grid with the flux limited 10kspec+30kZADE sample of tracer galaxies. The aperture is defined by the distance to the 5th nearest neighbour projected within $\pm$ 1000 \kms\ to redshift of the individual grid points. All three axis, distance from the survey centre in right ascension, distance from the survey centre in declination and distance (corresponding to redshift), are expressed in comoving \hh Mpc. The reconstructed overdensity field yields structures in a large range
of comoving  scales covering a spectrum of  different overdensities at
all redshifts reliably probed by  the 10k zCOSMOS survey. The colour scale of the $1+\delta_p$ values is presented below each cone covering 500 \hh Mpc. The positions of galaxies are marked with yellow dots.}
\end{figure}

\begin{figure}[tb]
\centering
\includegraphics[height=0.65\textheight]{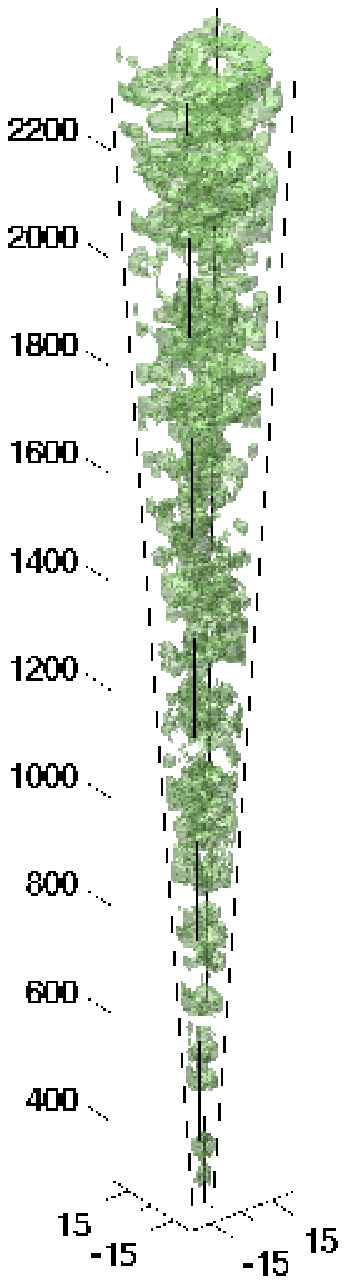}
\includegraphics[height=0.65\textheight]{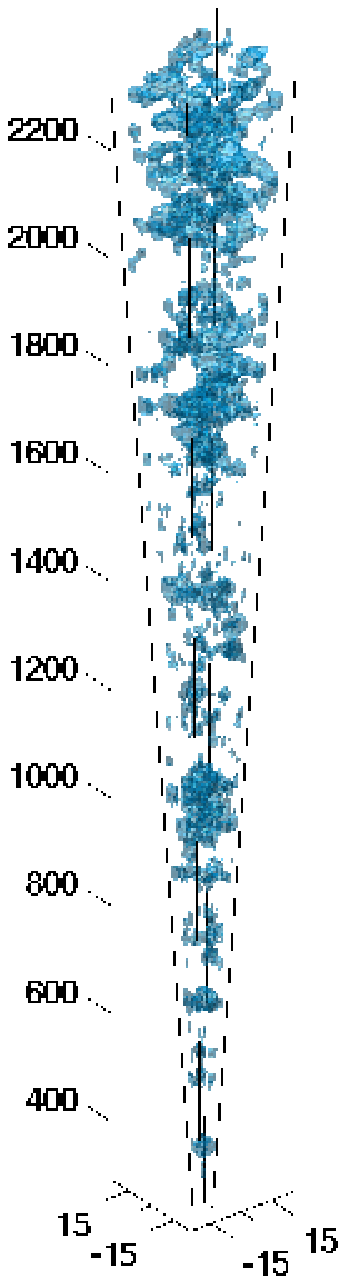}
\includegraphics[height=0.65\textheight]{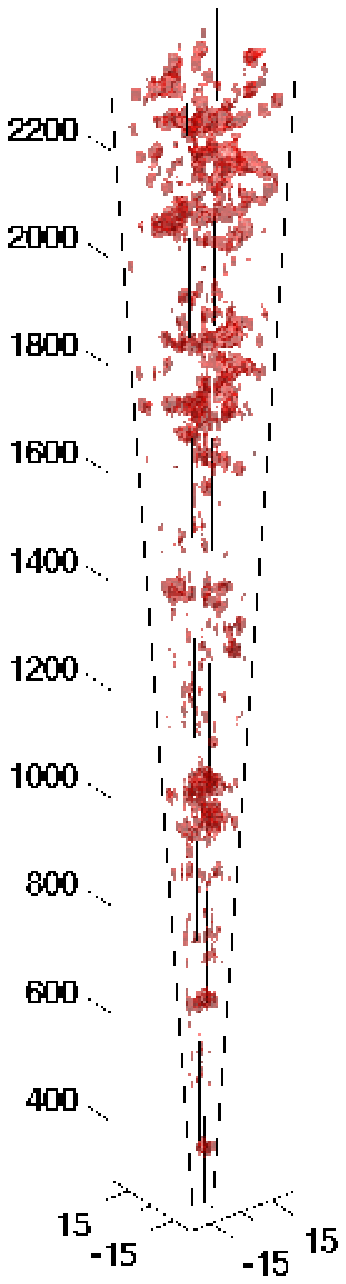}
\includegraphics[height=0.65\textheight]{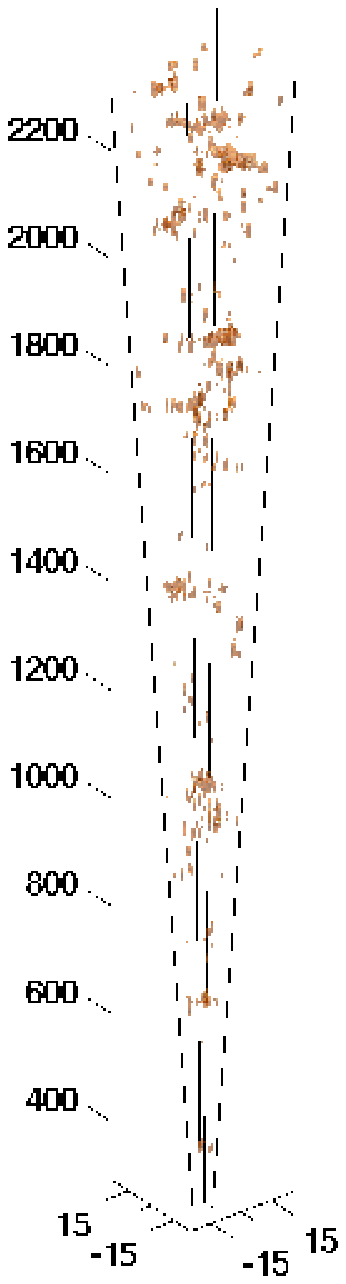}
\caption{\label{fig_10koverdense}Distribution of the overdense structures in the 10k zCOSMOS overdensity field. The structures are defined as isosurfaces enclosing regions with $1+\delta_p \ge$ 1.5, 3, 5 and 10 ($1 + \delta \ge$ 1.8, 5.2, 11.2 and 31.6) from the left to the right, respectively. Comoving distances in the radial direction correspond to $0.1<z<1$. The distribution of overdense structures matches the richness and complexity of the local cosmic web up to $z \sim 1$. The structures are delineated from the overdensity field reconstructed following the same scheme as in Figure~\ref{fig_10kgrayscale}. All three axis, distance from the survey centre in right ascension (left hand side axis), distance from the survey centre in declination (right hand side axis) and distance (corresponding to some redshift; vertical axis) are expressed in comoving \hh Mpc. The figures are compressed by a factor of $\sim 3.5$ in redshift direction.}
\end{figure}

\begin{figure}
\centering
\includegraphics[height=0.67\textheight]{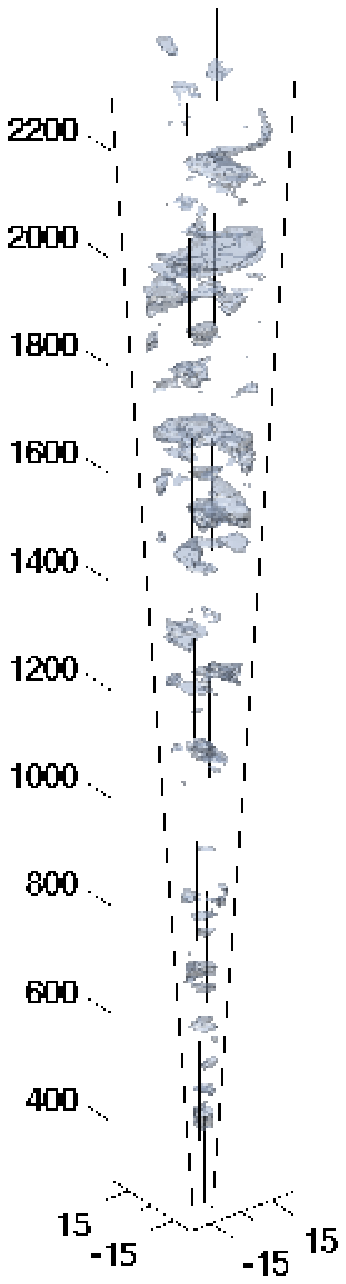}
\includegraphics[height=0.67\textheight]{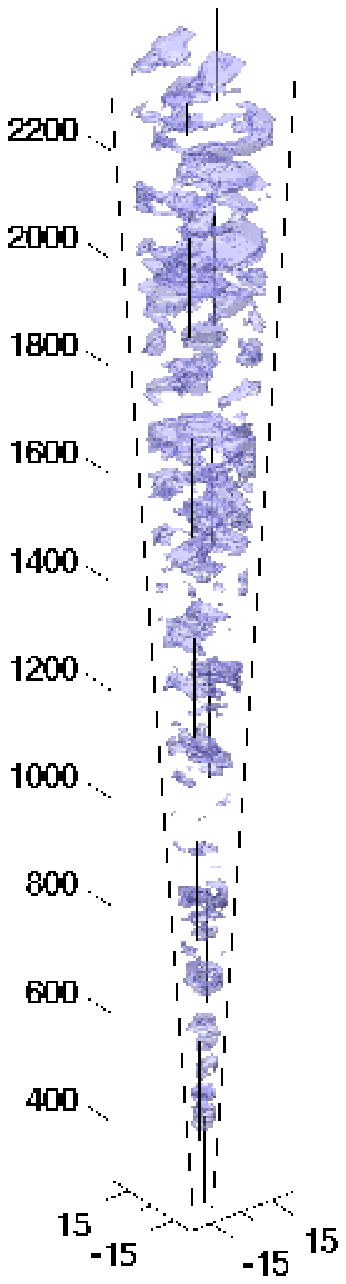}
\caption{\label{fig_10kunderdense}The 10k zCOSMOS underdense regions (voids).  The structures are defined as isosurfaces enclosing regions with $1 + \delta_p \le$ 0.15 (6.67 times smaller than $\bar{\rho}(z)$) and 0.25 (4 times smaller than $\bar{\rho}(z)$) in the left and the right cone, respectively. Comoving distances in the radial direction correspond to $0.1<z<1$. The structures are delineated from the overdensity field reconstructed following the same scheme as in Figure~\ref{fig_10kgrayscale}. All three axis, distance from the survey centre in right ascension (left hand side axis), distance from the survey centre in declination (right hand side axis) and distance (corresponding to some redshift; vertical axis) are expressed in comoving \hh Mpc. The figures are compressed by a factor of $\sim 3.5$ in redshift direction.}
\end{figure}

A visual representation of the zCOSMOS overdensity field  is presented in Figures~\ref{fig_10kgrayscale}, \ref{fig_10koverdense} and \ref{fig_10kunderdense}. For these presentations we use the overdensity field reconstructed on scales defined by the distance to the projected 5th nearest neighbour in the sample of flux limited  tracer  galaxies. Note that these imply that the actual smoothing scale is increasing with redshift (see Figure~\ref{fig_nndist}), and that different populations of galaxies are used in different redshifts to reconstruct the overdensity field. However, using the smoothing scale defined by the 5th nearest neighbour in the flux limited sample we are able to obtain the broadest possible dynamical range of the overdensity field and to preserve at best the variety of the structures in the zCOSMOS survey. As \citet{Strauss&Willick.1995} point out, a flux limited redshift survey ``...is useful for qualitative and cosmographical description of the structures that are seen, and in some sense shows the maximum amount of information in the redshift survey''.

The reconstructed overdensity field, presented in Figure~\ref{fig_10kgrayscale}, shows structures in a large range
of comoving  scales covering a spectrum of  different overdensities at
all redshifts reliably probed by  the 10k zCOSMOS survey. Galaxies are
distributed  into   cluster-like  structures,  surrounded   by  empty,
void-like   regions  up  to   $z=1$. A few points should be noted. First, the increased smoothness and extension of the
 structures in the overdensity field towards higher redshift is an artefact of the increased smoothing scale with redshift. Second, for the smallest scale structures the imprint of the filter $W$ used is also clearly visible, especially in Figure~\ref{fig_10kgrayscale}. Even though sampling in the zCOSMOS field is not uniform, the resolution of similar structures is the same, given that we use ZADE to take into account galaxies without spectroscopic redshift and keep their position on the sky.

Figure~\ref{fig_10koverdense} provides another view of the complex structure of this overdensity field, but compressed by a factor of $\sim 3.5$ in redshift direction. Here, we present the LSS delineated by the isosurfaces enclosing regions with $1+\delta_p \ge$ 1.5, 3, 5 and 10 ($1 + \delta \ge$ 1.8, 5.2, 11.2 and 31.6) going from the left to the right, respectively. The isosurface structures up to $1+\delta_p \ge 5$ are connected over  the transverse comoving scales  covering the full  zCOSMOS area,
and   coherent over hundred or  more comoving \hh  Mpc (or more than $\Delta z \sim 0.1$) in the
radial direction.

Three large structures dominate the zCOSMOS cosmic web at $z \le 1$. These structures are located at $z \sim 0.35$ (comoving distance $D_c \sim 980$  \hh Mpc) , $\sim 0.7$ ($D_c \sim 1800$ \hh Mpc) and $\sim 0.85$ ($D_c \sim 2100$ \hh Mpc) and they correspond to the peaks already visible in the number distribution of galaxies with redshift (see Figure~\ref{fig_nz}). The connectivity of the two largest high redshift structures extend over 200 \hh Mpc in the radial direction. Even though we are working with a flux limited sample of tracer galaxies, meaning that we are detecting only the brightest and presumably most massive galaxies, at $z \sim 0.9$ the smoothing scale for the overdensities $1+\delta_p \ge 3$ are still below or slightly larger than 1 \hh Mpc, see Figure~\ref{fig_nndist}. Based on the tests on the mock catalogues, which include the selection effects of the 10k zCOSMOS survey, the existence of these huge overdense structures at high $z$ is not an artefact of our reconstruction method. The other striking element in the zCOSMOS overdensity field is a very small number of $1+\delta_p \ge 3$ (and above) structures in $0.4 \lesssim z \lesssim 0.6$ ($1100 \lesssim D_c \lesssim 1580$ \hh Mpc) except for an overdense sheet at $z \sim 0.53$ ($D_c \sim 1420$ \hh Mpc).

Complementary to the large positively overdense structures, we identify in the zCOSMOS $1+\delta_p$ field also the structures enclosed by the overdensity values of $1 + \delta_p =0.15$ ($1 + \delta = 0.06$) and $\delta_p=0.25$ ($\delta = 0.13$) which contain only regions with $1 + \delta_p$ lower than these (the underdense regions, Figure~\ref{fig_10kunderdense}). The identified regions are at least 6.67 and 4 times less dense than the mean density, respectively. These regions with low galactic density do not show strong clustering in either the transversal (RA-DEC) or radial (redshift) direction. Also, they appear to be more homogeneously distributed over the zCOSMOS redshift range, even though they are less present in the redshift ranges of the three most overdense structures, being particularly absent at $z \sim 0.35$ ($D_c \sim 980$ \hh Mpc).

\begin{figure}
\includegraphics[width=0.24\textwidth]{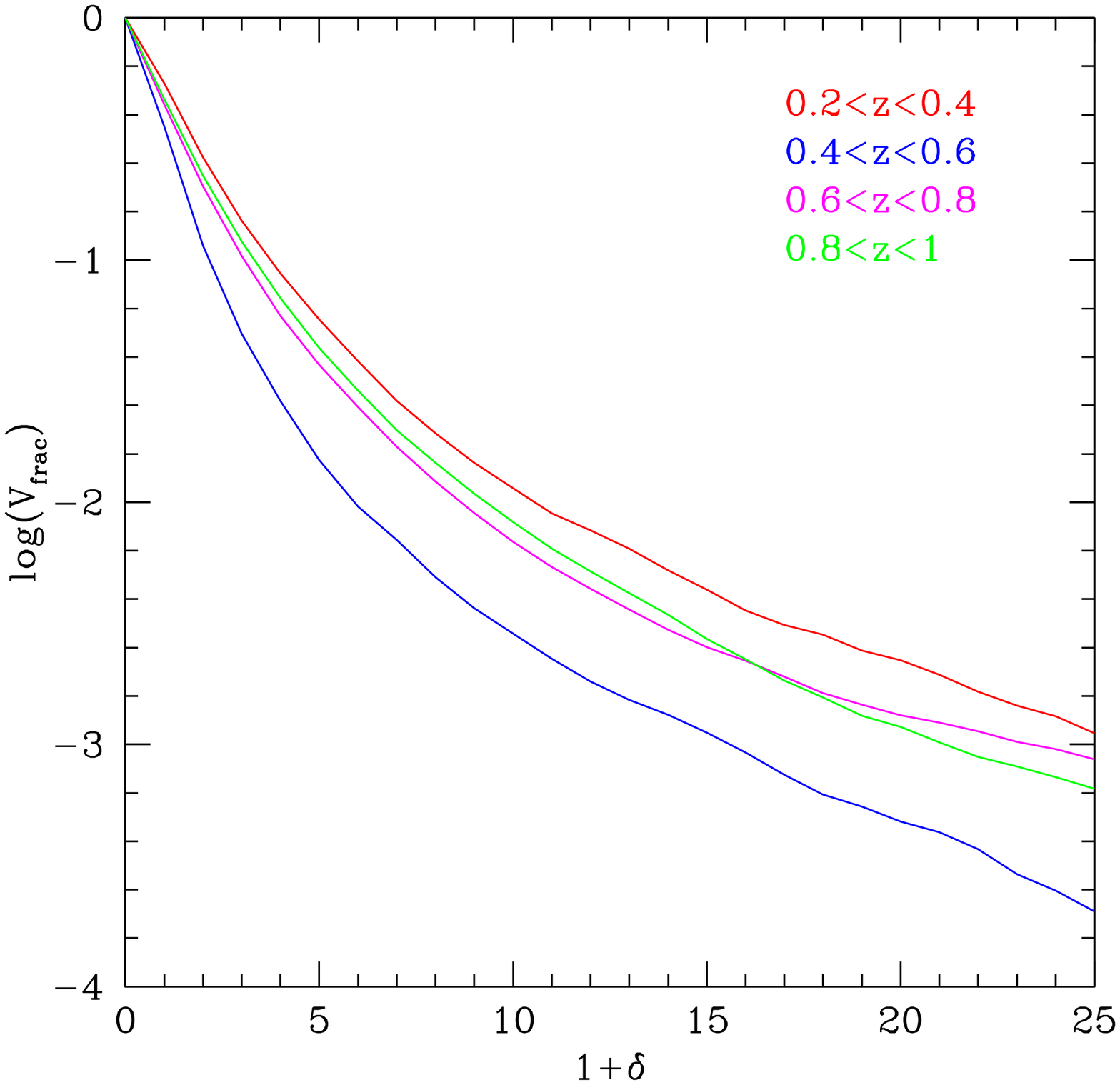}
\includegraphics[width=0.24\textwidth]{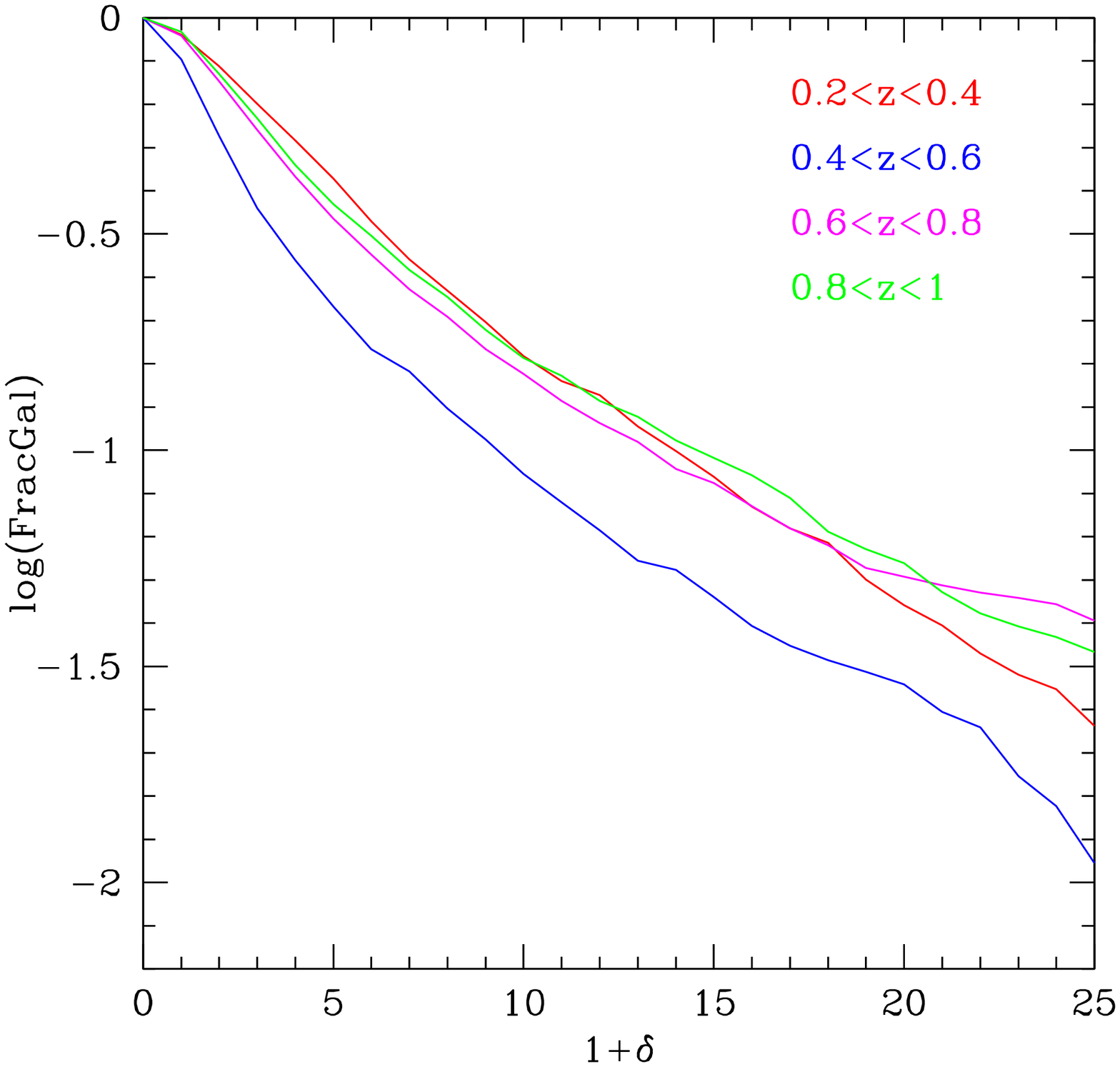}
\includegraphics[width=0.24\textwidth]{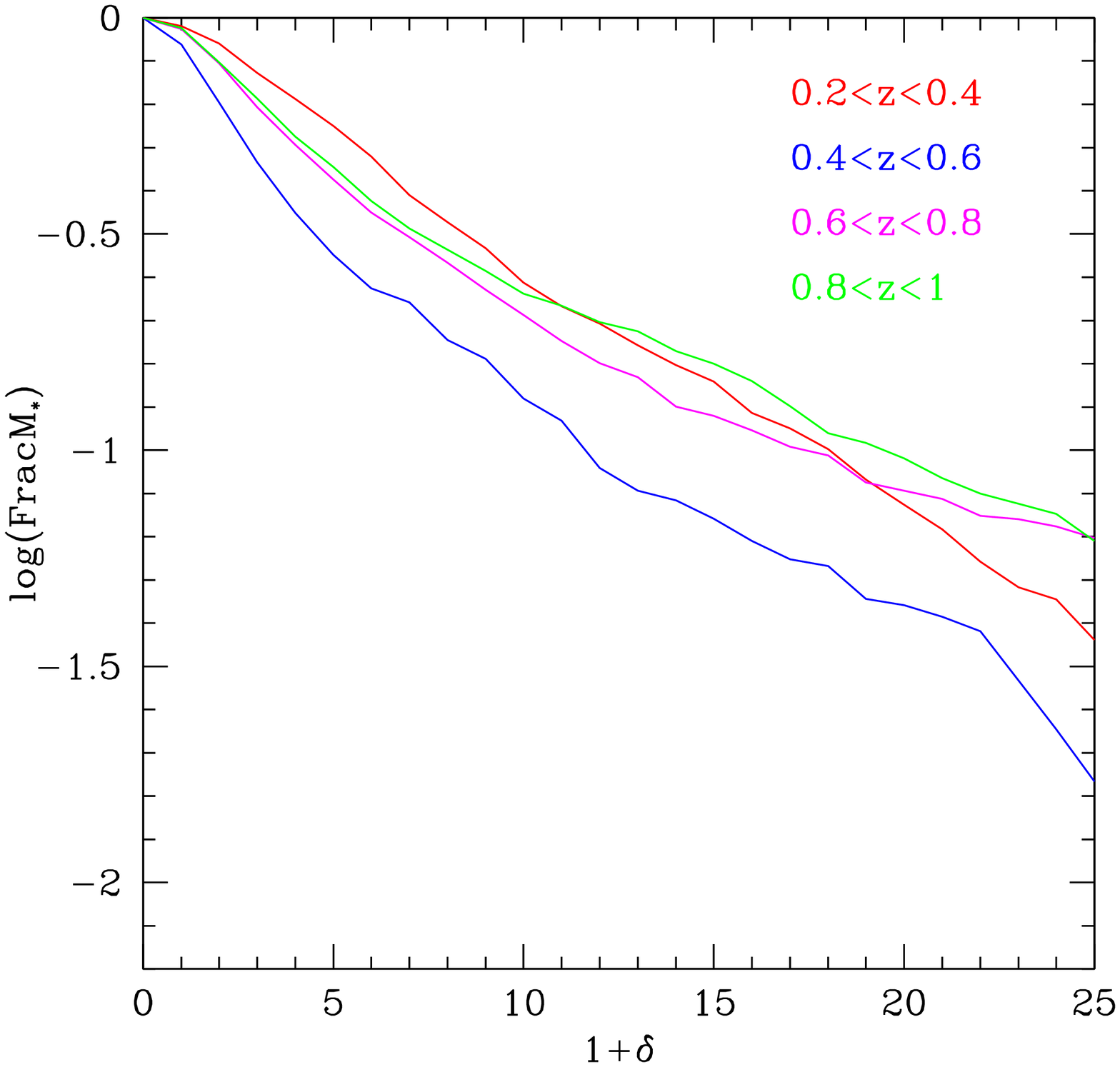}
\includegraphics[width=0.24\textwidth]{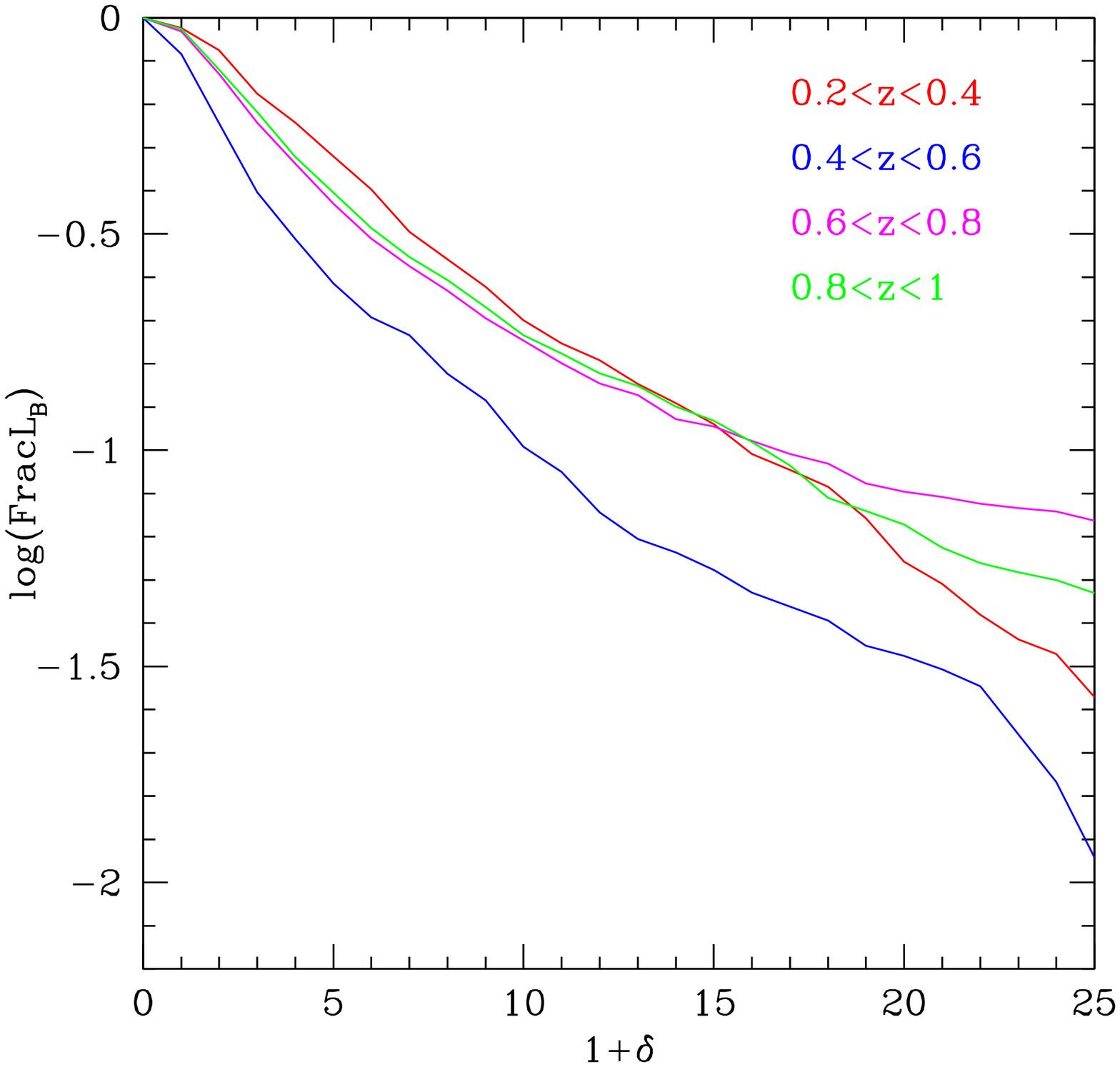}
\caption{\label{fig_fractionsinodens}Fractions of volume, galaxies, stellar masses and B-band luminosities enclosed in the structures defined by the given overdensity $1+\delta_p$ values, going from the left to the right. These fractions are presented in a narrow redshift slices $\Delta z = 0.2$, starting from $z=0.2$. The resulting fraction are dominated by cosmic variance. However, it is clear that the overdense structures are much more important in terms of their baryonic content than in the terms of the volume which they occupy. Note the different scale of the y-axis in the first panel.}
\end{figure}

Clearly, the amount of LSS in zCOSMOS field varies with redshift. Quantitatively, this cosmic variance of the observed structures is shown in the left panel in Figure~\ref{fig_fractionsinodens}, where we plot the fraction of the volume of the zCOSMOS survey enclosed within the isosurface structures of a given $1+\delta_p$ value. We split the sample in four $\Delta z =0.2$ slices, starting from $z=0.2$. The fraction of the volume occupied by the structures with at least $1+\delta_p$ values is rapidly decreasing with $1+\delta_p$. The two higher redshift slices are statistically more representative (they occupy larger volumes), and we see that for the slices above $z=0.6$ about 44\% of the volume is in the structures with $1+\delta_p \ge 1$, and about 10\% of the survey volume is in the structures with $1 + \delta_p \ge 3$. The cosmic variance is particularly noticeable between the two lower redshift slices. At $1+\delta_p=3$ and $1+\delta_p=15$, the  volume fractions in those structures in the two lower redshift slices are different by a factor of $\sim 3$ and $\sim 4$, respectively.

The fraction of galaxies residing in the isosurface overdensity structures is much higher than the volume fraction of these structures. Taking the two higher redshift slices as statistically more representative, we conclude that about 50\% of the galaxies reside in structures with $1+\delta_p \ge 3.5$, while about 50\% of the stellar mass and about 50\% of the B-band luminosity at a given $z$ inhabits the structures with $1+\delta_p \ge 4$ (three right panels in Figure~\ref{fig_fractionsinodens}).  The overdense structures are few tens time more important in terms of the fractions of their baryonic content, than in terms of the volume fractions which they occupy.

\subsection{Comparison to the other LSS measures}

The high resolution of the zCOSMOS-bright spectra allows also to identify virialised groups of galaxies with velocity dispersion of $\sim$ 250 \kms. \citet{Knobel.etal.inprep} have applied the friends of friends and the Voronoi based group finding algorithms to the zCOSMOS galaxies with the high confidence redshift. In total, their optimal group catalogue contains 800 groups with at least 2 detected members up to redshift of 1 in the zCOSMOS volume. From these, 151 groups have at least 4 members detected. Tests on mocks show that the vast majority of all of the identified groups will be virialised objects.

The other tracer of the virialised structures is the hot baryonic gas detected in the X-rays. \citet{Finoguenov.etal.2007} carried out the identification of the X-ray clusters in the COSMOS field using the XMM-Newton observations. The updated catalogue, improved with new additional X-ray data and using the spectroscopic zCOSMOS redshifts, contains 218 X-ray clusters detected with high confidence \citep{Finoguenov.etal.inprep}.

\begin{figure}
\centering
\includegraphics[width=0.43\textwidth]{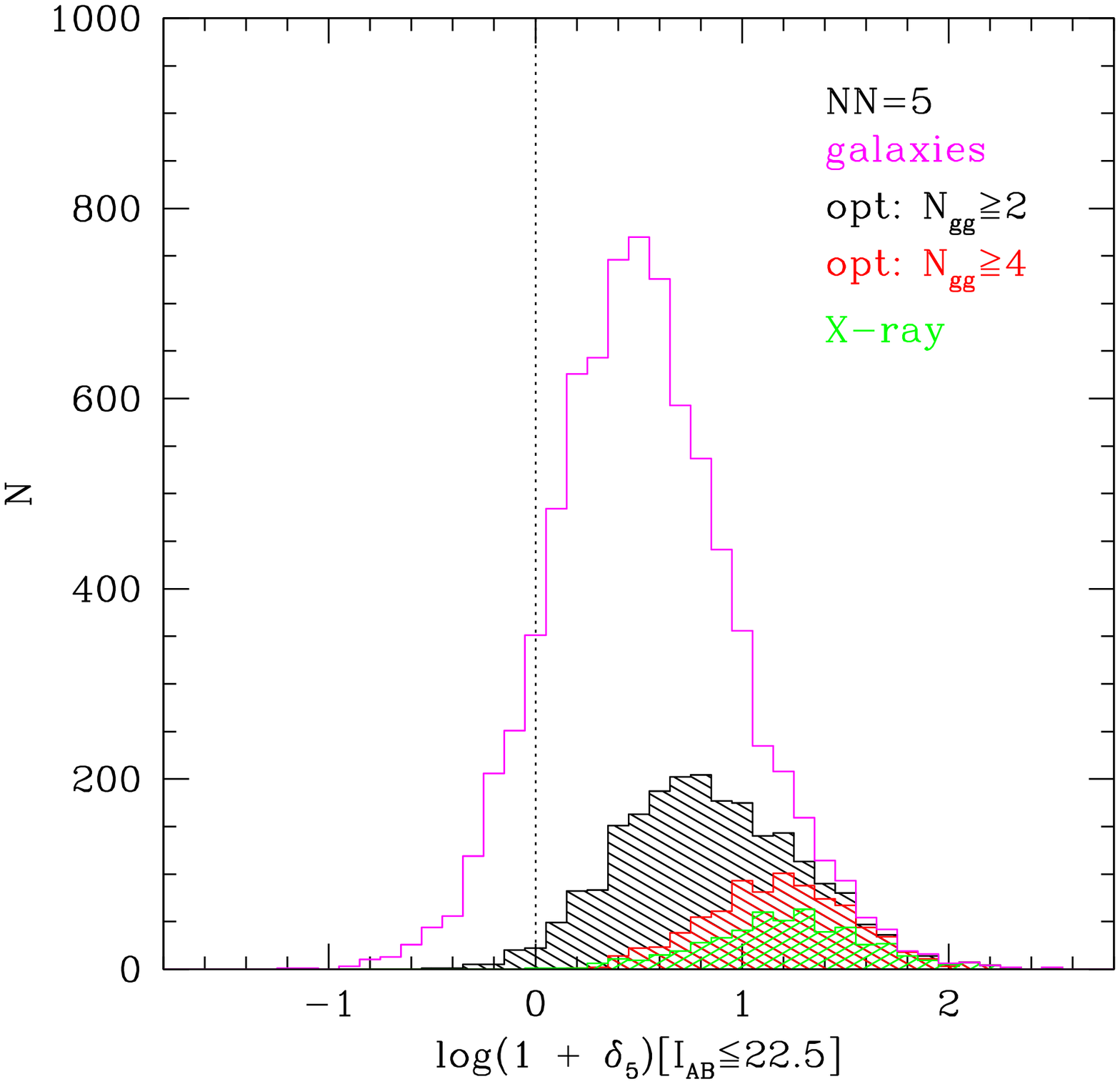}
\includegraphics[width=0.43\textwidth]{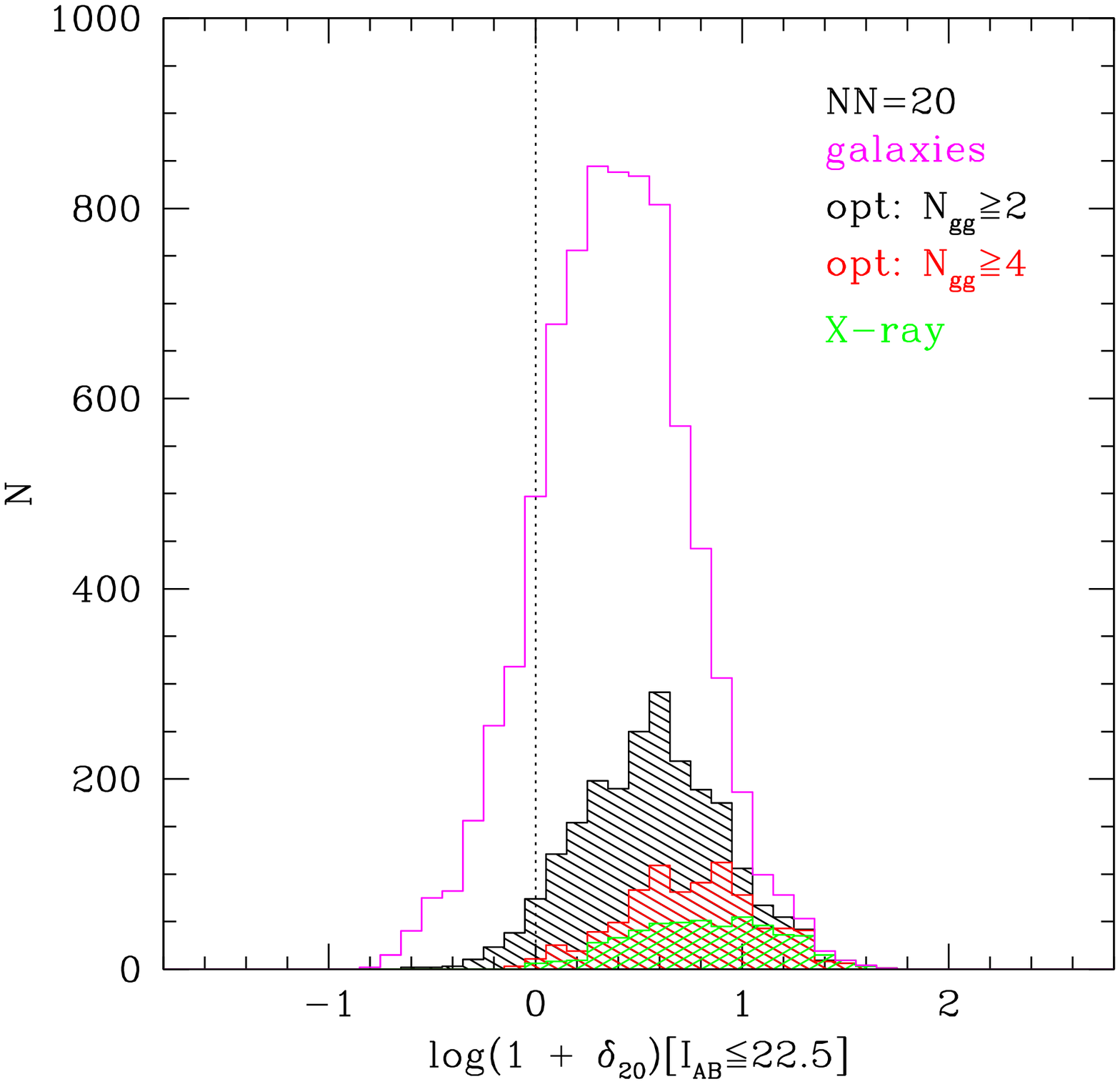}
\caption{\label{fig_comp_odensgalvirstr_nnf}Comparison of the overdensities reconstructed
  at the  positions  of all zCOSMOS  galaxies  and those galaxies residing in the virialised structures. The histogram of the overdensities at the positions of zCOSMOS galaxies is presented in magenta. The shaded histograms represent the distribution of overdensities centred on galaxies which reside in the virialised structures: X-ray clusters (green shaded histogram), the optical groups with at least 2 detected members (black shaded histogram) and the optical groups with at least 4 detected members (red shaded histogram). The histograms overlap each other.}
\end{figure}

\citet{Guzzo.etal.2007} used projected densities (reconstructed with the photometric redshifts), X-ray surface brightness and one of the first weak-lensing convergence maps to describe the extendend structure at $z \sim 0.7$ in the COSMOS field. A comparison of the overdensities reconstructed
  at the  positions  of all zCOSMOS  galaxies  and of those galaxies residing in the virialised structures is  presented  in
Figure~\ref{fig_comp_odensgalvirstr_nnf}. The majority  of galaxies 
 detected in the  X-ray clusters  or the
richer optical groups reside in extremely overdense regions. The galaxies defining the poorer groups extend to much lower overdensities than the rich $N\ge4$ optical groups and X-ray clusters, and they almost completely avoid the most overdense regions.

\begin{figure}
\centering
\includegraphics[height=0.67\textheight]{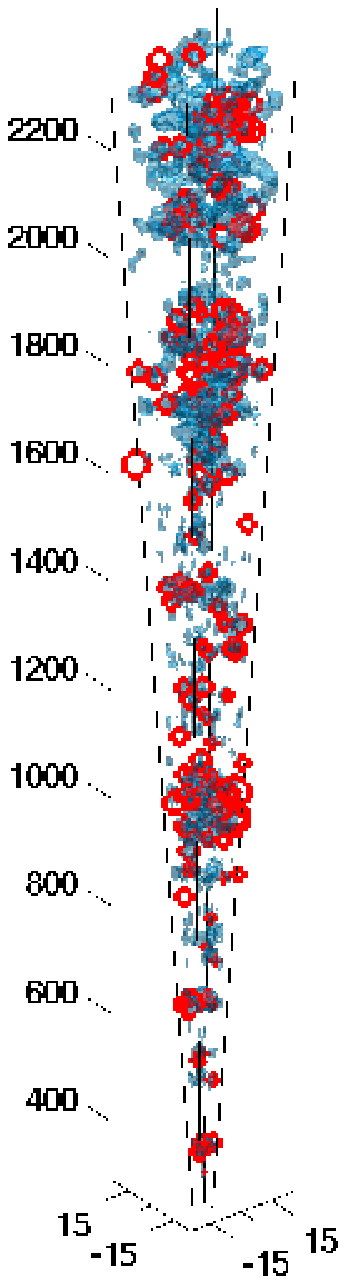}
\includegraphics[height=0.67\textheight]{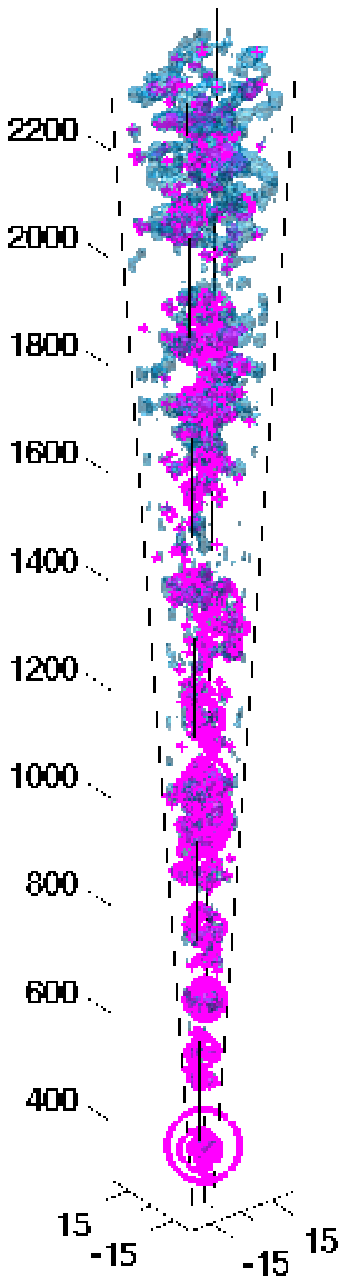}
\caption{\label{fig_complss}Comparison of the zCOSMOS density field to the estimates of the bound structures. On the left and right figures is presented the zCOSMOS density field at $1+\delta_p \ge 3$ obtained with the 10k+30kZADE flux limited sample and the apertures defined by the distance to the 10th nearest neighbour projected within $\pm$ 1000 \kms. On top of the overdensity field are overlayed X-ray detected clusters (red circles, on the left) and optical groups with at least three detected members (magenta circles, on the right). Each circle is centred at the position of the defined structures. Radii of the circles are scaled in $RA-DEC$ plane to the X-ray luminosity  for the X-ray clusters and to the number of detected objects in the optical groups.}
\end{figure}

In  addition,  in   Figure~\ref{fig_complss}  we  present  the  visual
comparison  of the zCOSMOS overdensity  field with  the  bound structures: X-ray clusters \citep[left;][]{Finoguenov.etal.inprep} and zCOSMOS groups with at least 3 members \citep[right;][]{Knobel.etal.inprep}.  There  is  an  overall  good  correspondence  of  the
overdense regions and the bound structures in the zCOSMOS volume. Almost none virialised structure is detected in regions with $1+\delta_p < 3$.

More detailed insight into the spatial distribution of the overdense and virialised structures is presented in the panels of Figure~\ref{fig_slices}. We use the same data as in the previous figure, but now the overdensity field is projected in redshift slices of $\Delta z = 0.025$ width, starting from redshift 0.2. We use the cut of $1+\delta_p \ge 6.67$ to define the overdense regions, higher than  in   Figure~\ref{fig_complss}. For a contrast, we also plot the equivalently underdense regions defined by isosurfaces with $1 + \delta_p \le 0.15$. The X-ray structures and the optically defined groups with at least 3 detected members reside inside of the LSS defined by the chosen overdensity value in most of the cases in the whole redshift range $0.2<z<1$. However, there are some virialised structures which do not live in these most overdense regions. In fact, as we have already seen in   Figures~\ref{fig_comp_odensgalvirstr_nnf} and  ~\ref{fig_complss}, the virialised structures trace also the less overdense regions. Moreover, the apparent overlap between some of the virialised structures and underdense regions in Figure~\ref{fig_slices} is only due to the projection effects.

\subsection{Comparison of the 10k zCOSMOS LSS to the LSS in the mock catalogues}

The overdensity field reconstructed using the galaxies detected in the
zCOSMOS survey is highly complex,  resembling the network of the local
cosmic web up  to the highest redshift $z=1$  probed.  We compare here
the zCOSMOS  overdensity  field to  the  overdensity  fields  in the  ``10k+30kZADE'' mock
catalogues.  The mock  catalogues  and  the density  field reconstruction
procedure on  the mocks were discussed in Section~\ref{sec_mocks}. For
the exact  comparison of the data  and the mock  overdensity field, we
reconstruct the  overdensity field  of the 10k  zCOSMOS galaxies  on the same
grid as was used for the mock catalogues, defined by $\Delta RA = \Delta
DEC = 2$ arcmin and $\Delta z = 0.002$ and using the   flux   limited  tracers   of
galaxies.  We did not calculate the edge
corrections  for all  the  mock  catalogues (which  is  a very time consuming
process computationally), and instead compare  the
overdensity  fields uncorrected  for  edge effects for both the data and the mocks. Given that we are only interested in comparing the
structures in the  overdensity fields in the data  with  structures
in  the mocks, and  not to  draw any  scientific conclusions  from the
distributions  of the  detected structures, this should not matter.

In  Figure~\ref{fig_overd_gridradez}  we  show the  three  dimensional
distribution of  the $1+\delta_p=3$ isosurface in  the overdensity field
estimated on the angular grid of  the 10k zCOSMOS  galaxies and  of  the  mock
catalogues.  The visual inspection of the LSS defined by the $1+\delta_p=3$ isosurfaces leaves  the impression that there  are more large structures in the real data than in the mock catalogues.

For a quantitative comparison  of the overdensity  field of  the real
zCOSMOS   and  mock   catalogues   we  implement   a  volume   filling
statistic. We  calculate the fraction  of the survey volume  in which
the overdensity  value is above  a given threshold.  As a reference,  we use the  value obtained by averaging the individual
statistics  from the  12  mock  catalogues. We  use the  standard
deviation of the 12 mock results as an error estimate, which is
dominated by the cosmic variance.

First,  we compare
the overall  distribution of the survey volumes in  structures of a given $\delta_p$ value in
the $0.2<z<1$ redshift range using the overdensity field reconstructed in the apertures defined  by the distance to the
10th  nearest  neighbour   of  the  flux   limited  tracer
galaxies (Figure~\ref{fig_volfracz0210}, left). Even though we use the angular grid to reconstruct the density field, we count the comoving volumes (in [\hh Mpc]$^3$) of the individual cells to properly calculate these volume fractions. There is an excellent agreement, within $1 \sigma$ errors, between the fractions of the volumes in the isosurface structures  in the data and in the average mock.

We carry out  the same analysis dividing the redshift range in four  intervals: $0.2<z<0.4$, $0.4<z<0.6$, $0.6<z<0.8$ and $0.8<z<1$ (Figure~\ref{fig_volfracz0210}, right). While the overdensity distributions of the data in $0.2<z<0.4$ and $0.6<z<0.8$  redshift ranges are in relatively good agreement with the mock results, the redshift interval $0.4<z<0.6$ is underdense in the data with respect to the mocks, while the situation is reversed in the redshift interval $0.8<z<1$. Obviously, at any redshift bin and at a given $\delta_p$, the real data are affected by the cosmic variance, and we can not see the growth in the cosmic structure over redshifts, as it is visible in the mock curves.

We follow up on this with a more detailed comparison of the distribution of ``sizes'' (volumes) of the structures above a given overdensity in the data and in the individual mock catalogues. We calculate the size of a structure by adding comoving volumes of the connected grid cells with overdensity above a chosen value. The grid cells need to have at least one common side (either in RA, DEC or z direction) to be considered connected. The results are presented in Figure~\ref{fig_volfracstruct} for $1+\delta_p \ge 3$. The individual plots correspond to the survey volume fractions contained within the structures of at least the indicated size. It is noticeable that there is not a single mock catalogue which contains as much volume as the real data in the largest structures in $0.2<z<1$. As a check we also recalculate the same statistics for the data when using the mean density estimated following  the same  smoothing scheme as applied to the mocks (even though this smoothing is not ideal for the data). In this case, the difference between the data and the mocks is even larger.

Carrying the same analysis in the narrower redshift intervals we realise that the majority of this difference is accounted for by the large structure  in $0.8<z<1$. Already, when we limit the redshift range to $0.3<z<0.9$, the data are not so different from the mocks. To conclude, except for the large structure in the highest redshift bin $0.8 < z< 1$, the fraction of the volume within the structures of a given size in the data and in the mock catalogues is in reasonable agreement. At the current state, given that there is a disagreement in only one $\Delta z$ slice, it is difficult to say whether the large structure in $0.8<z<1$ reflects only a cosmic variance, or it is an unusual object for the current cosmology.

\section{Summary}

We have used the  first $\sim$ 10,000 spectra from  the zCOSMOS bright survey
to reconstruct the  density field in the survey volume up to $z=1$. We use a new  method for the
reconstruction, which is based on  the combination of the high quality
spectroscopic redshifts and the ZADE-modified photometric redshifts of
galaxies without spectroscopic redshifts. Our method enables us to
reliably reconstruct a broader range of environments  than it would
be possible by using only galaxies with spectroscopic redshifts.

We  use the  weighted counts  of  tracer galaxies  within  various
apertures to measure the galaxy environments in the zCOSMOS volume. Given the variety of the scientific applications, we carry out the reconstruction in different ways. We construct flux and volume limited samples of tracer galaxies, specify fixed and
adaptive apertures (characterised by distance to the Nth nearest neighbour), define three dimensional and projected distances.  We
weight  tracer galaxies  in  three different  ways: with  unity,
B-band luminosity and stellar mass.

We present in detail the density field reconstructed on the grid filling the zCOSMOS volume up to $z = 1$. The apertures are defined by the distance to the Nth nearest neighbour (5th, 10th and 20th) projected within $\pm$ 1000 \kms\ to
the redshift of a point where the density is being measured. We use the flux limited tracer galaxies, as this sample of tracers allows to reconstruct the density field with the broadest dynamical range and with the greatest detail of the structures. The reconstructed  zCOSMOS overdensity field consists of cluster-like structures, surrounded by void-like regions, showing  a complexity
 of the cosmic web up to $z \sim  1$. 

The regions in the density field enclosed by the $1+\delta_p \ge3$, or higher cutoff values, are well traced by the virialised objects in the zCOSMOS volume, i.e. X-ray clusters and optical groups with at least 3 detected members. The galaxies defining the poorer groups (with 2 or 3 detected members) are found to live also in much lower overdensities than the rich $N\ge4$ optical groups and X-ray clusters.

Further on, we have compared the LSS in the zCOSMOS data and in the mock catalogues. There is an excellent agreement of the fractions of the volume enclosed in structures of all sizes above a given overdensity between the data and the mocks in $0.2<z<1$. However, for the chosen $1+\delta_p=3$ value, there is more volume enclosed in the large structures in the data than in any of the used mocks; we want to stress that this difference is mainly driven by the existence of a very large structure in the zCOSMOS, centred at $z \sim 0.9$ and extending in radial direction over $\Delta z \sim 0.2$.

\section{Acknowledgments}

This work has been supported in part by a grant from 
the Swiss National Science Foundation and by grant ASI/COFIS/WP3110I/026/07/0.
We thank M.G. Kitzbichler and S.D.M. White for providing the mock catalogues \citep{Kitzbichler&White.2007}.





\clearpage

\begin{figure}
\centering
\includegraphics[width=0.3\textwidth]{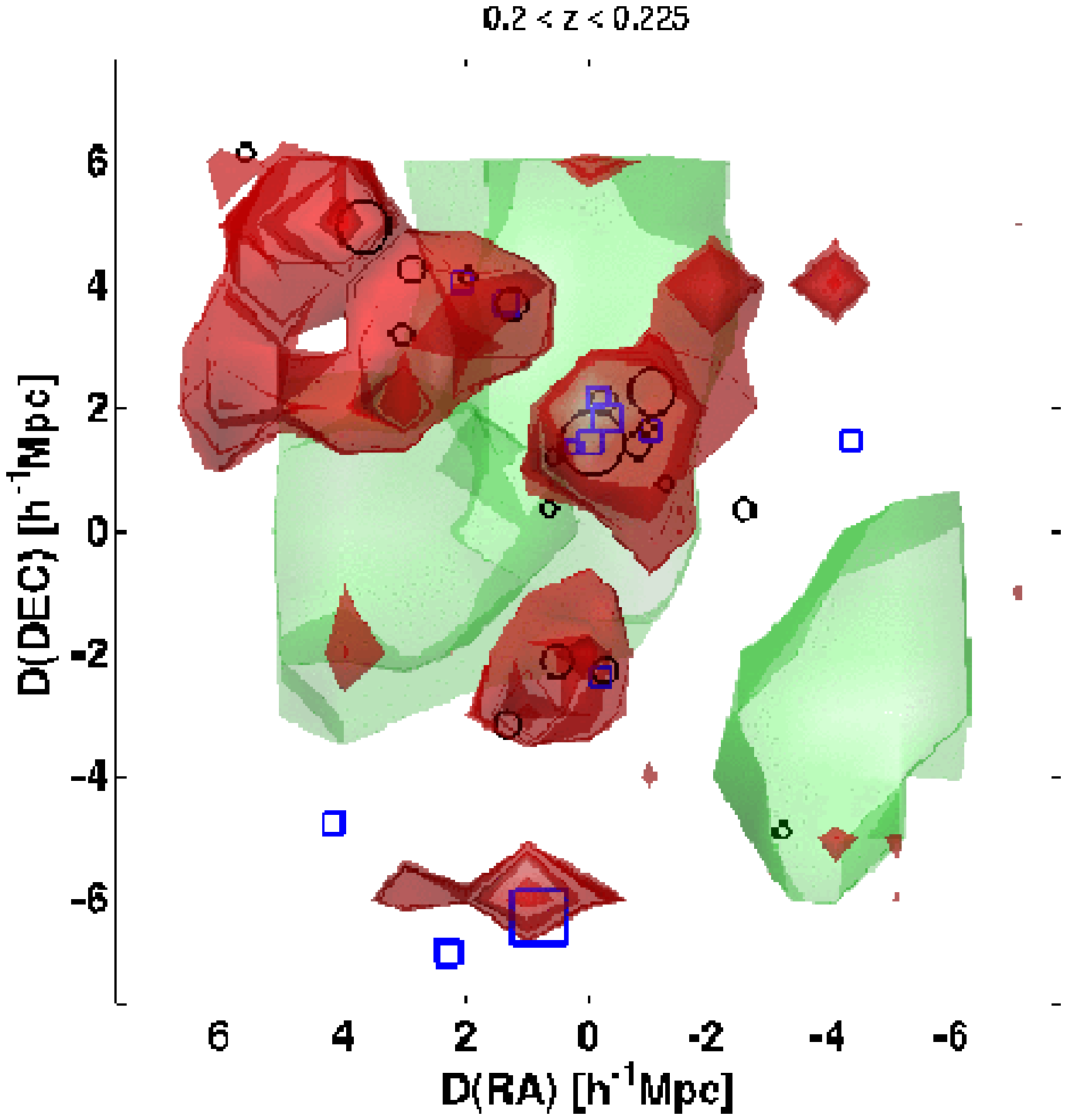}
\includegraphics[width=0.3\textwidth]{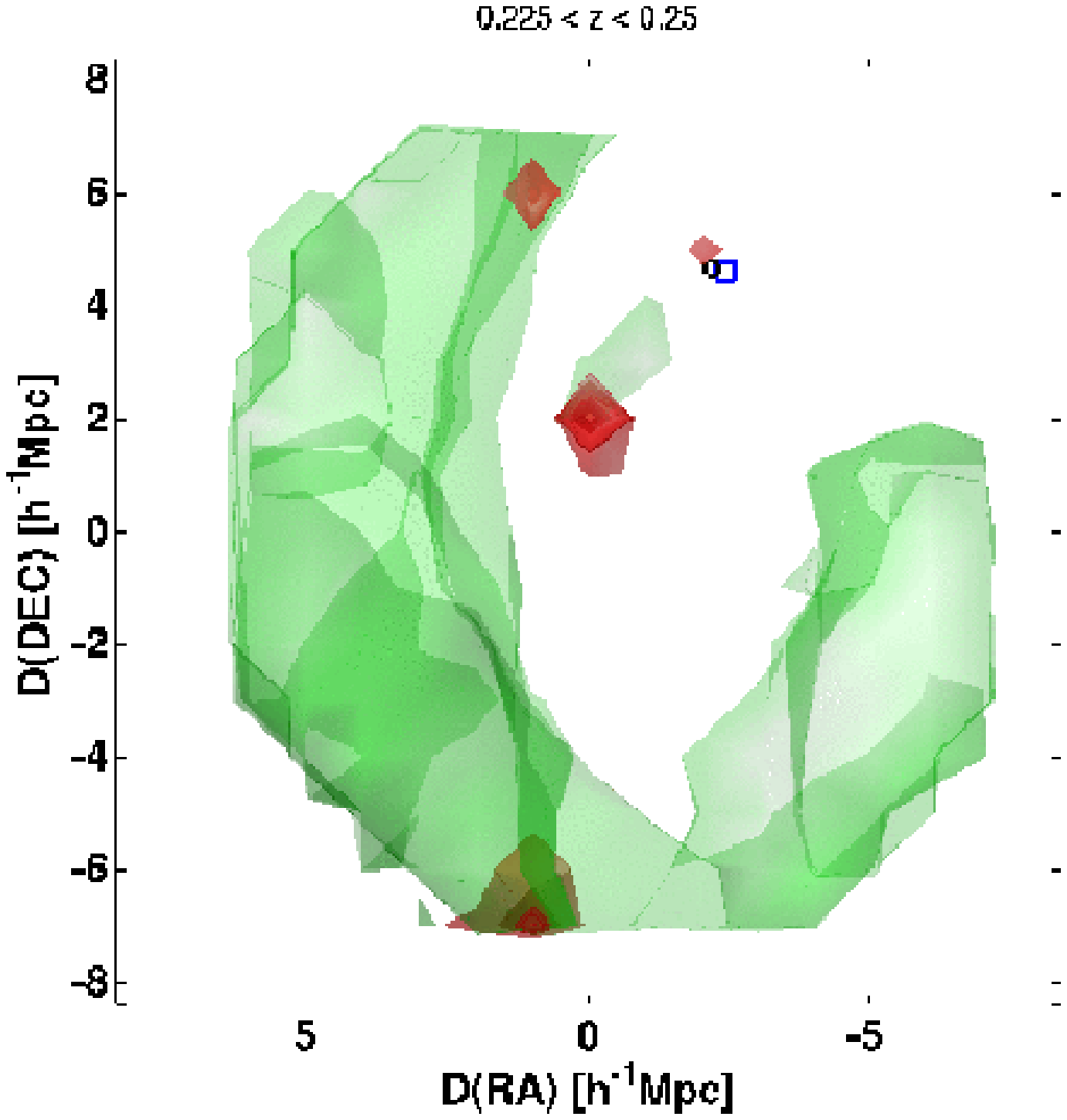}
\includegraphics[width=0.3\textwidth]{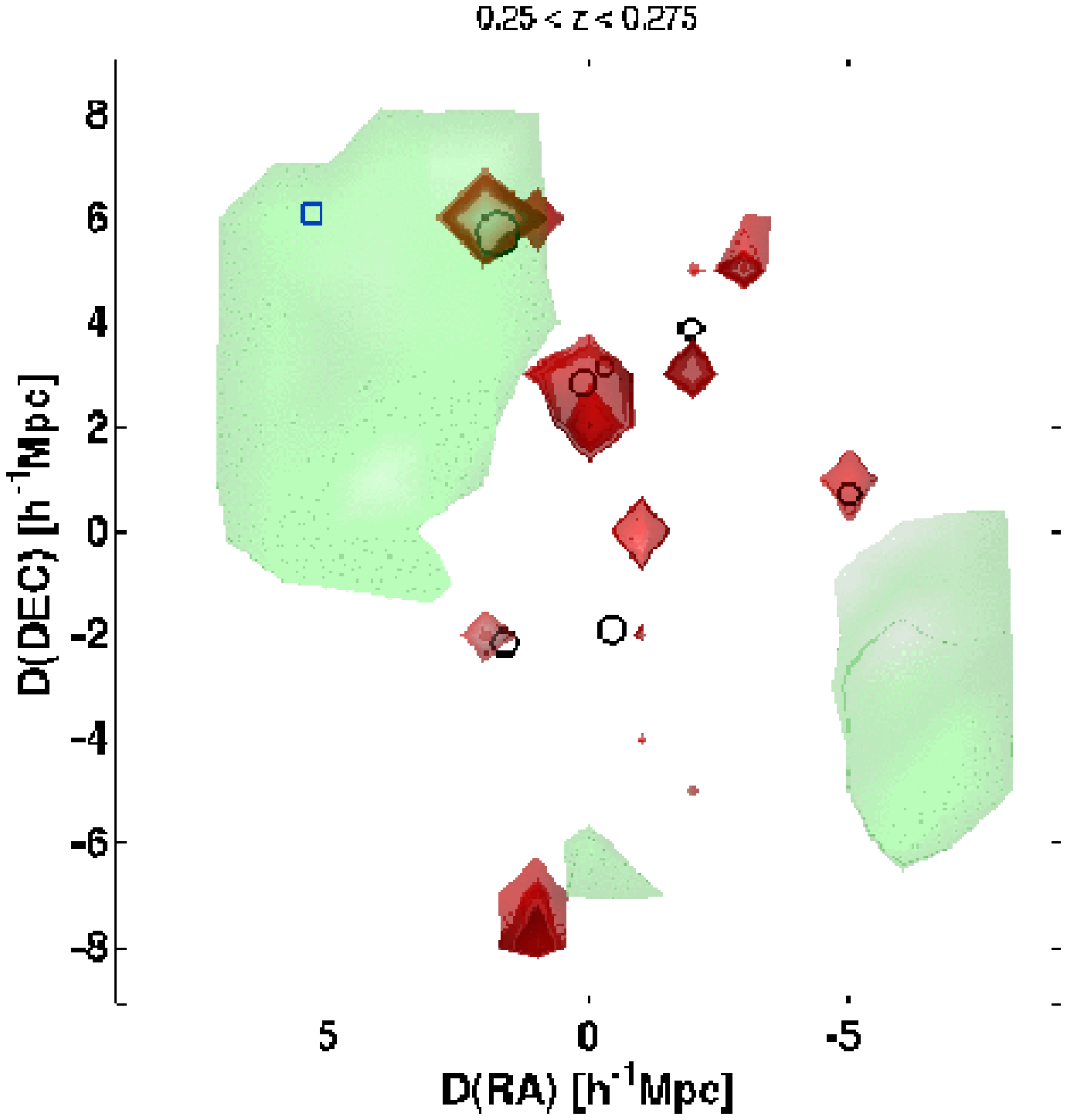}
\includegraphics[width=0.3\textwidth]{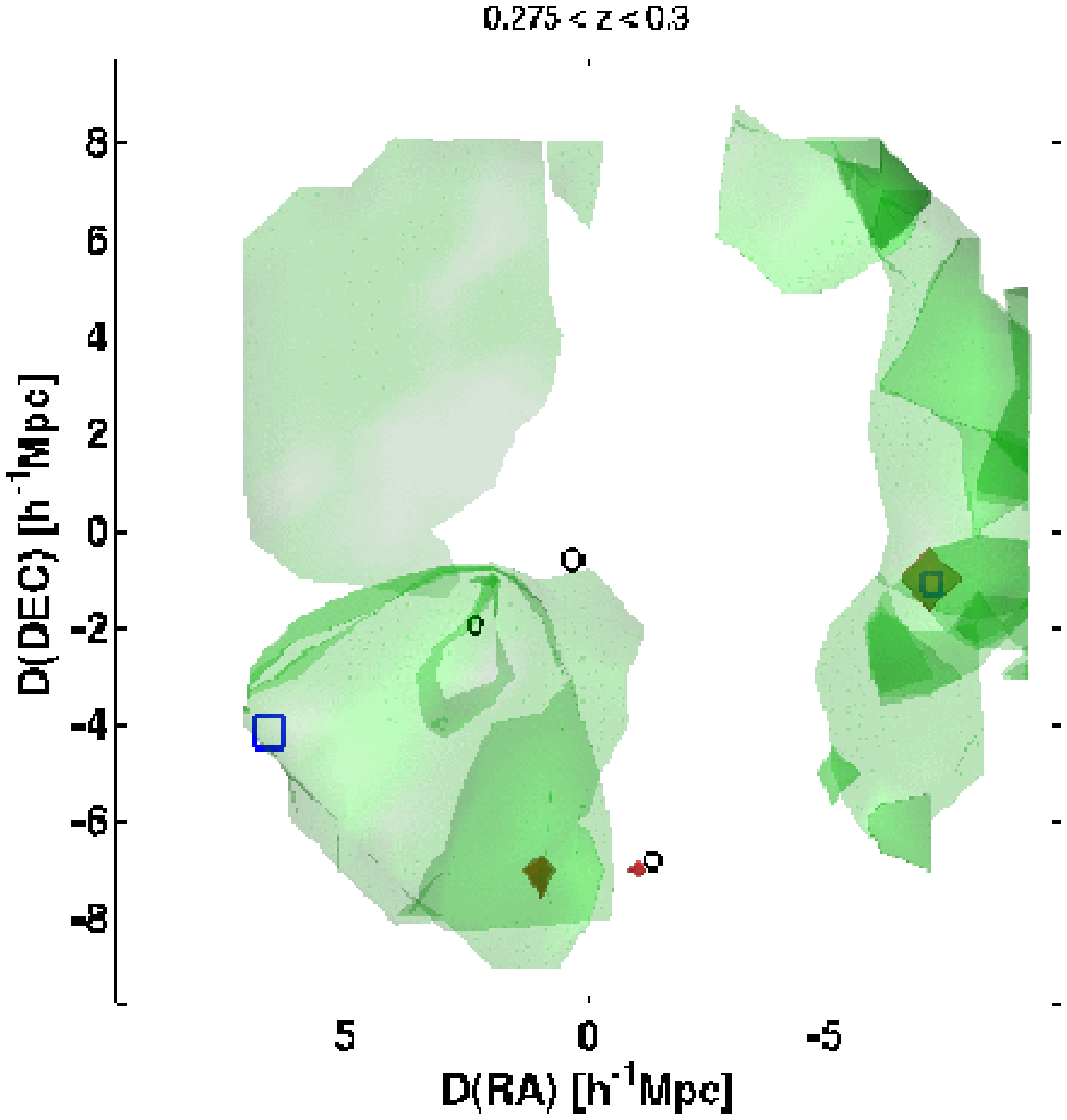}
\includegraphics[width=0.3\textwidth]{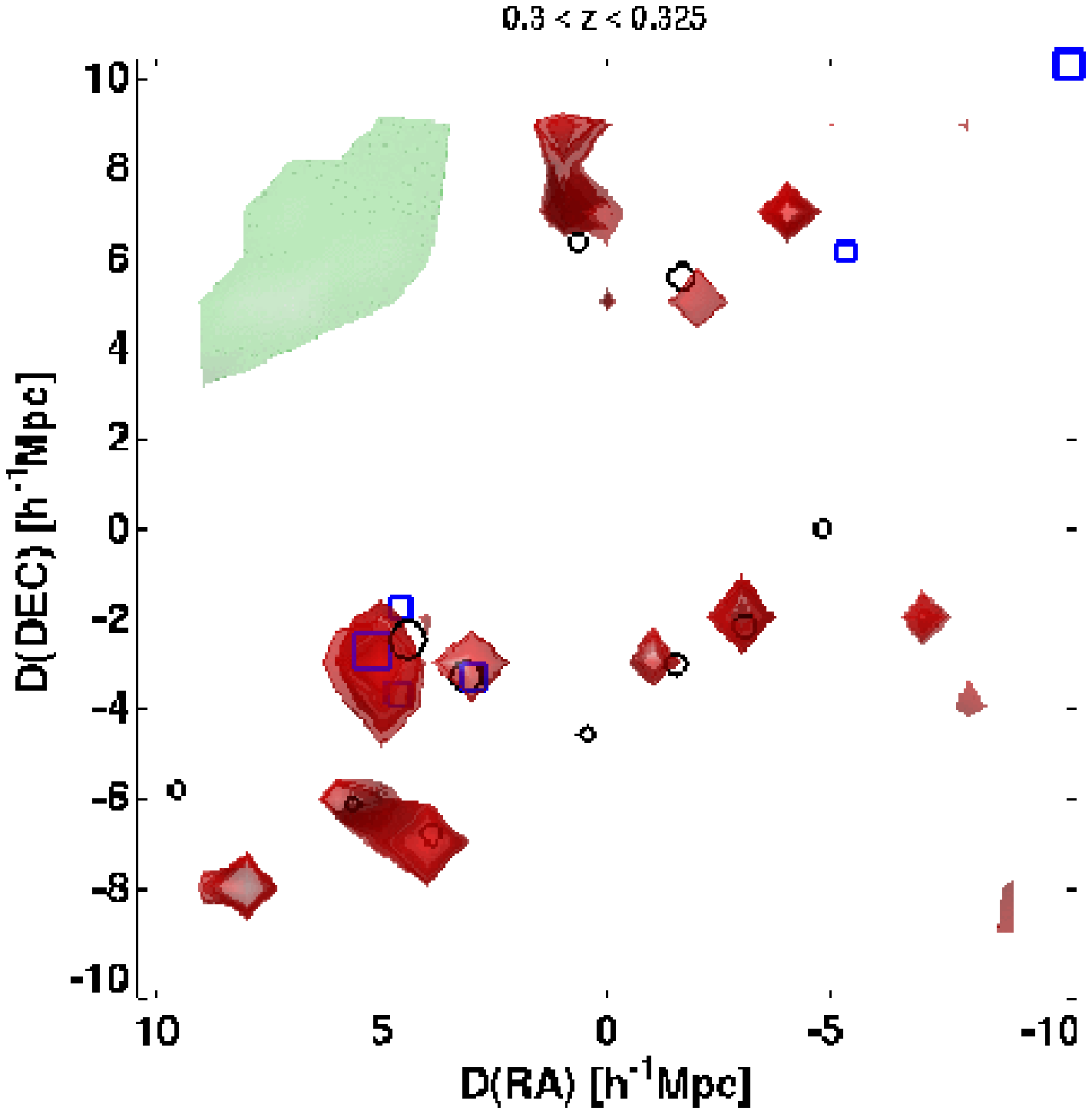}
\includegraphics[width=0.3\textwidth]{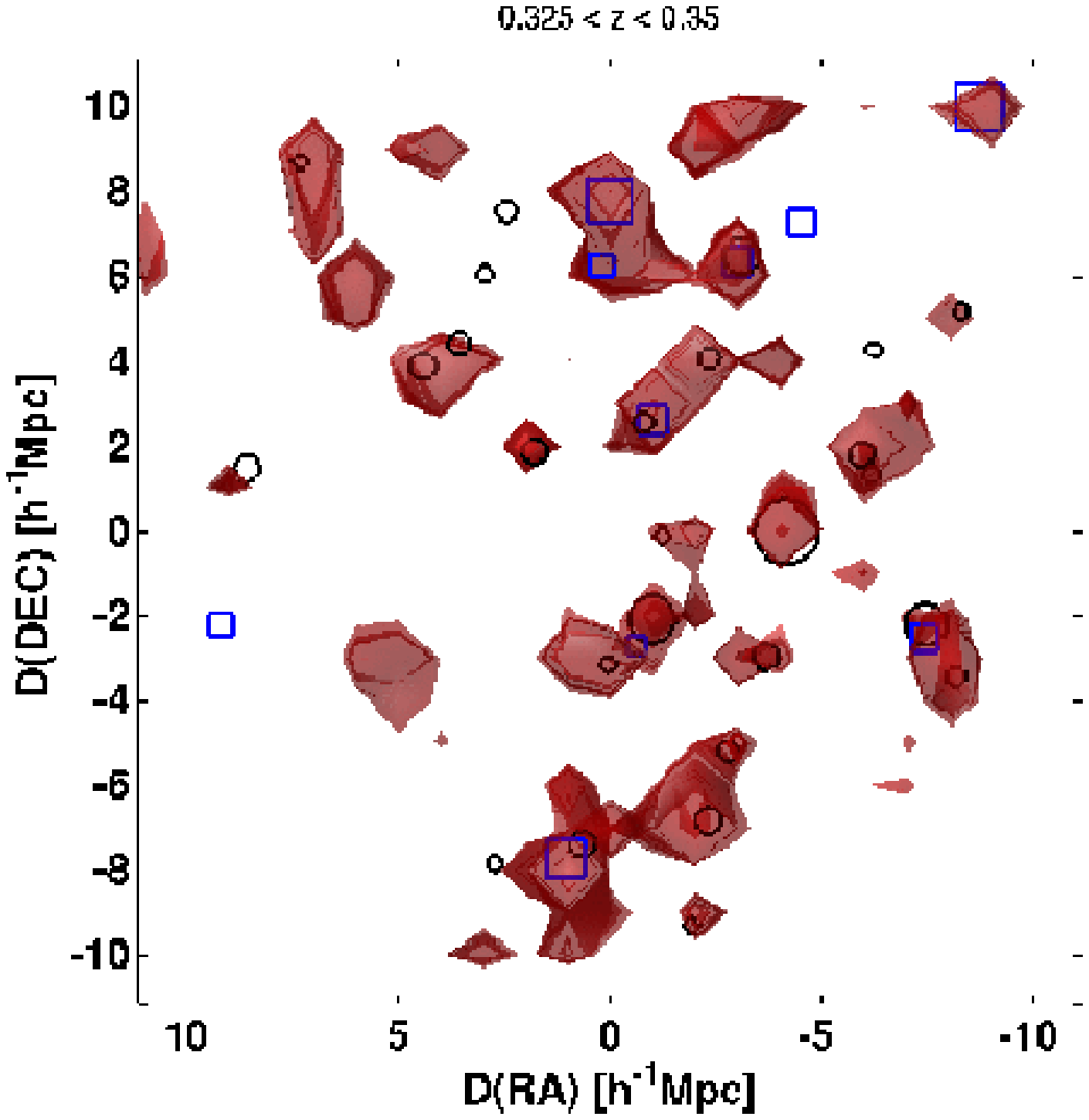}
\includegraphics[width=0.3\textwidth]{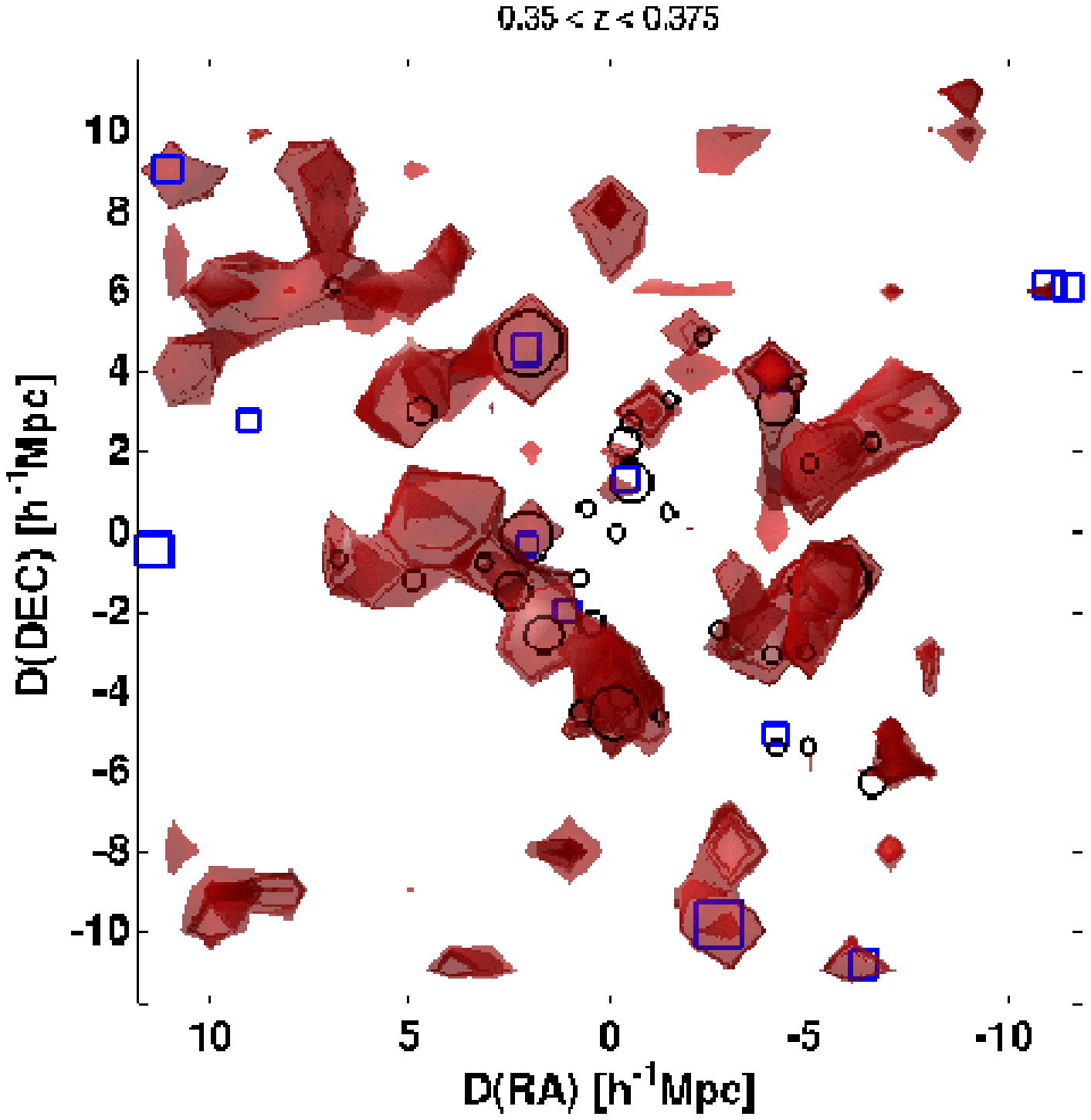}
\includegraphics[width=0.3\textwidth]{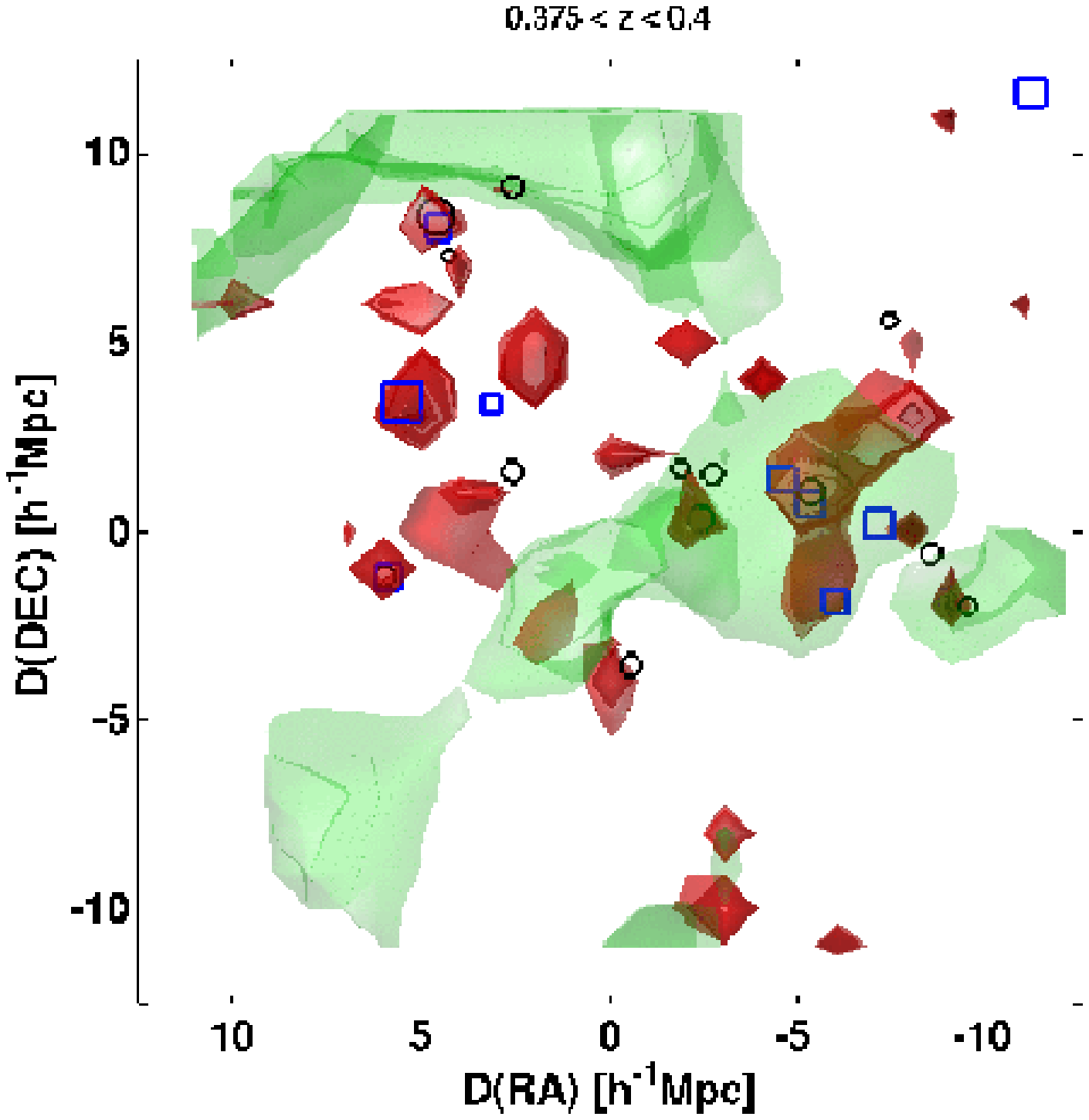}
\includegraphics[width=0.3\textwidth]{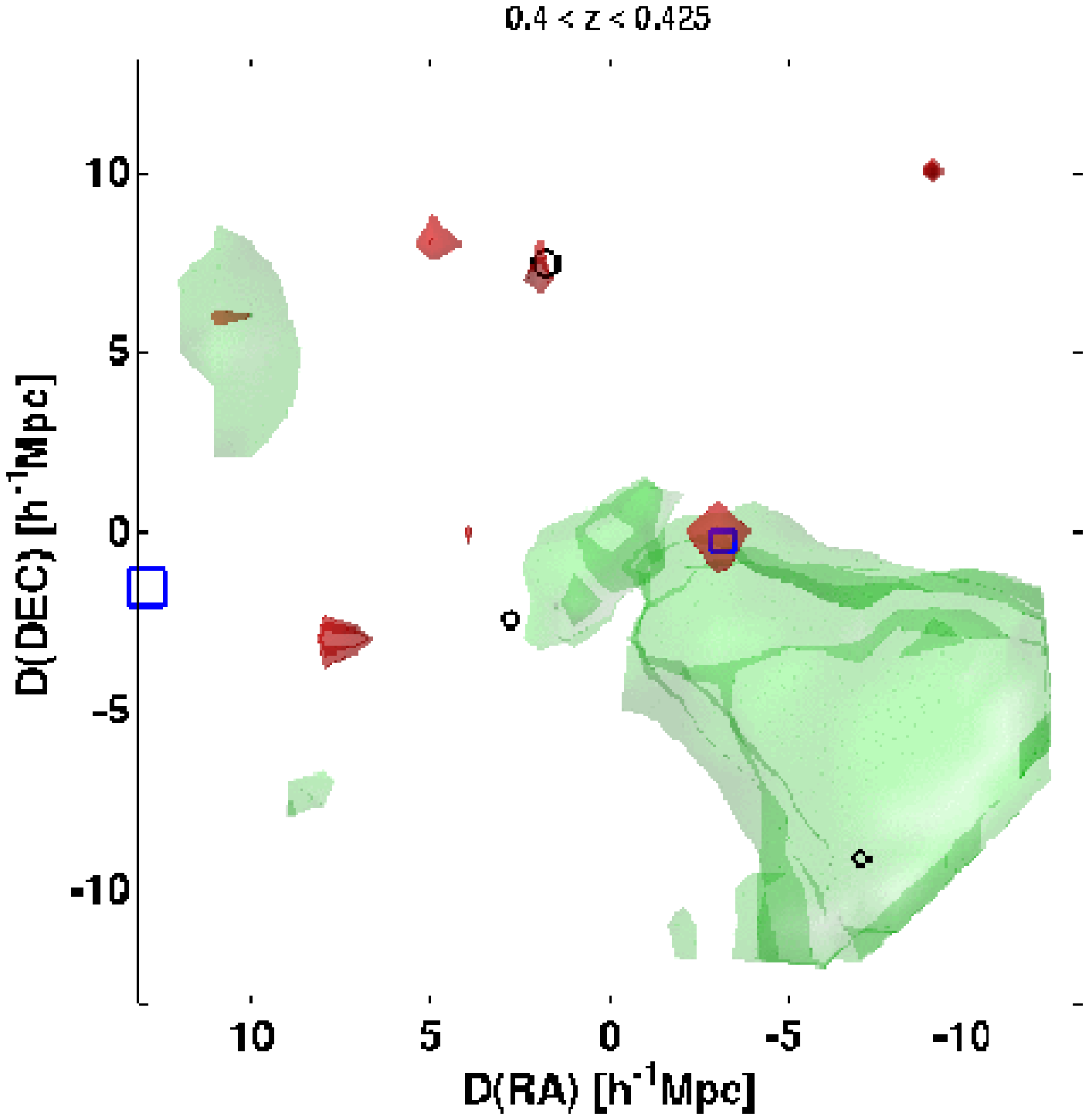}
\caption{\label{fig_slices}The overdensity field of galaxies projected in redshift slices of 0.025 width, covering the redshift interval from 0.2 to 1. The overdensity field is reconstructed using the flux limited sample of galaxies within the apertures defined by the 5th nearest neighbour projected within $\pm$ 1000 \kms. Only structures defined by the isosurface contours of $1+\delta_p \ge 6.67$ (red) and $1 + \delta_p \le 0.15$ (green) are shown. As a comparison, the optical groups in the same redshift bins with al least 3 detected members are overplotted as circles and X-rays clusters as squares. The sizes of the symbols to mark the positions of the virialised structures are scaled as in Figure~\ref{fig_complss}. The redshift slices $0.3-0.375$, $0.675-0.75$ and $0.875-1$ are dominated by the overdense structures. Large, $RA-DEC$-extended underdense structures are detected in the $0.45-0.5$, $0.525-0.55$, $0.575-0.6$, $0.625-0.675$ and $0.8-0.825$ redshift slices.}
\end{figure}

\clearpage

\addtocounter{figure}{-1}
\begin{figure}
\centering
\includegraphics[width=0.3\textwidth]{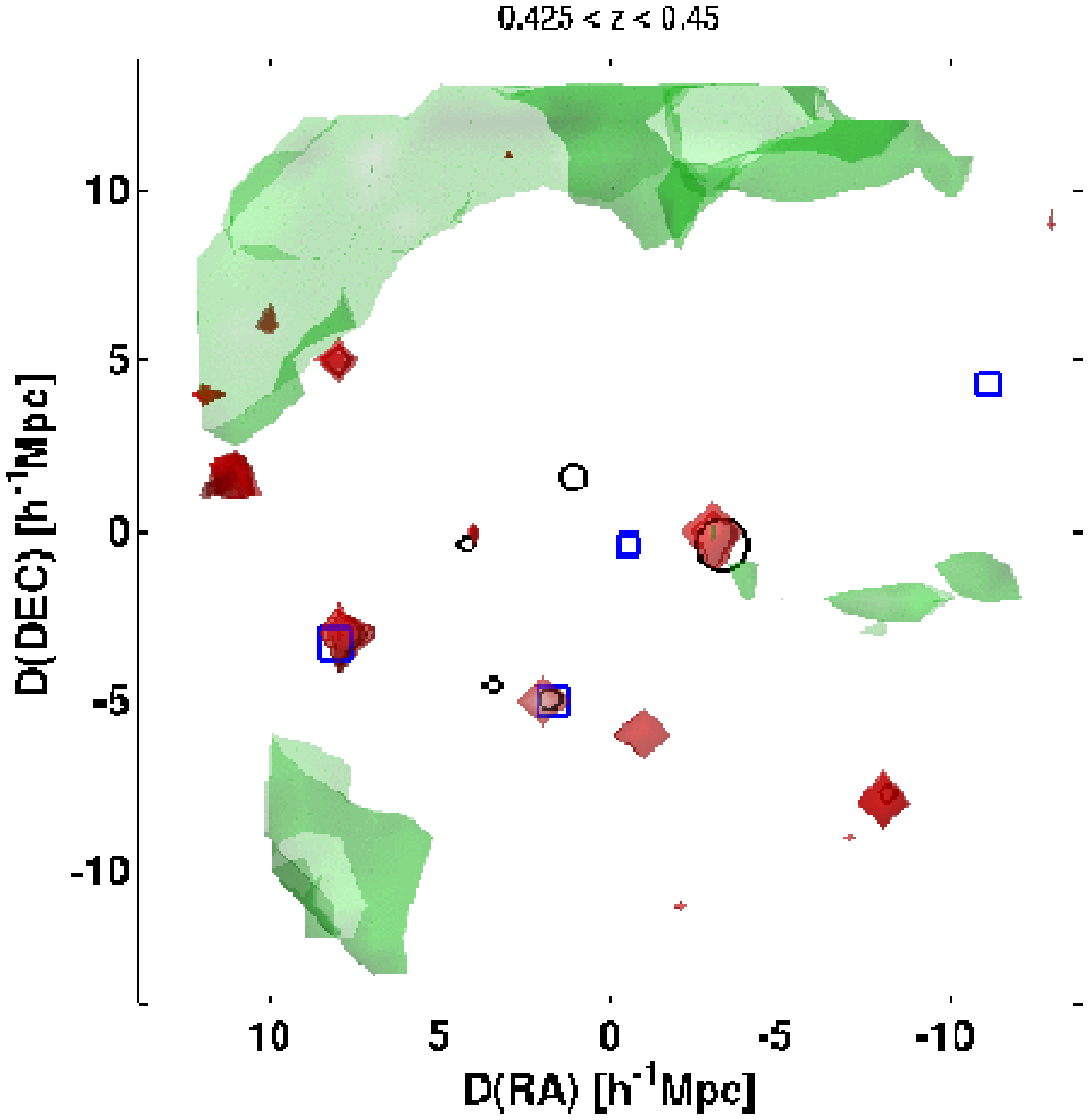}
\includegraphics[width=0.3\textwidth]{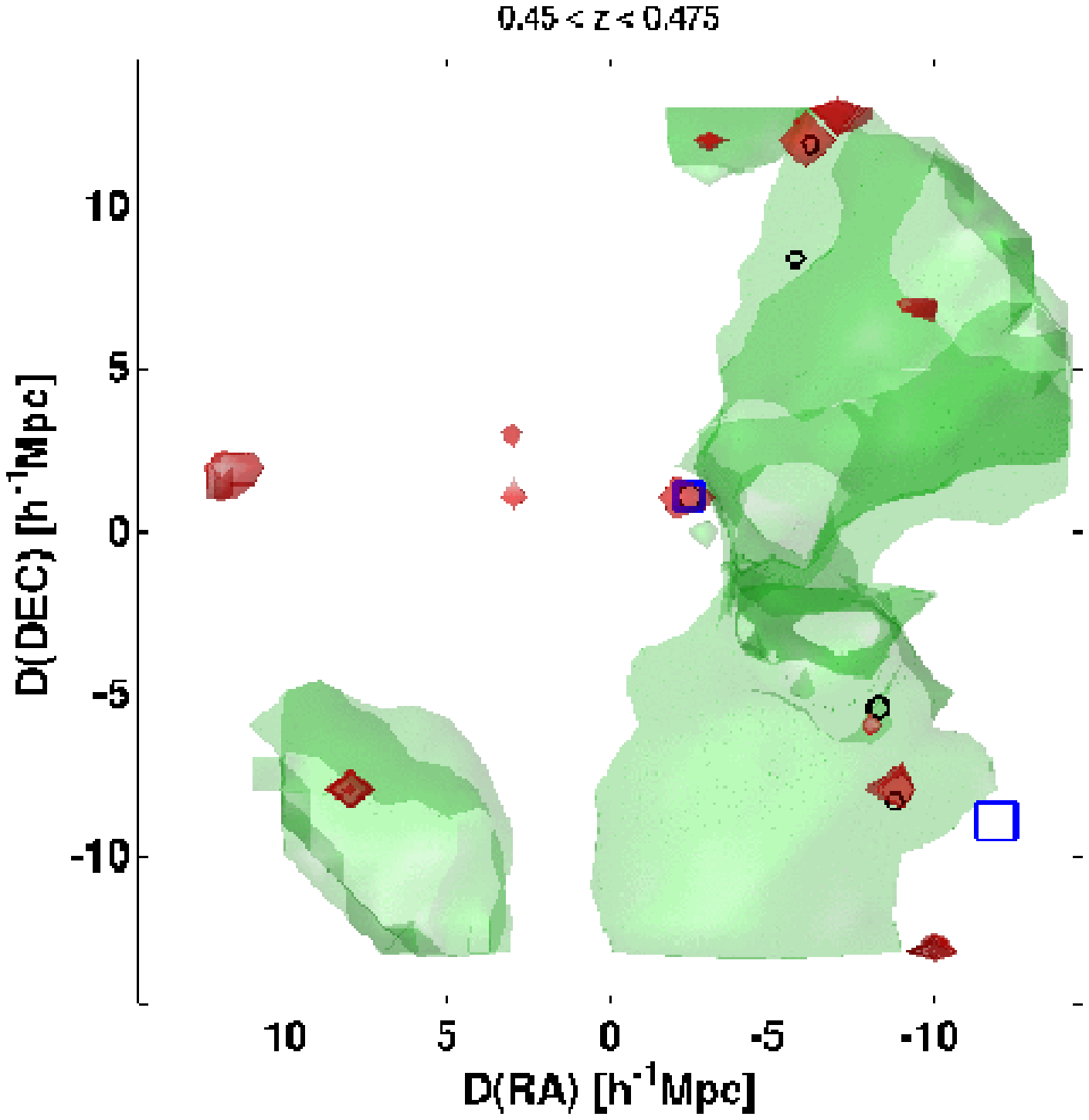}
\includegraphics[width=0.3\textwidth]{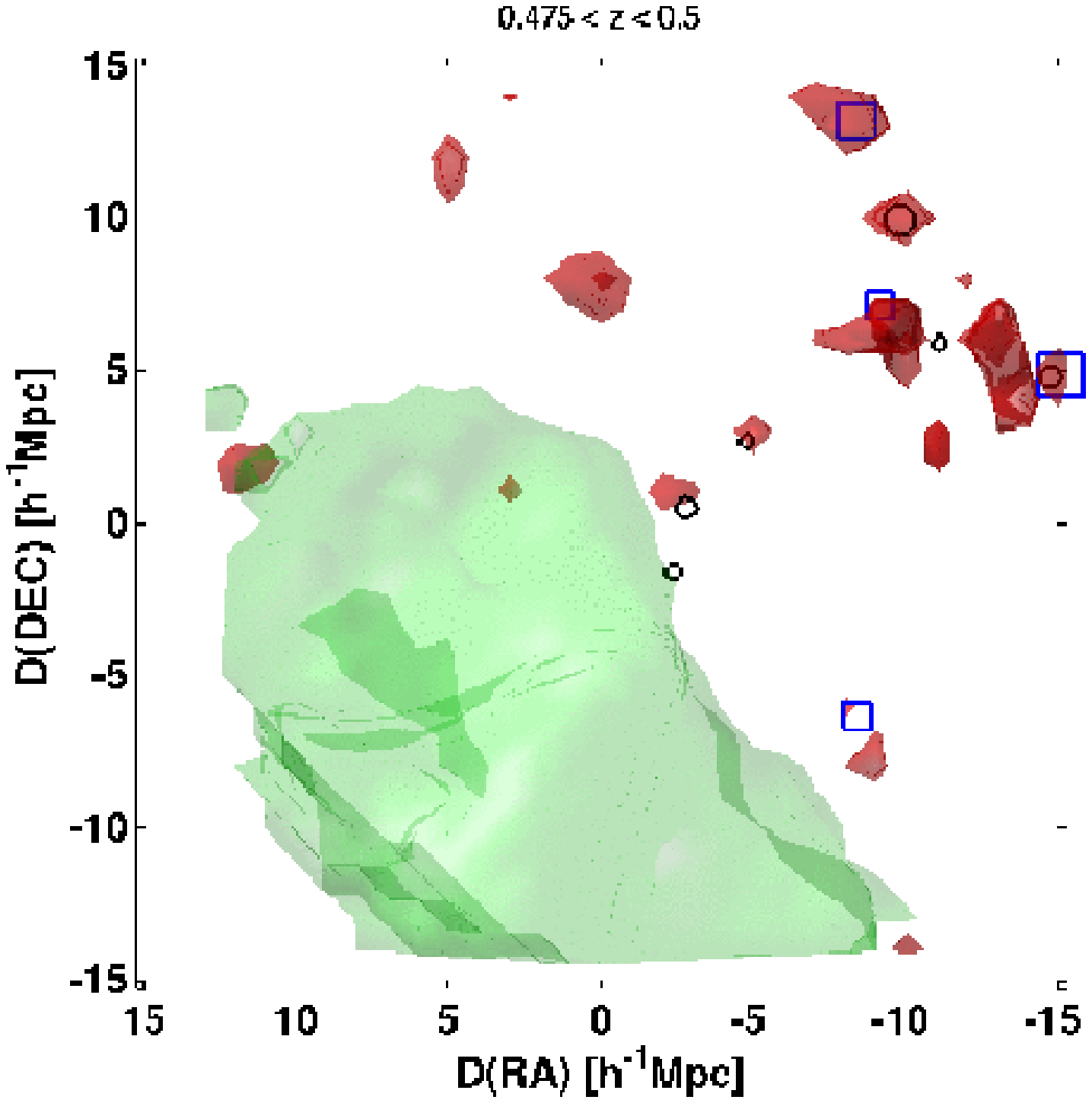}
\includegraphics[width=0.3\textwidth]{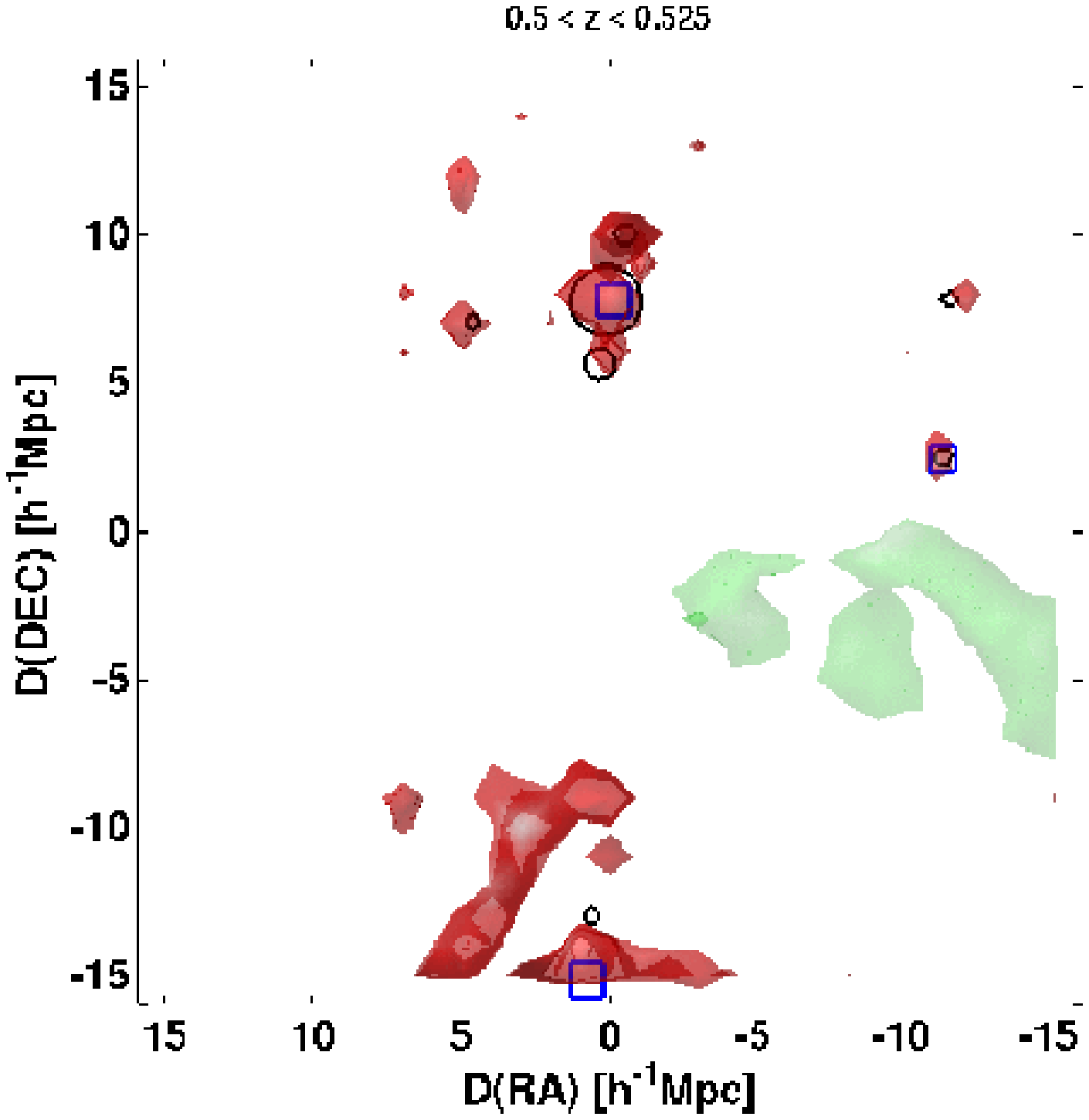}
\includegraphics[width=0.3\textwidth]{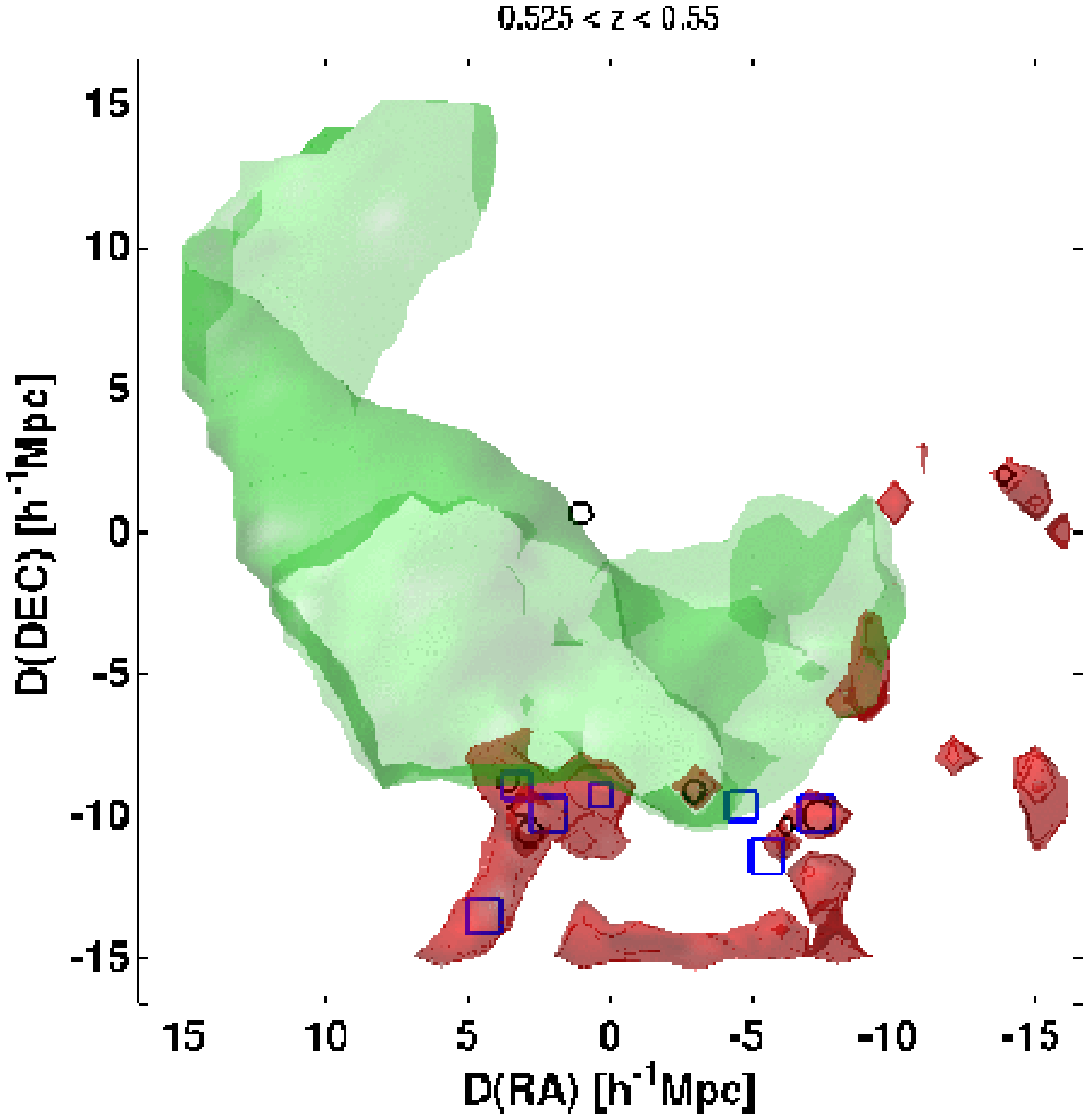}
\includegraphics[width=0.3\textwidth]{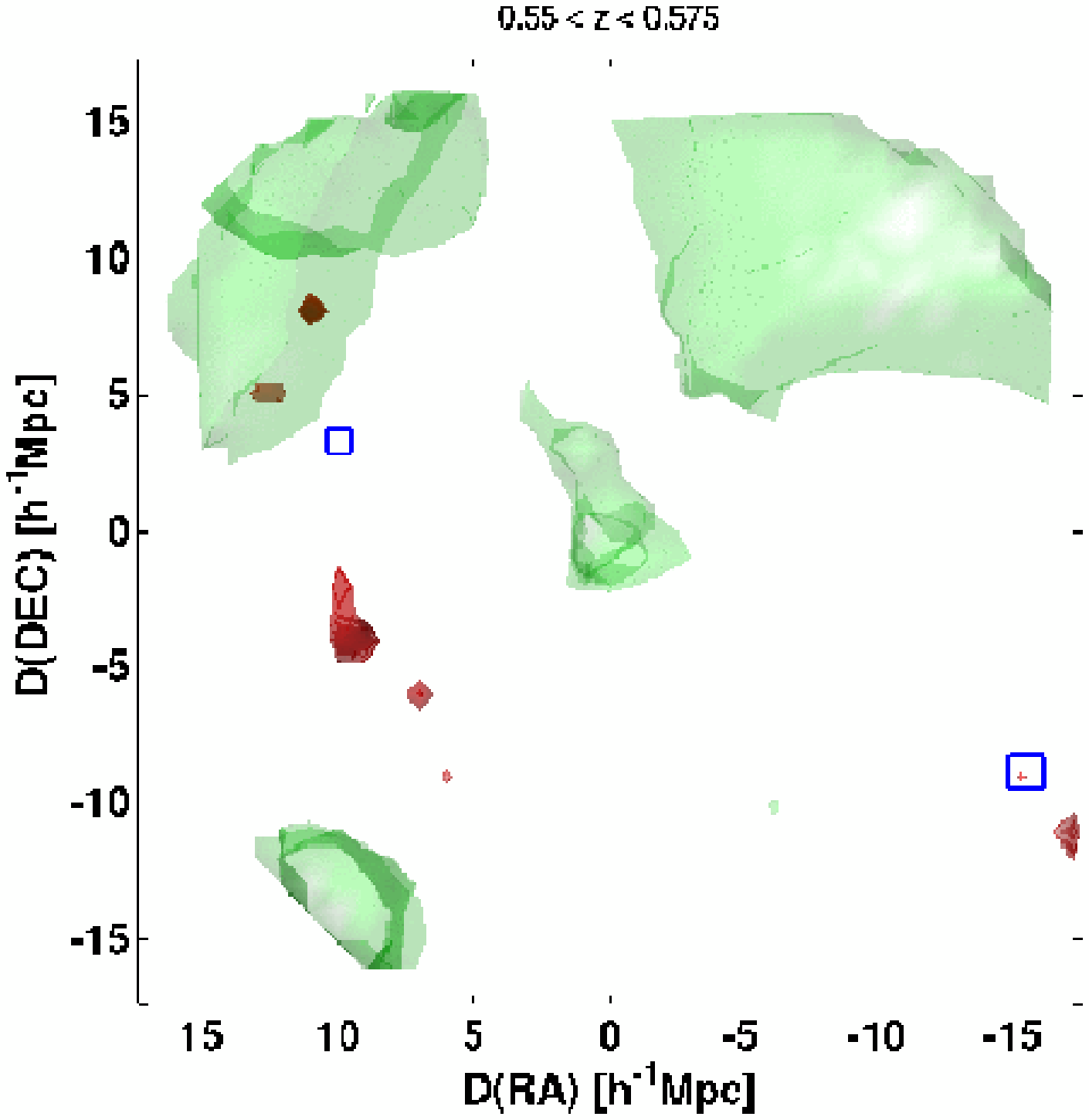}
\includegraphics[width=0.3\textwidth]{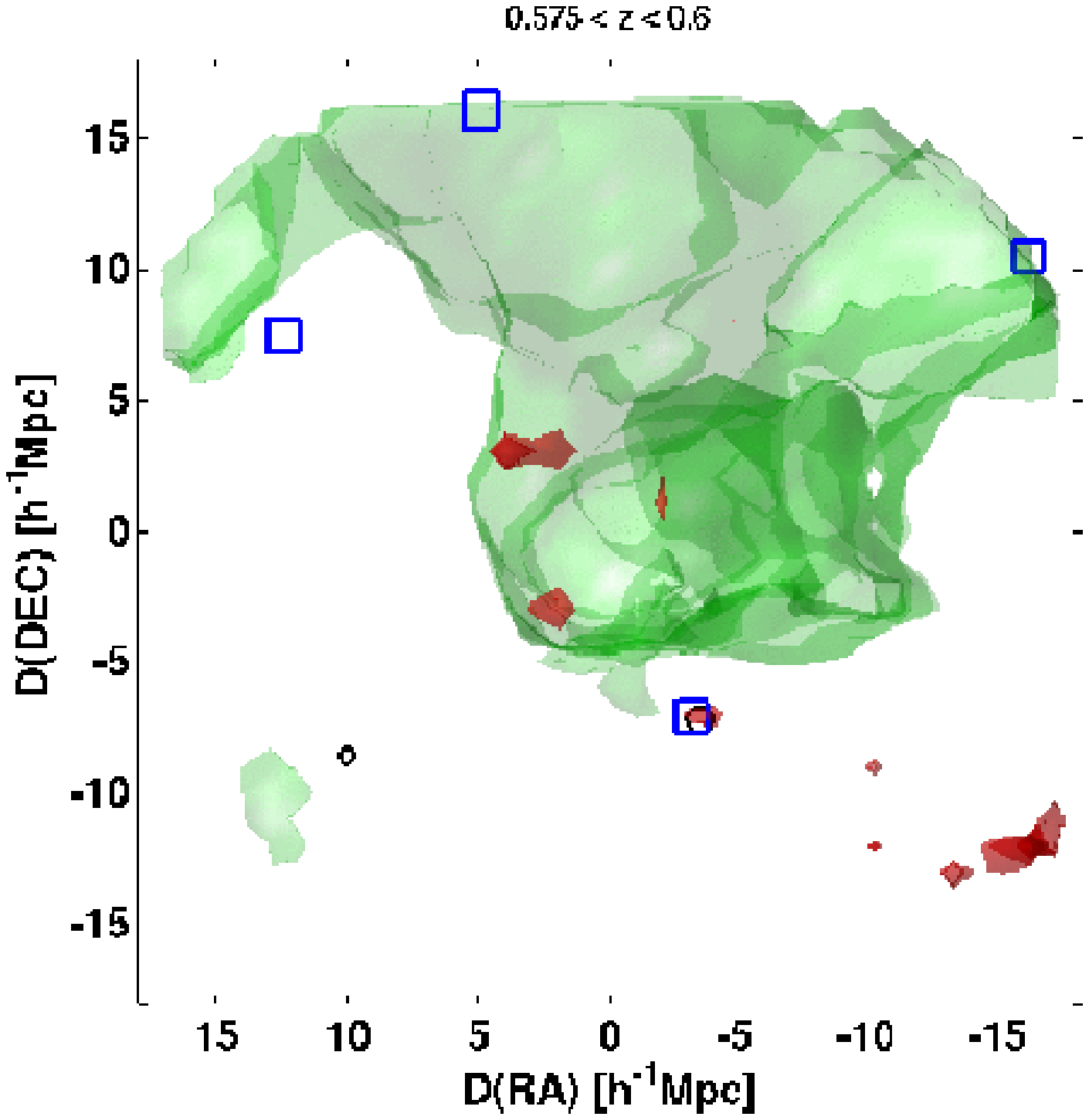}
\includegraphics[width=0.3\textwidth]{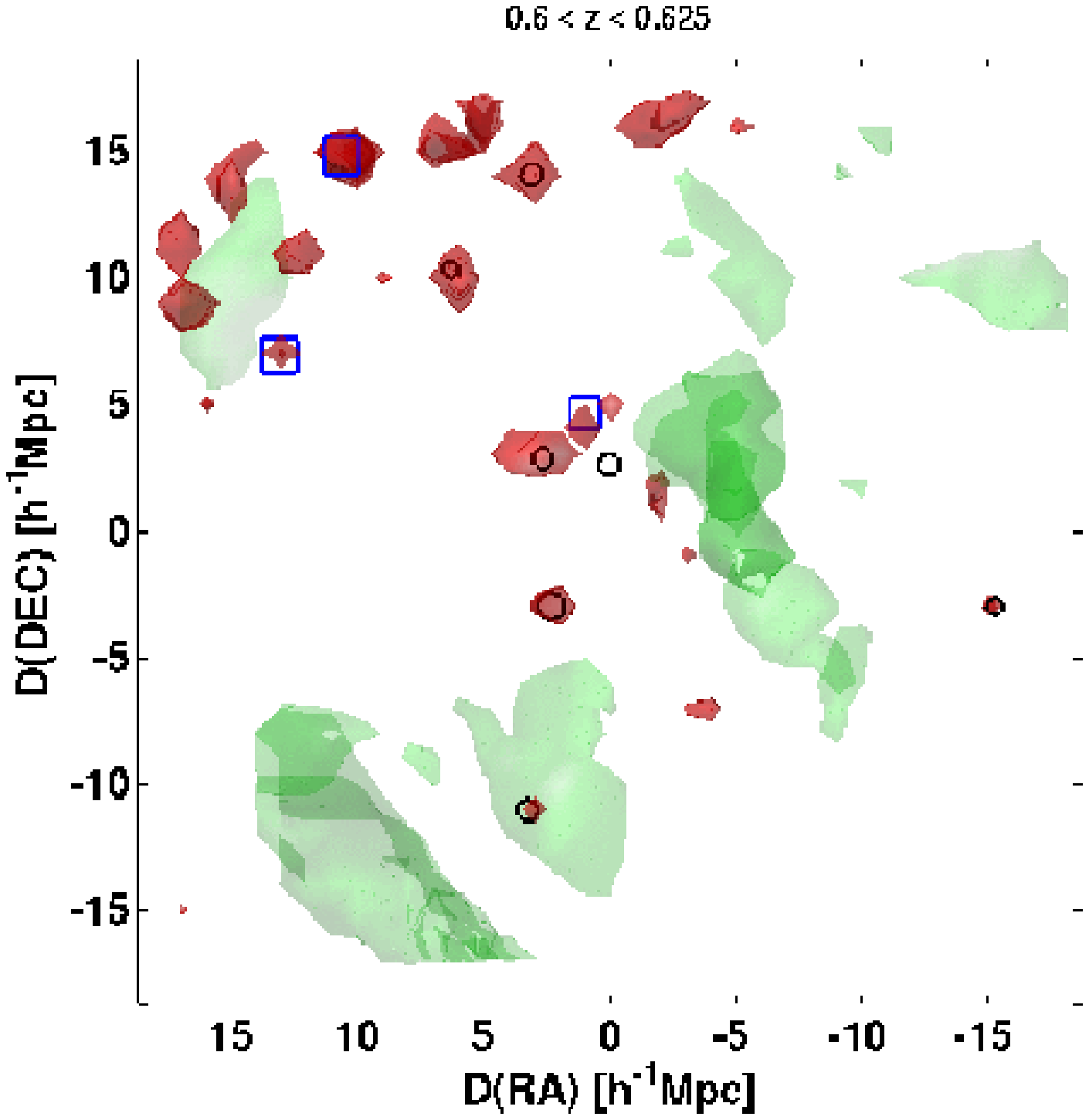}
\includegraphics[width=0.3\textwidth]{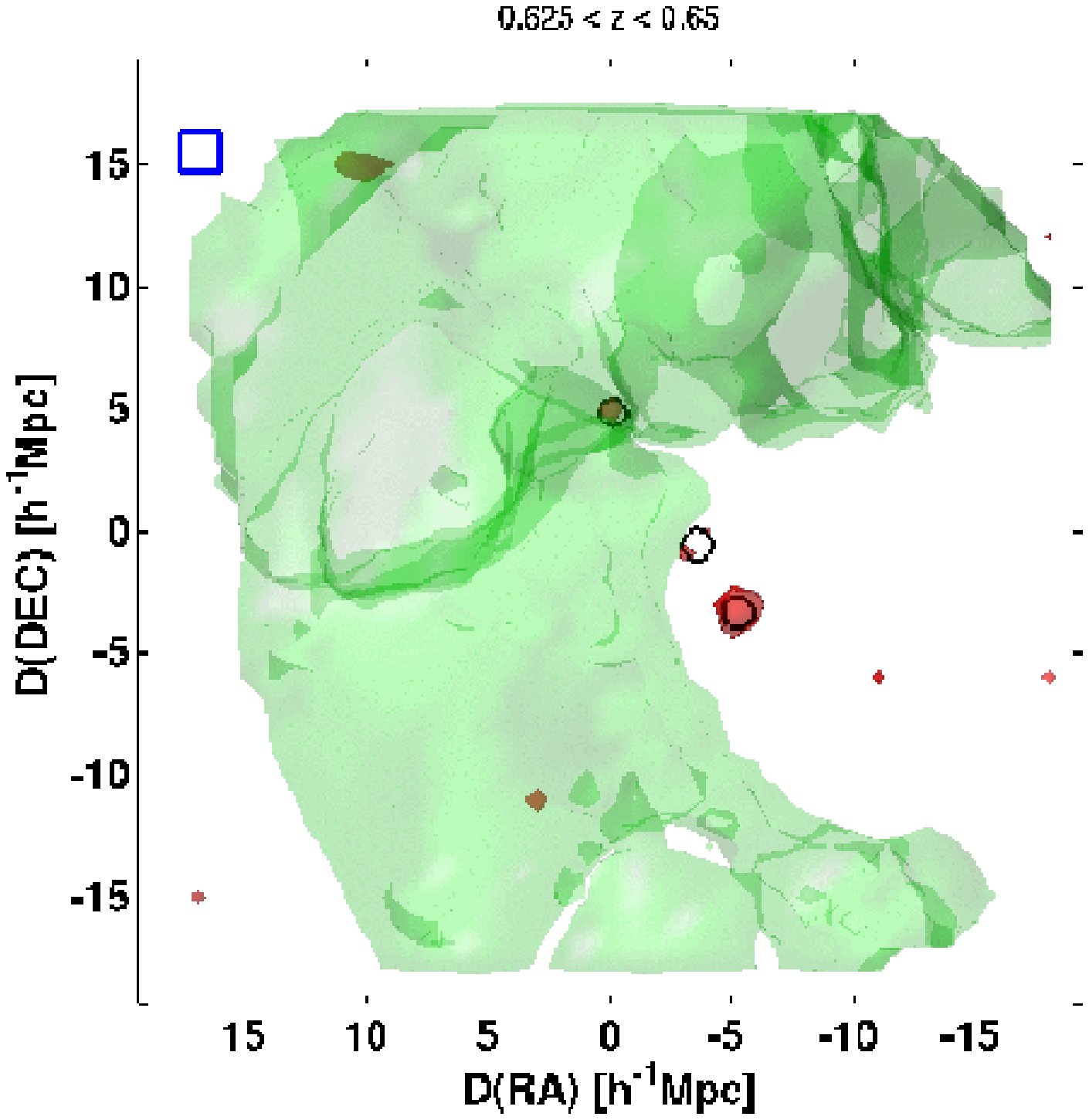}
\caption{Continued.}
\end{figure}

\clearpage

\addtocounter{figure}{-1}
\begin{figure}
\centering
\includegraphics[width=0.3\textwidth]{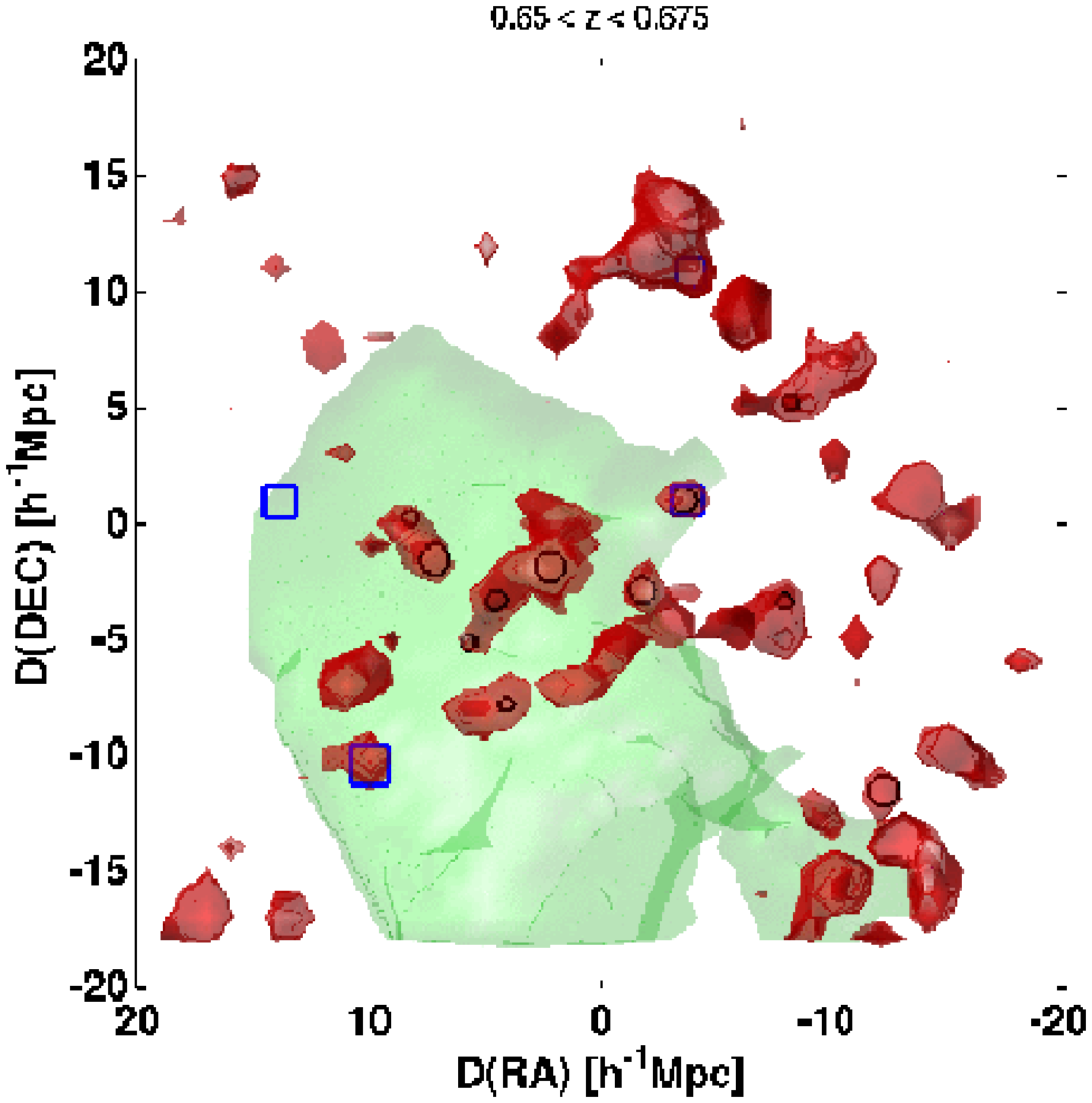}
\includegraphics[width=0.3\textwidth]{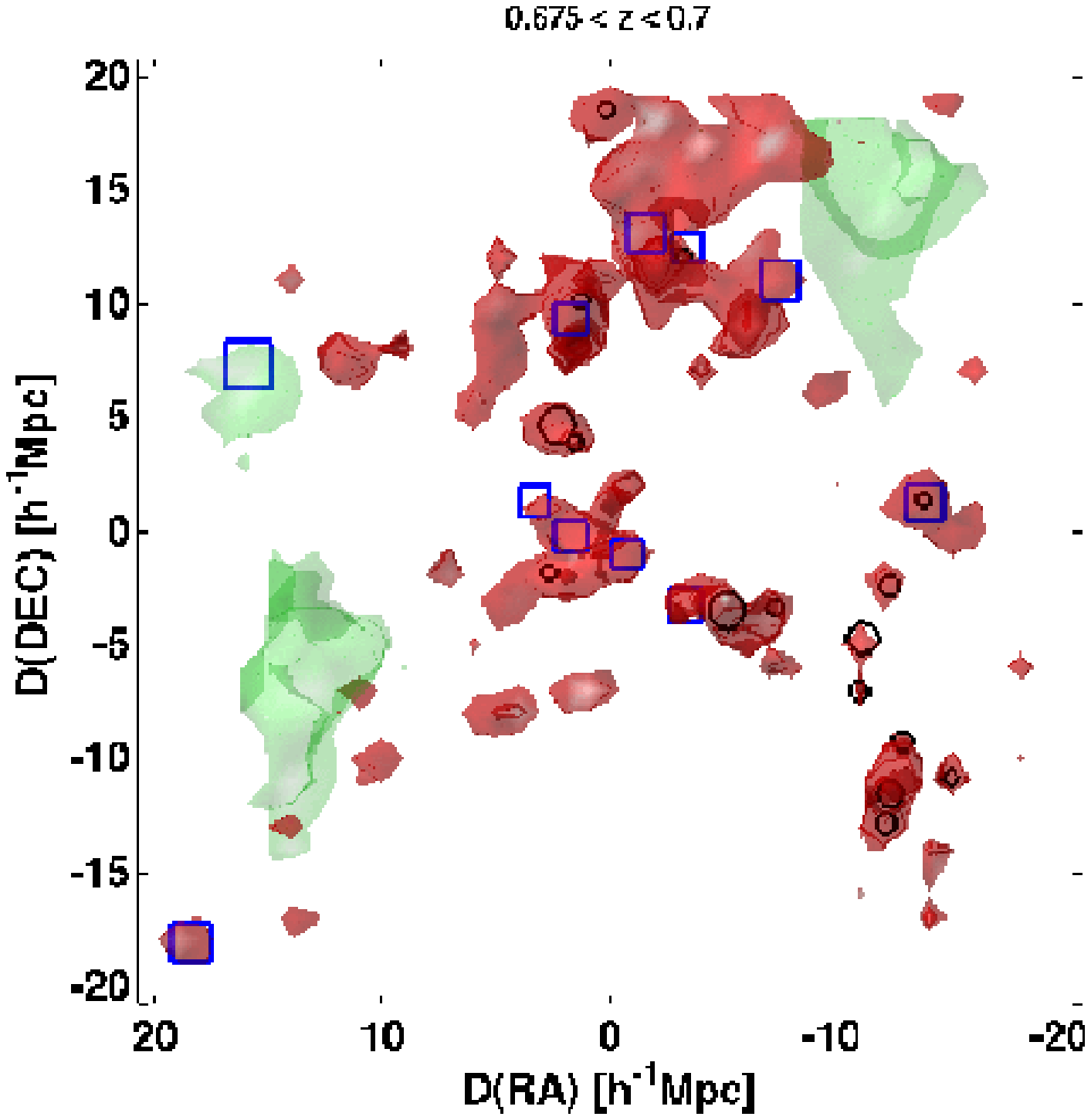}
\includegraphics[width=0.3\textwidth]{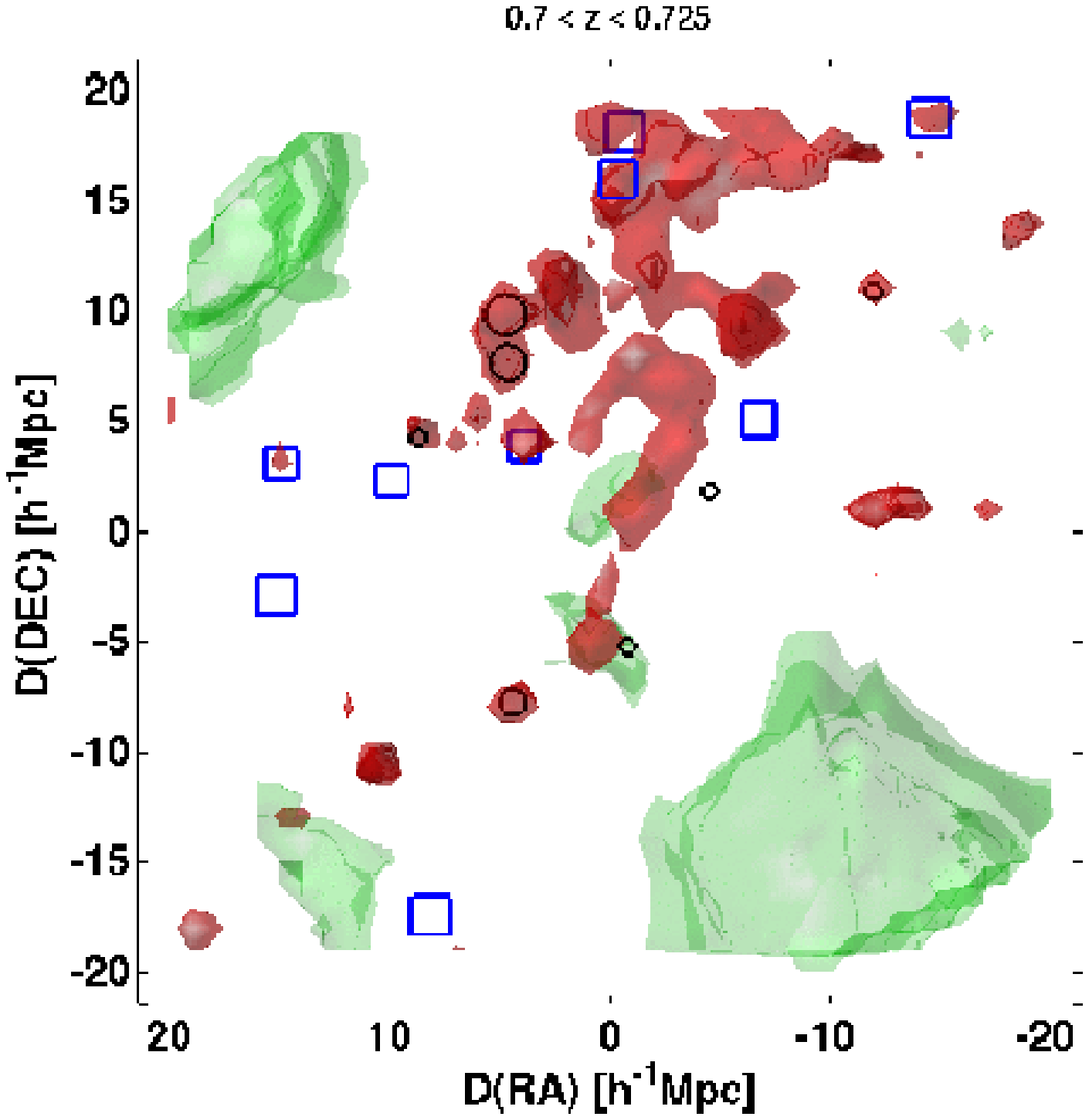}
\includegraphics[width=0.3\textwidth]{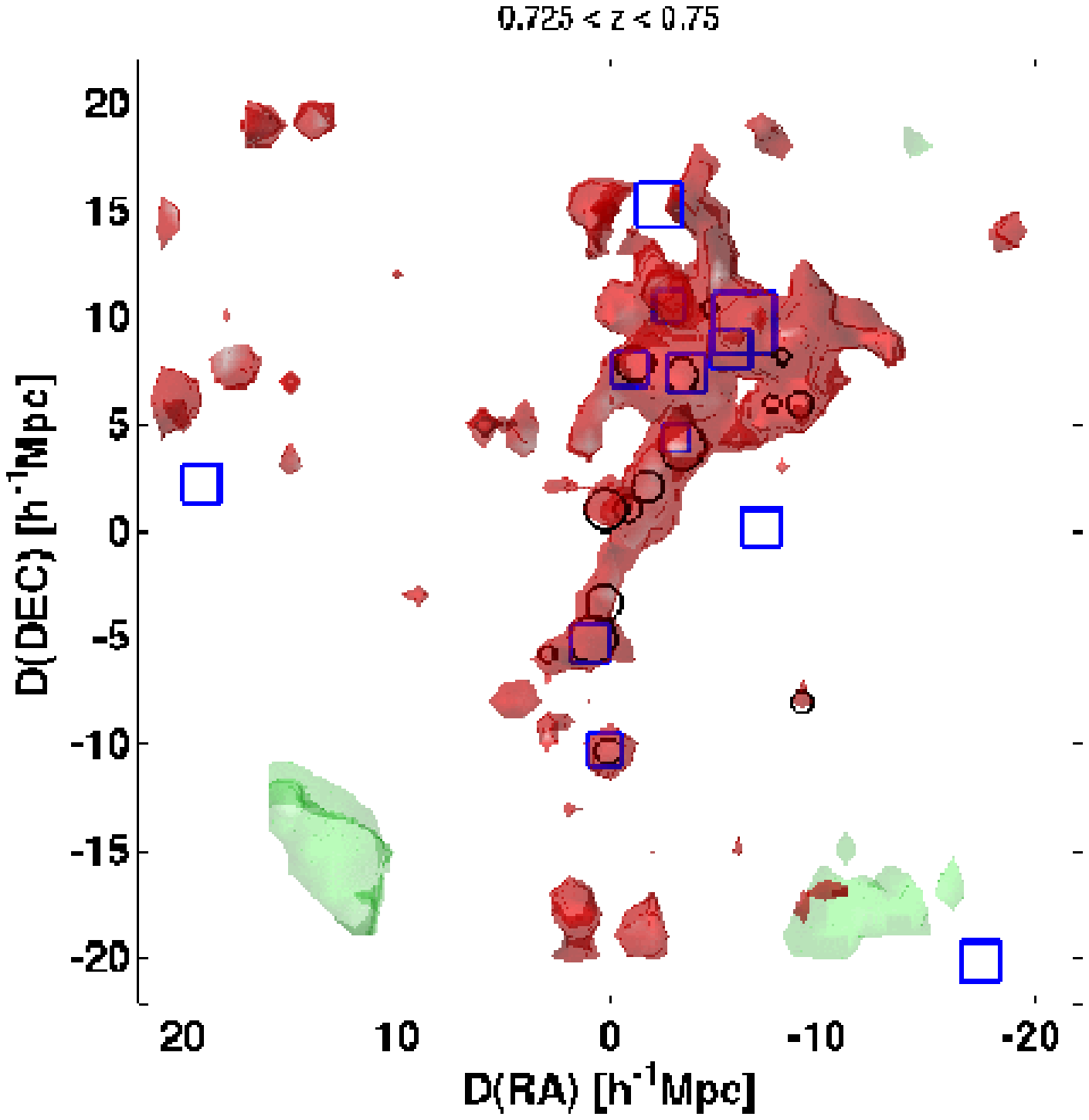}
\includegraphics[width=0.3\textwidth]{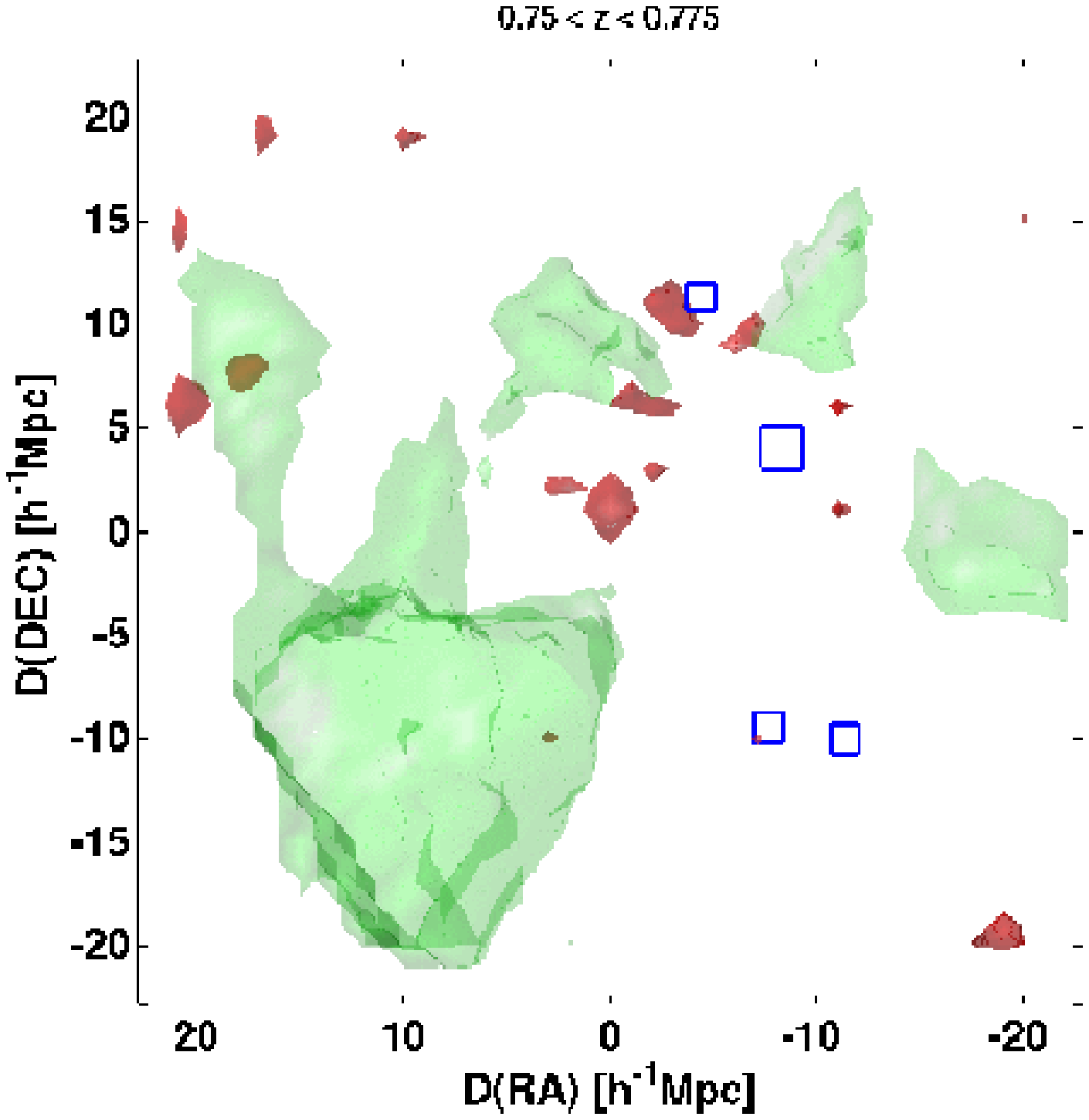}
\includegraphics[width=0.3\textwidth]{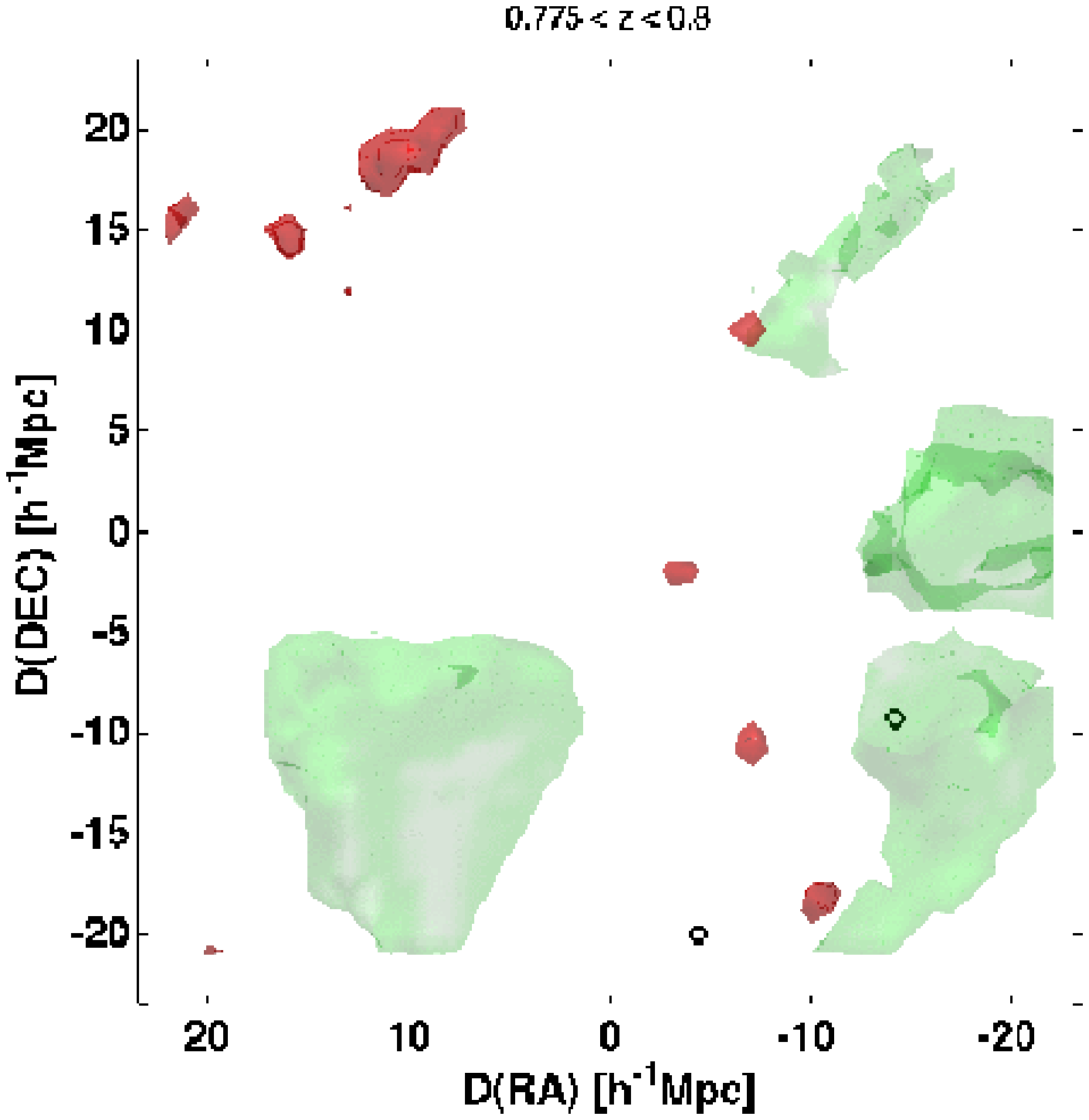}
\includegraphics[width=0.3\textwidth]{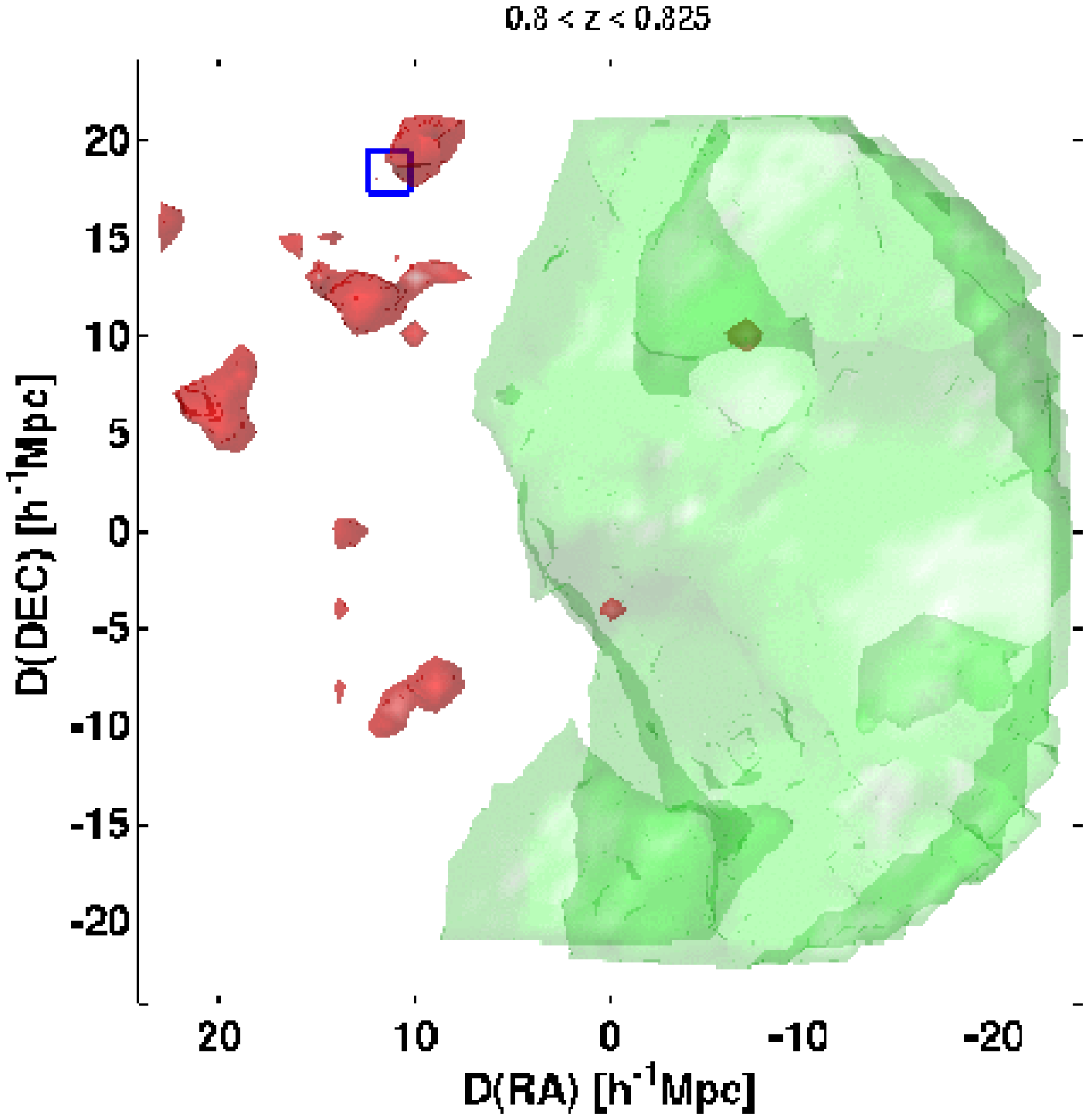}
\includegraphics[width=0.3\textwidth]{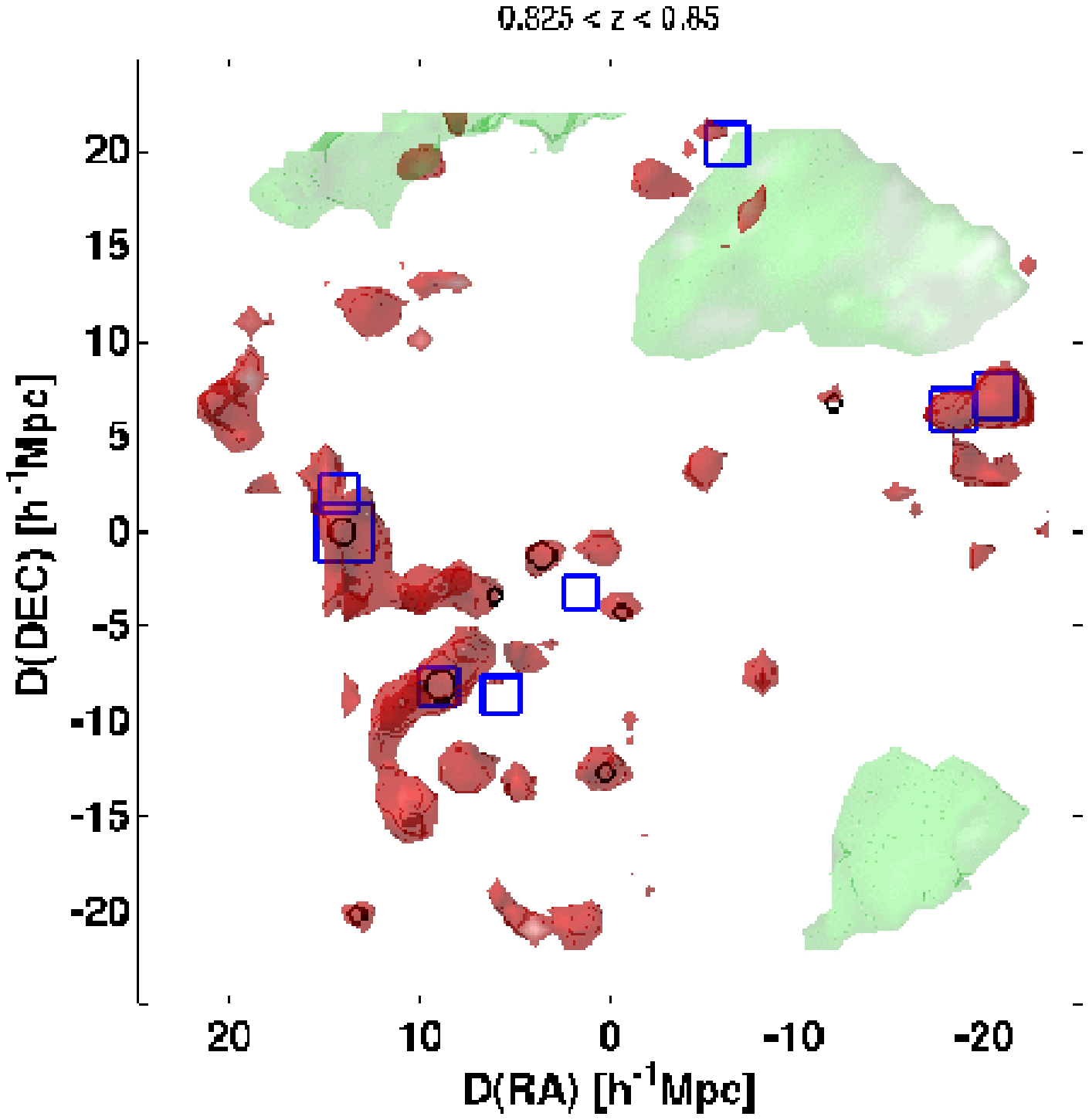}
\includegraphics[width=0.3\textwidth]{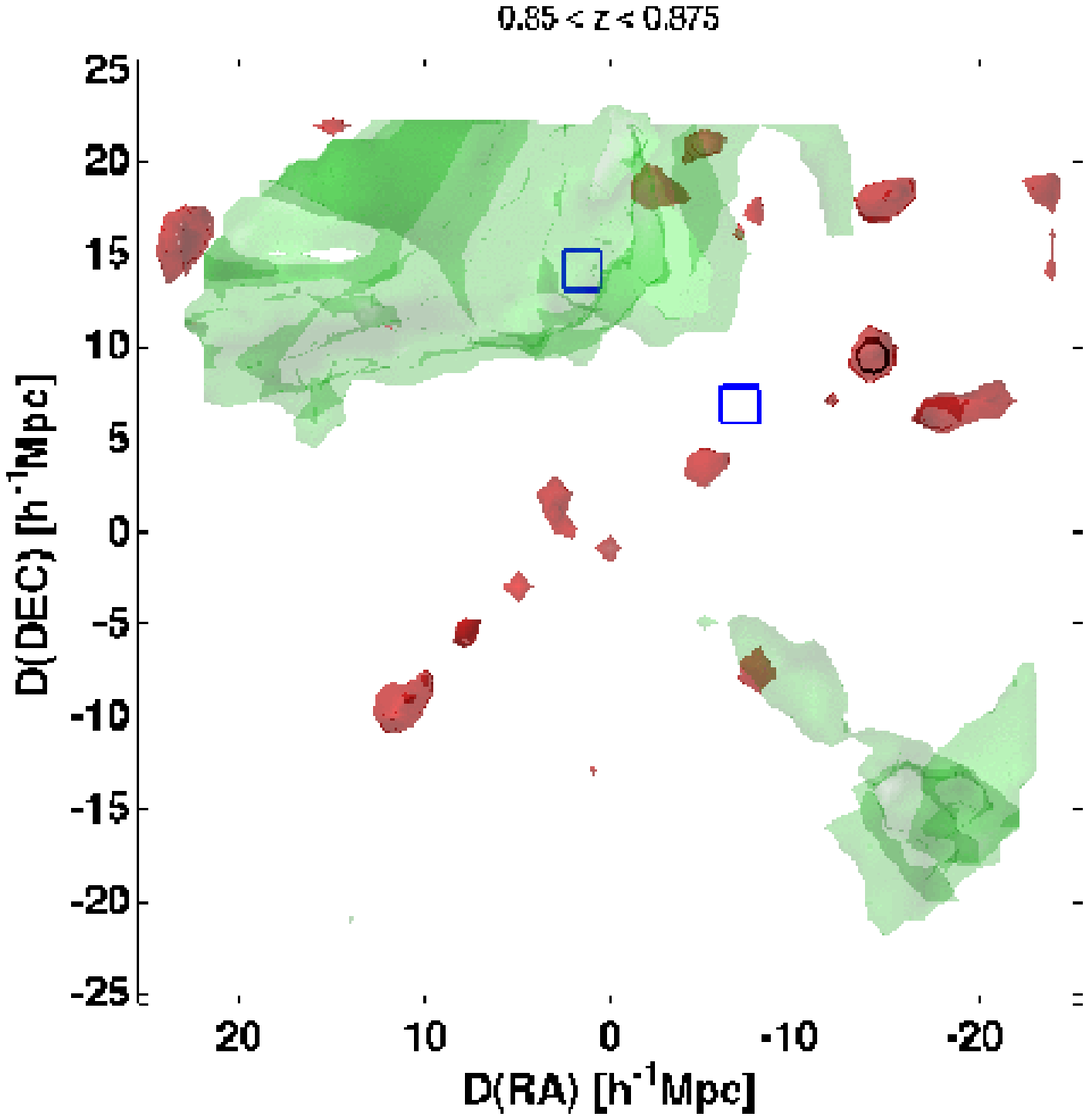}
\caption{Continued.}
\end{figure}

\clearpage

\addtocounter{figure}{-1}
\begin{figure}
\centering
\includegraphics[width=0.3\textwidth]{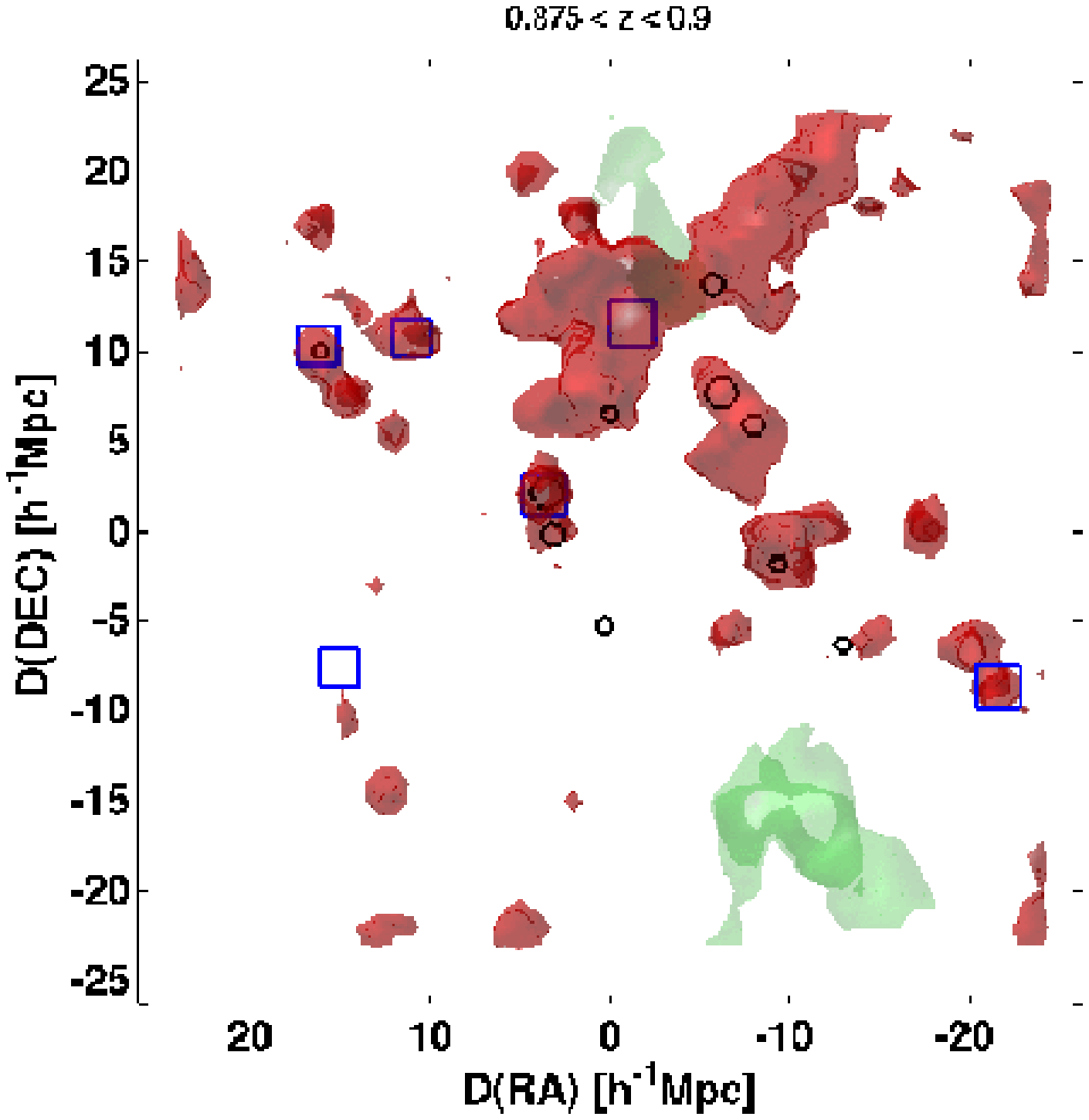}
\includegraphics[width=0.3\textwidth]{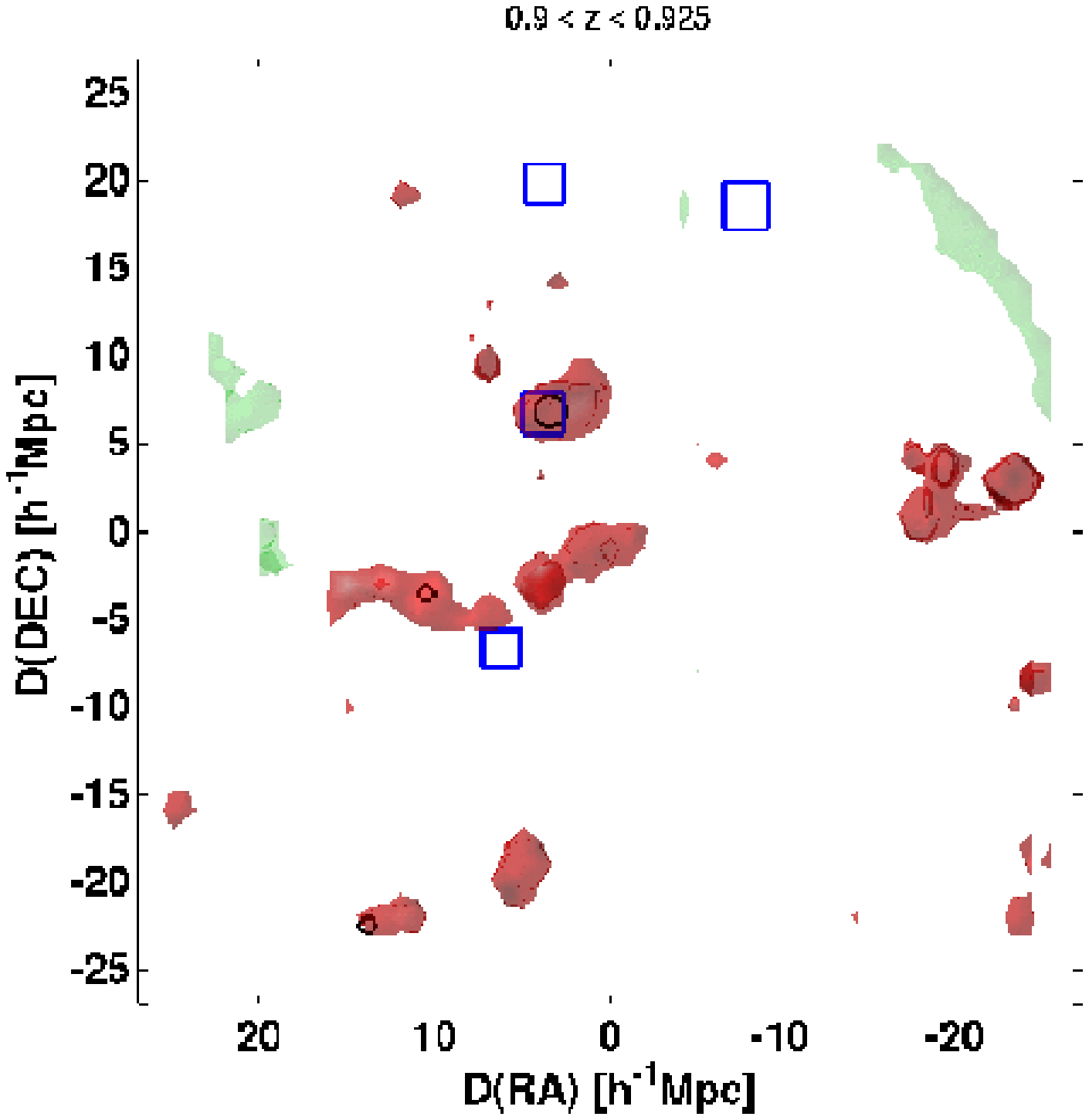}
\includegraphics[width=0.3\textwidth]{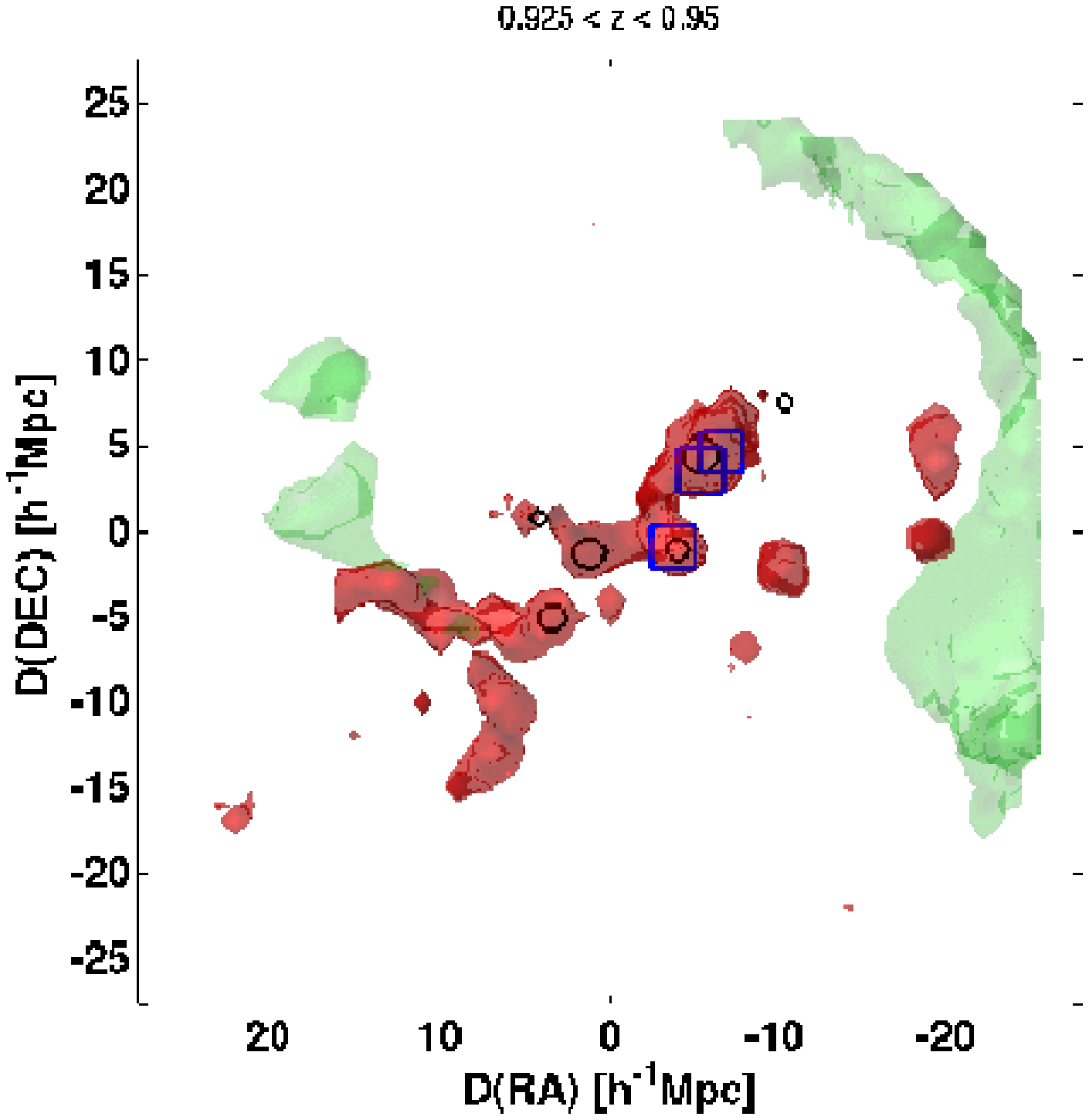}
\includegraphics[width=0.3\textwidth]{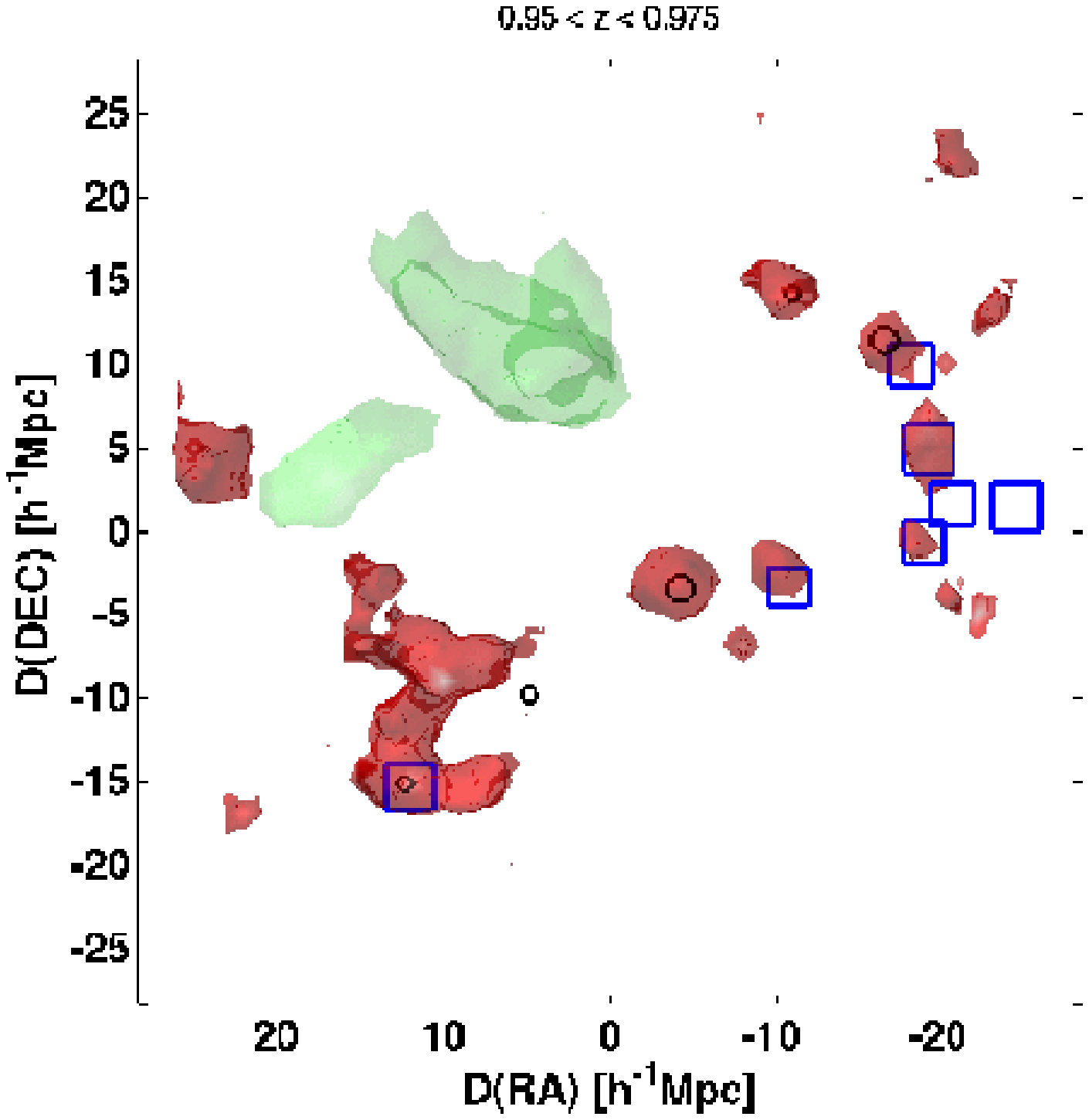}
\includegraphics[width=0.3\textwidth]{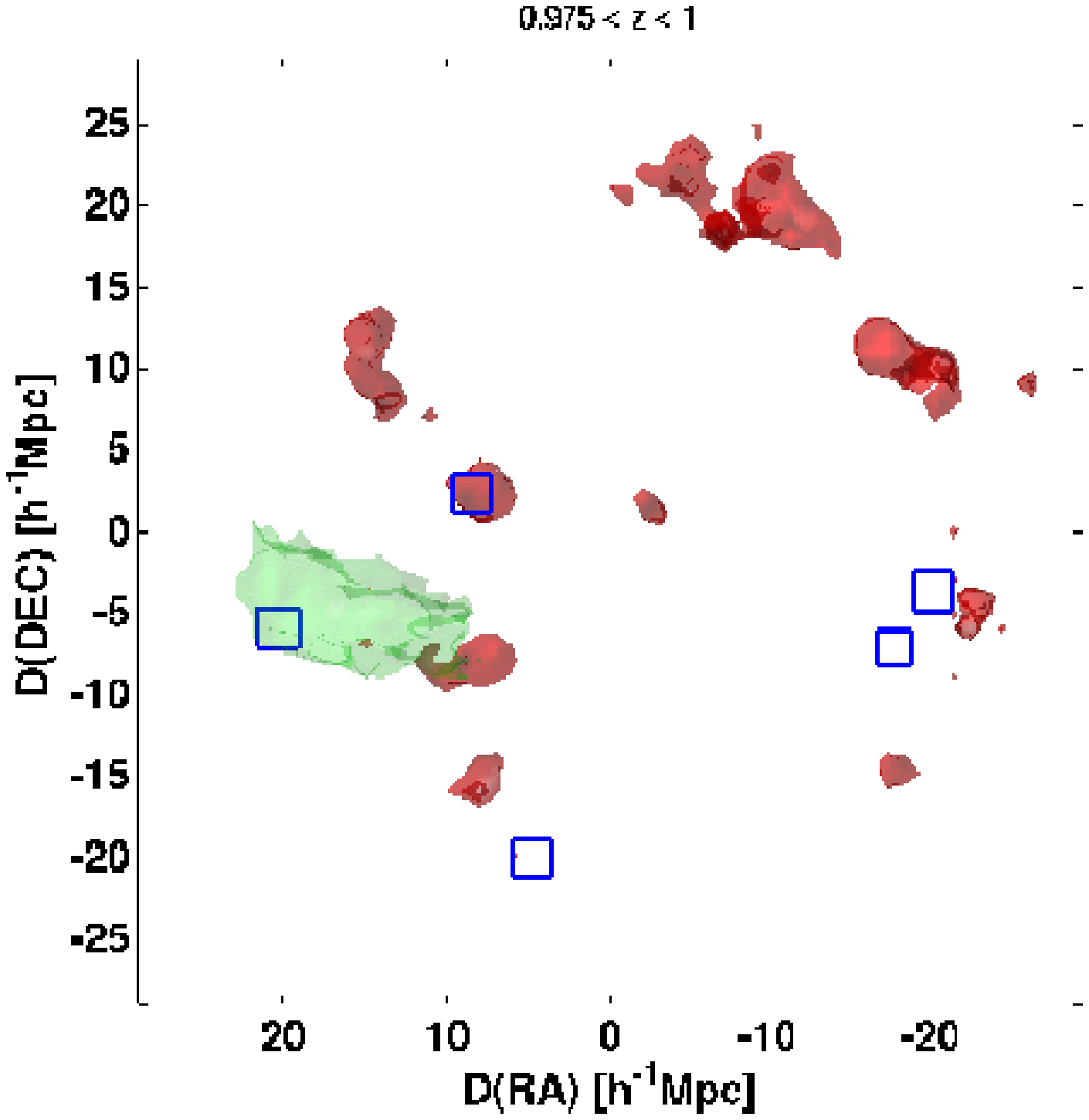}
\caption{Continued.}
\end{figure}

\clearpage

\begin{figure}
\includegraphics[width=0.19\textwidth]{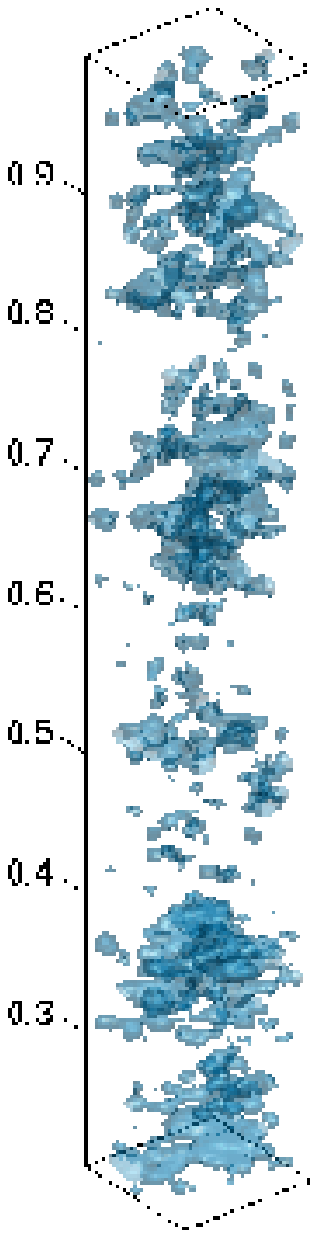}
\includegraphics[width=0.19\textwidth]{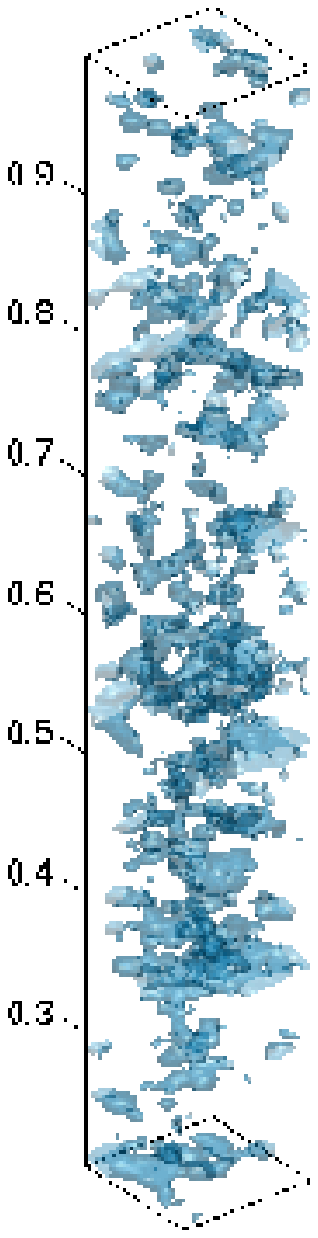}
\includegraphics[width=0.19\textwidth]{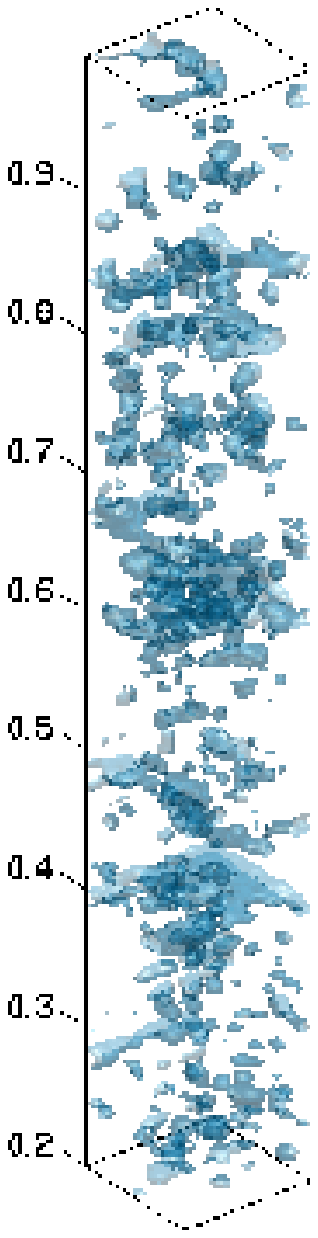}
\includegraphics[width=0.19\textwidth]{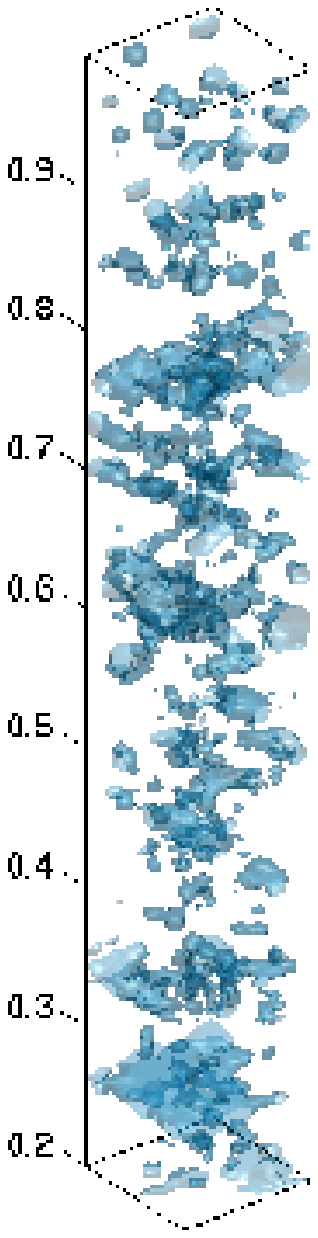}
\includegraphics[width=0.19\textwidth]{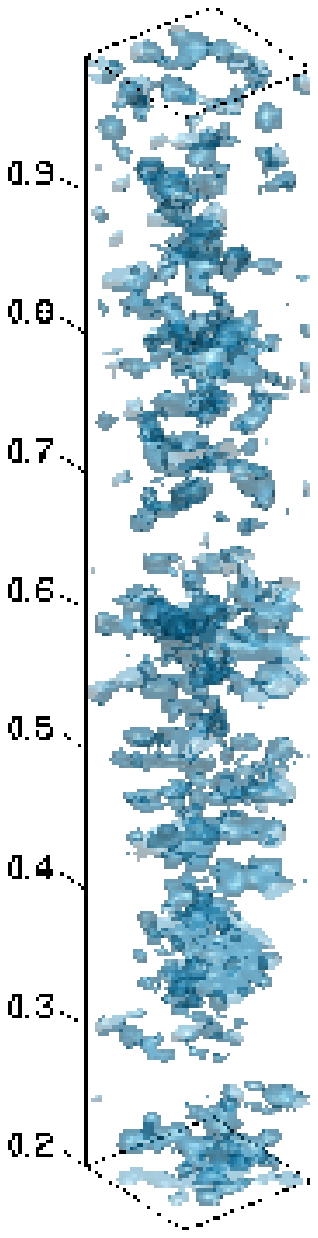}
\caption{\label{fig_overd_gridradez}Comparison of the $1+\delta_p=3$ isosurfaces in the overdensity field reconstructed using the 10k zCOSMOS sample of flux limited tracer galaxies and the equivalent sample of galaxies in the mock catalogues. First left hand figure: overdensity field reconstructed using the real 10k+30kZADE zCOSMOS galaxies. Rest of figures: overdensity field reconstructed using the mock 10k+30kZADE samples. In every point the overdensity field is reconstructed by counting the number of objects with an integrated ZADE-modified probability distribution in the apertures defined by the distance to the 10th nearest neighbour projected within $\pm$ 1000 \kms of the redshift of the grid point. The grid is regular in $\Delta RA = \Delta DEC = 2$ arcmin and $\Delta z = 0.002$. The overdensity values are presented without the edge correction. The high complexity of the cosmic web is noticeable in all figures. It appears that there are more structures on large scales ($\sim$ 10 \hh Mpc), visible above $z \sim 0.8$, in the real data than in the mock catalogues. To highlight the structures in the figures, we omit the axis. The transversal axis are RA and DEC, covering the zCOSMOS area $\sim$ 1 deg$^2$, and the vertical axis is redshift in the range $0.2<z<1$.}
\end{figure}

\clearpage

\addtocounter{figure}{-1}
\begin{figure}
\centering
\includegraphics[width=0.19\textwidth]{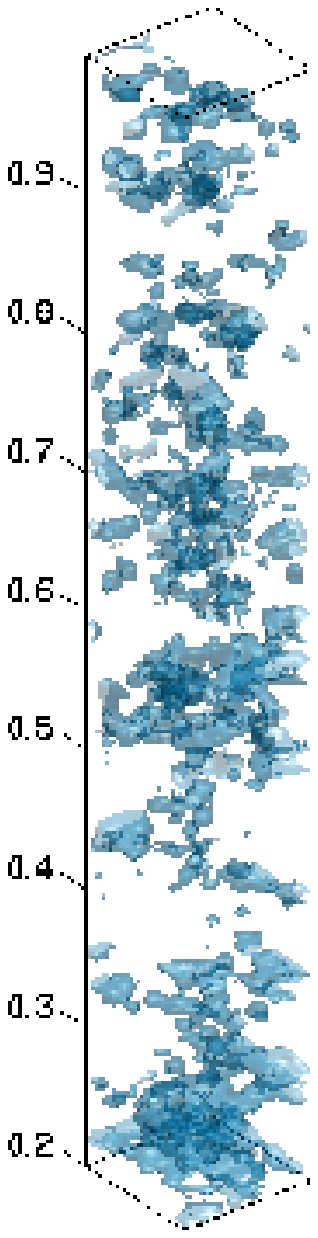}
\includegraphics[width=0.19\textwidth]{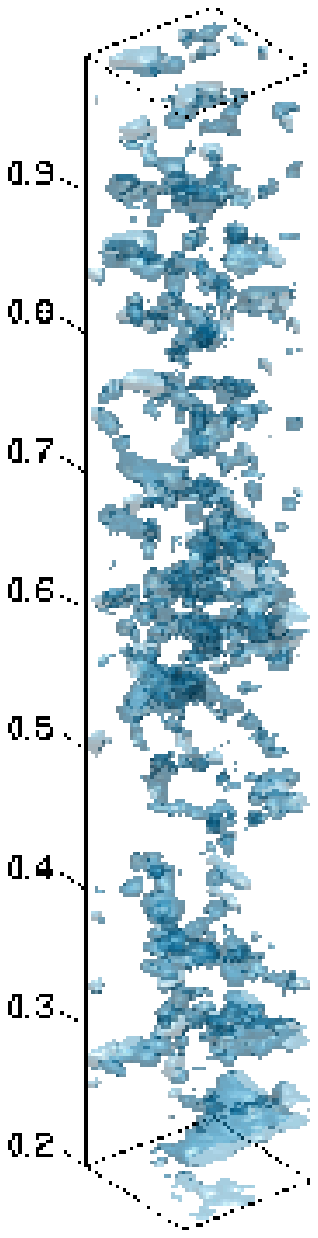}
\includegraphics[width=0.19\textwidth]{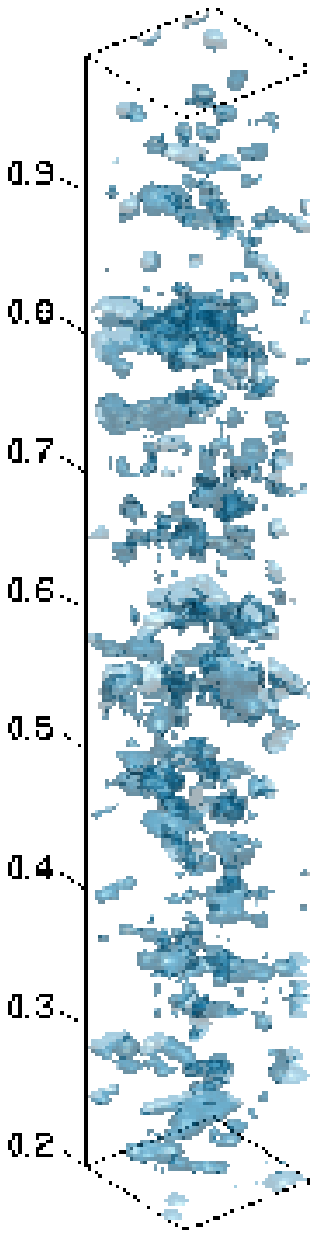}
\includegraphics[width=0.19\textwidth]{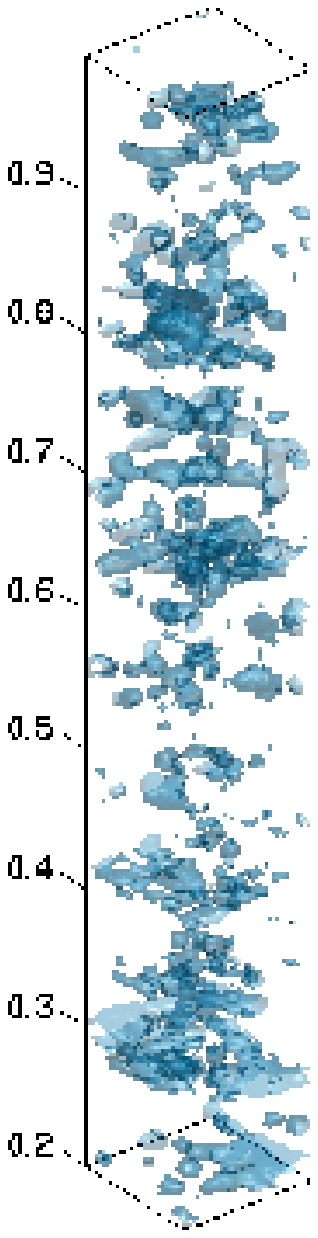}

\includegraphics[width=0.19\textwidth]{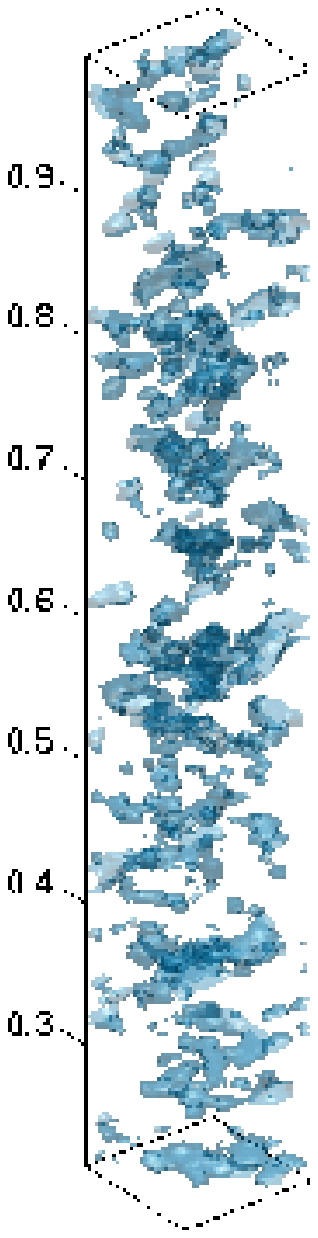}
\includegraphics[width=0.19\textwidth]{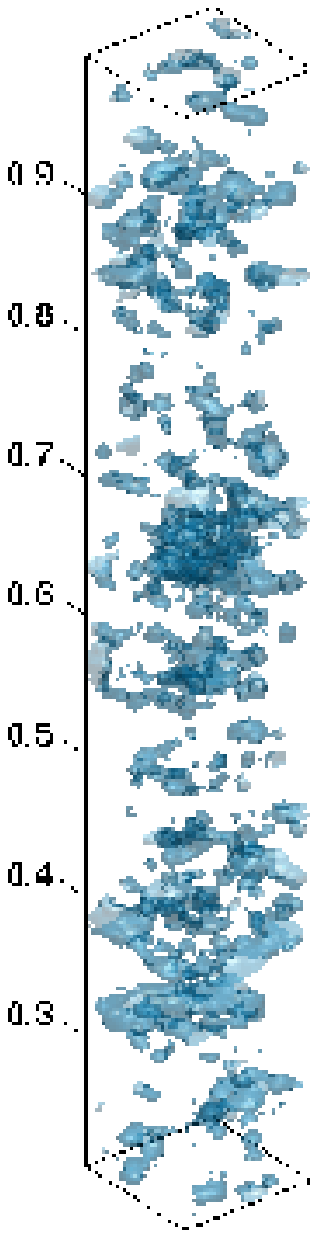}
\includegraphics[width=0.19\textwidth]{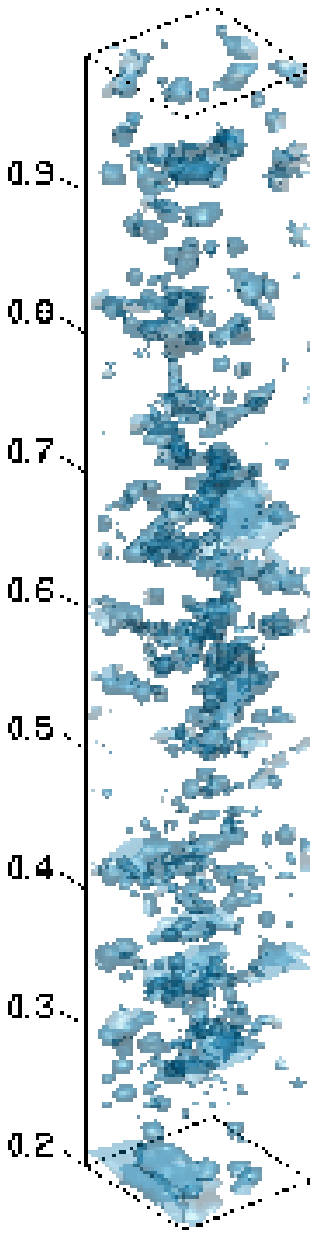}
\includegraphics[width=0.19\textwidth]{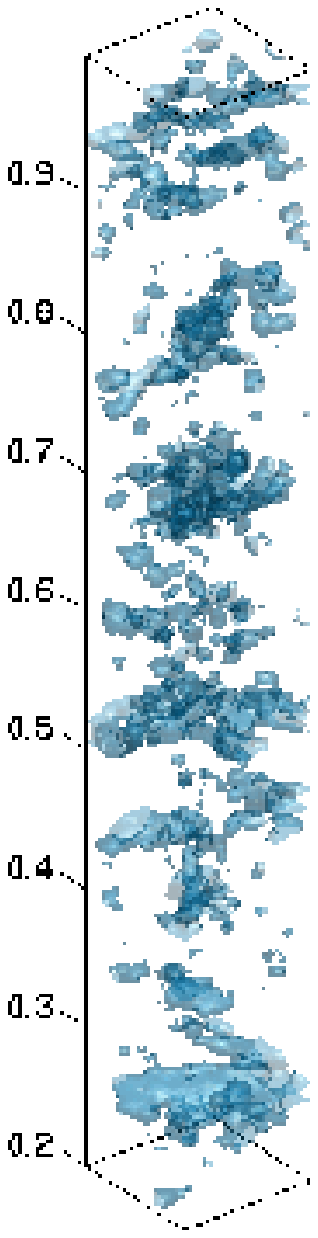}
\caption{Continued.}
\end{figure}

\begin{figure}
\centering
\includegraphics[width=0.43\textwidth]{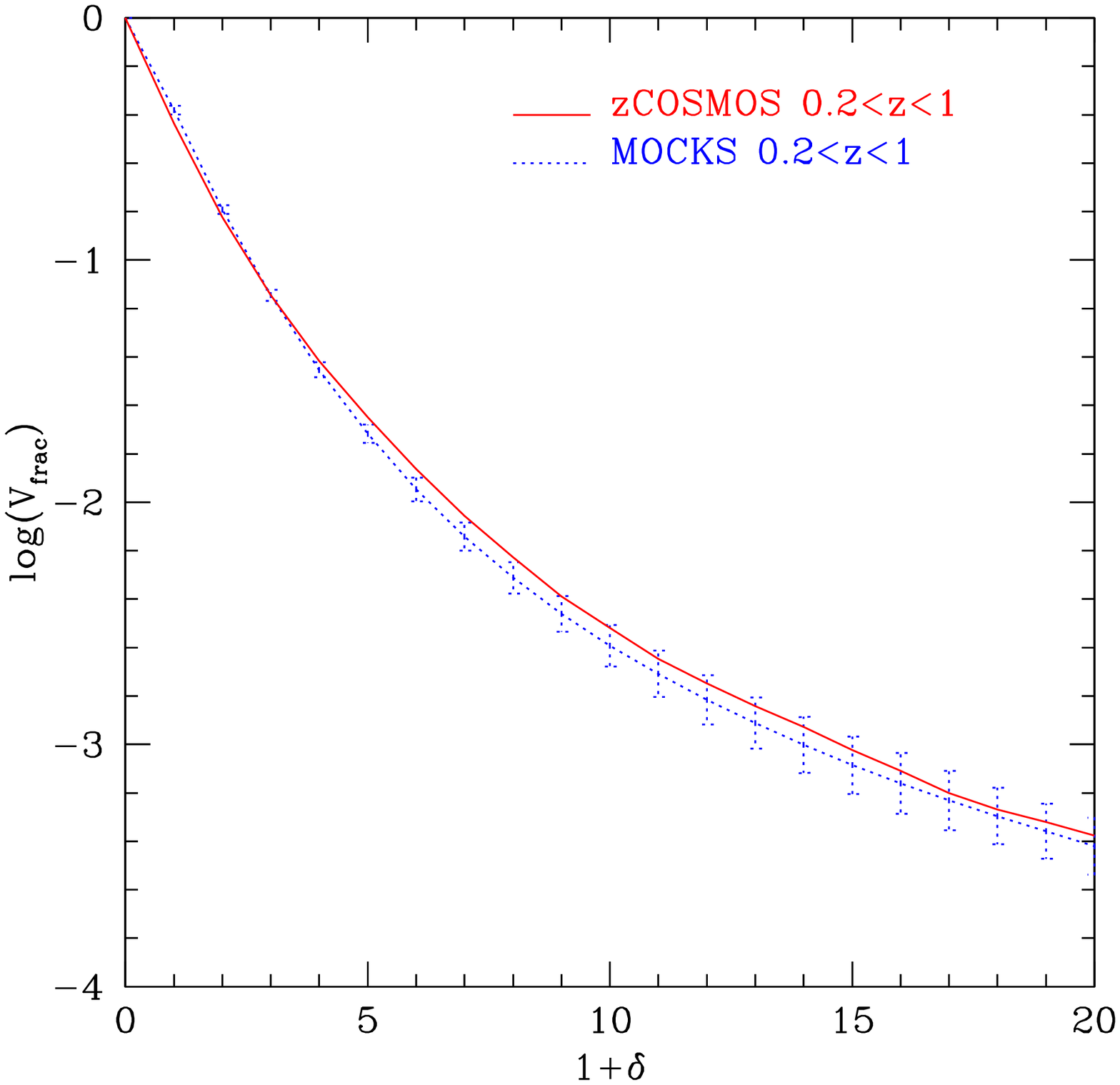}
\includegraphics[width=0.43\textwidth]{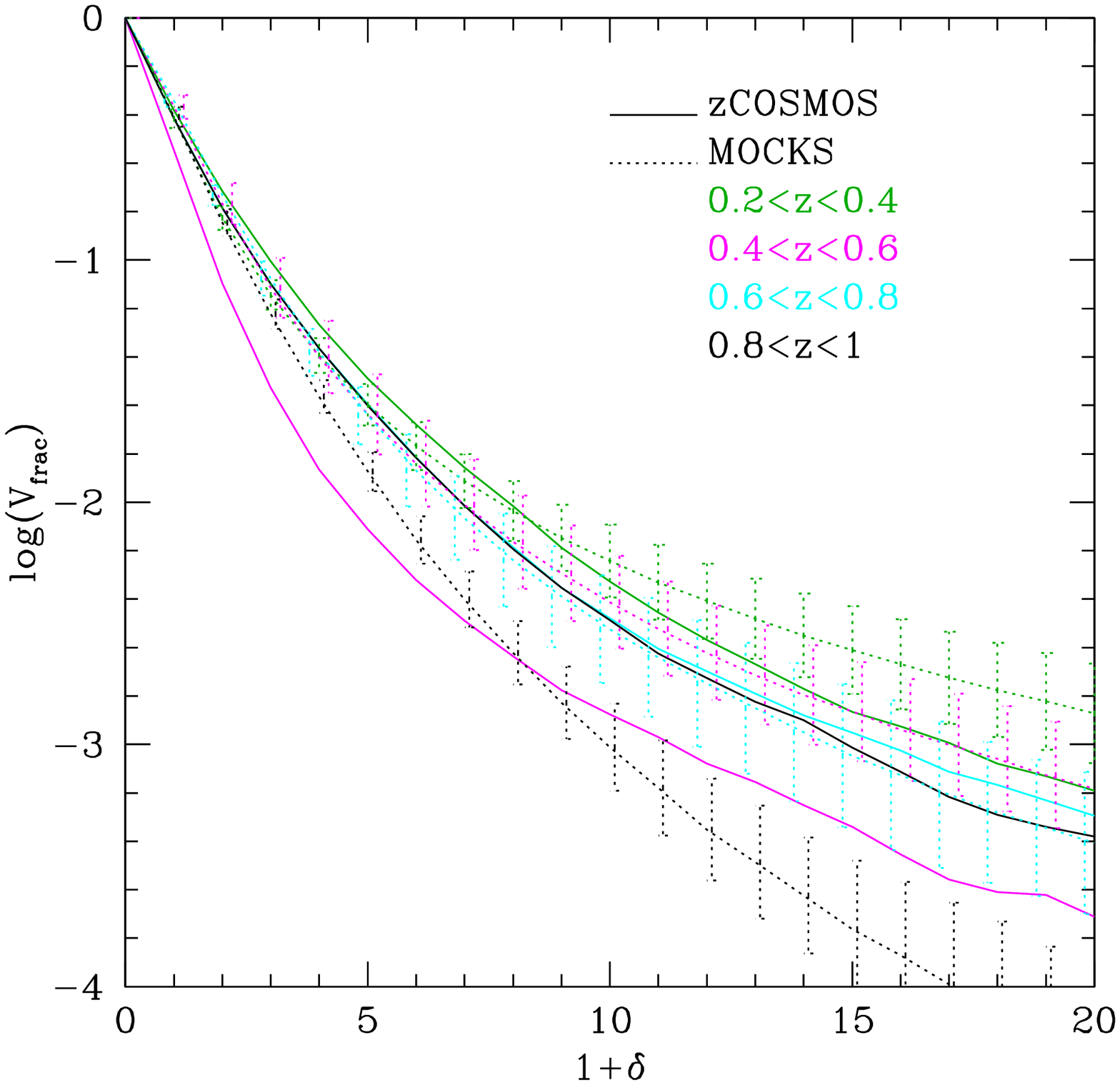}
\caption{\label{fig_volfracz0210}Fraction of the volume with the overdensity $\delta_p$ above a given value. The aperture in the overdensity field is defined using the distance to the 10th nearest neighbour projected to the redshift of the grid point within $\pm$ 1000 \kms.  A mock value is obtained by averaging results obtained from the individual 12 mock catalogues. The error is calculated as the standard deviations from the individual mock catalogue results. The continuous lines are for the real data, the dotted lines are for the corresponding mock catalogues. Volumes are estimated in [\hh Mpc]$^3$.
Left: Statistics in $0.2<z<1$.  The dashed red lines are for the real data, the continuous  blue lines are the corresponding mock catalogues. 
Right: Statistics in four redshift bins: $0.2<z<0.4$ (green), $0.4<z<0.6$ (magenta), $0.6<z<0.8$ (cyan) and $0.8<z<1$ (black).}
\end{figure}

\begin{figure}
\centering
\includegraphics[width=0.43\textwidth]{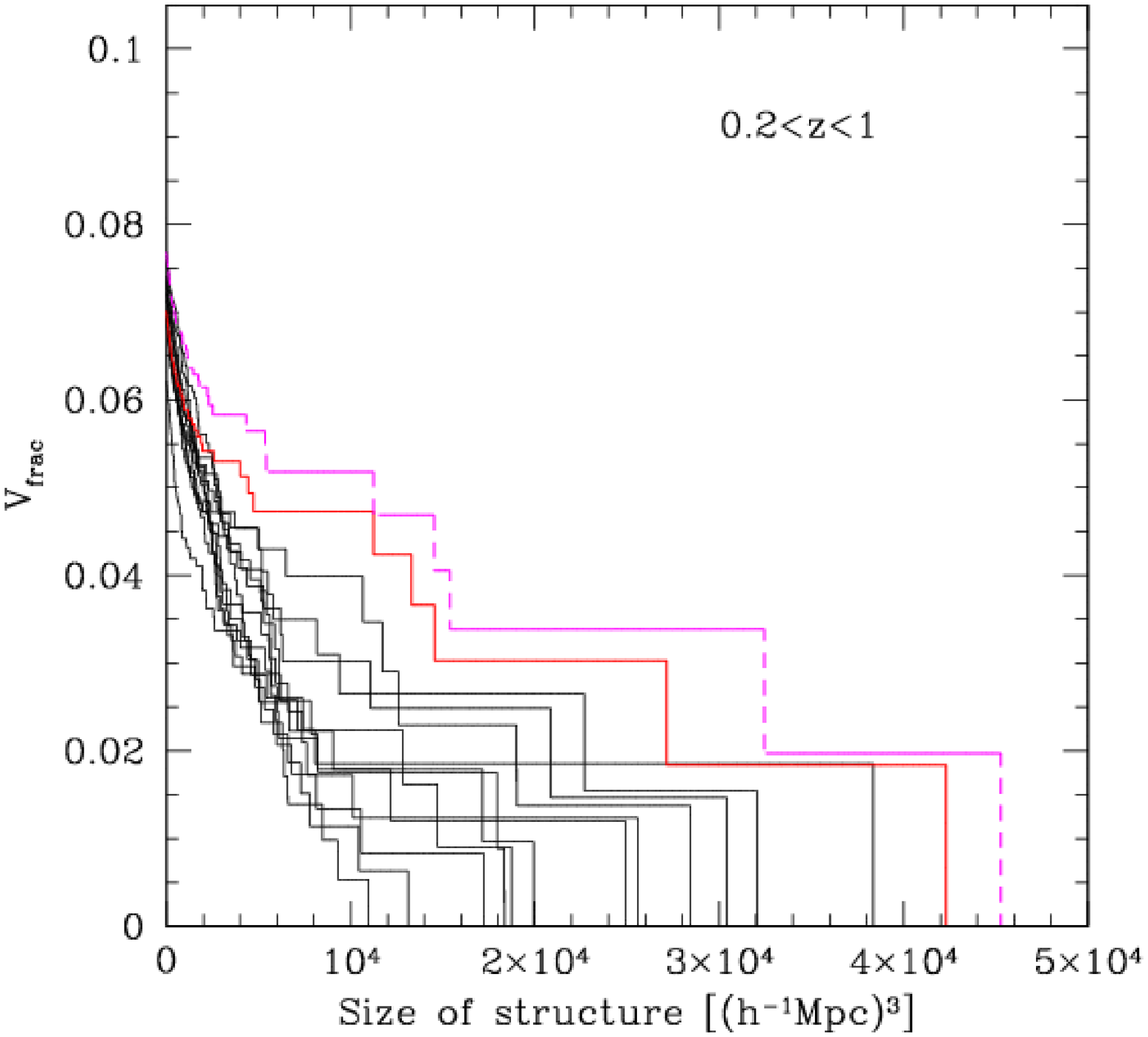}
\includegraphics[width=0.43\textwidth]{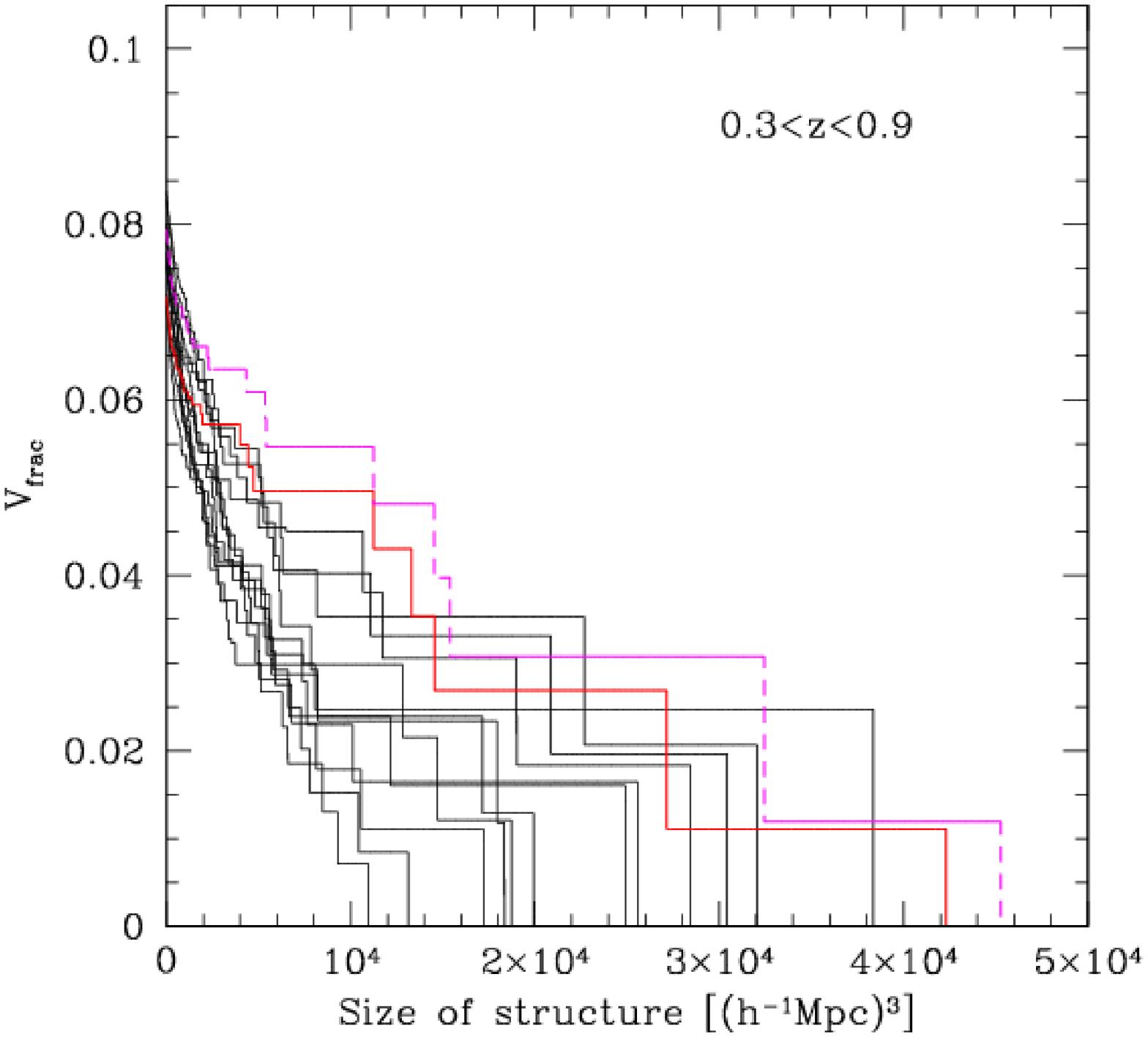}
\caption{\label{fig_volfracstruct}A comparison of the sizes of the structures above $1+\delta_p \ge 3$ in the data [red: $V/V_{max}$ smoothing to obtain N(z), magenta: smoothing equivalent to the smoothing applied to the mock N(z)] and in the individual mock catalogues (black curves). The curves correspond to the volume fractions contained within the structures of at least the size indicated on the x-axis, where size is measured in [\hh Mpc]$^3$. There is not a single mock catalogue which contains as much volume as the real data in the large structures in $0.2<z<1$. When we limit our statistics to $0.3<z<0.9$, the data results falls within the statistics outlined by the mocks. Difference between the data and the mocks at the large sizes is dominated by the structure at $z \sim 0.9$.}
\end{figure}

\end{document}